\documentclass[longauth]{aa}

\usepackage{graphicx}
\usepackage{natbib}
\usepackage{scalerel}
\usepackage{comment}
\usepackage{adjustbox}
\usepackage{booktabs}
\usepackage{ulem}
\usepackage{siunitx}

\usepackage[table]{xcolor}

\usepackage{txfonts}

\usepackage{hyperref}
\hypersetup{
    colorlinks=true,
    linkcolor=blue,
    filecolor=magenta,      
    urlcolor=blue,
    citecolor=blue
}
\makeatletter
\renewcommand*\aa@pageof{, page \thepage{} of \pageref*{LastPage}}
\makeatother

\usepackage[utf8]{inputenc}

\usepackage[switch, modulo]{lineno}

\usepackage{euclid}

\newcommand{\xic}{\xi_{\rm m}}
\newcommand{\xit}{\xi_{\rm cc}}
\newcommand{\xin}{\xi_{\rm nn}}
\newcommand{\xil}{\xi_{\ell\ell}}
\newcommand{\xitn}{\xi_{\rm cn}}
\newcommand{\xitl}{\xi_{\rm c\ell}}
\newcommand{\xiln}{\xi_{\ell \rm n}}

\newcommand{\rt}{R_{\rm c}}
\newcommand{\rn}{R_{\rm n}}
\newcommand{\rl}{R_{\ell}}
\newcommand{\rc}{R_{\rm m}}

\newcommand{\fc}{f_{\rm tot}}

\newcommand{\bx}{\mathbf{x}}
\newcommand{\br}{\mathbf{r}}

\newcommand{\aabf}[1]{\textcolor{magenta}{#1}}
\renewcommand{\aabf}[1]{#1}

\begin{document}

\title{\Euclid preparation}
\subtitle{
LXXXIII. 
The impact of redshift interlopers on the two-point correlation function analysis}

\newcommand{\orcid}[1]{}                           
\author{Euclid Collaboration: I.~Risso\orcid{0000-0003-2525-7761}\thanks{\email{ilaria.risso@ge.infn.it}}\inst{\ref{aff1},\ref{aff2}}
\and A.~Veropalumbo\orcid{0000-0003-2387-1194}\inst{\ref{aff1},\ref{aff2},\ref{aff3}}
\and E.~Branchini\orcid{0000-0002-0808-6908}\inst{\ref{aff3},\ref{aff2},\ref{aff1}}
\and E.~Maragliano\orcid{0009-0009-6004-4156}\inst{\ref{aff3},\ref{aff2}}
\and S.~de~la~Torre\inst{\ref{aff4}}
\and E.~Sarpa\orcid{0000-0002-1256-655X}\inst{\ref{aff5},\ref{aff6},\ref{aff7}}
\and P.~Monaco\orcid{0000-0003-2083-7564}\inst{\ref{aff8},\ref{aff9},\ref{aff7},\ref{aff10}}
\and B.~R.~Granett\orcid{0000-0003-2694-9284}\inst{\ref{aff1}}
\and S.~Lee\orcid{0000-0002-8289-740X}\inst{\ref{aff11}}
\and G.~E.~Addison\orcid{0000-0002-2147-2248}\inst{\ref{aff12}}
\and S.~Bruton\orcid{0000-0002-6503-5218}\inst{\ref{aff13}}
\and C.~Carbone\orcid{0000-0003-0125-3563}\inst{\ref{aff14}}
\and G.~Lavaux\orcid{0000-0003-0143-8891}\inst{\ref{aff15}}
\and K.~Markovic\orcid{0000-0001-6764-073X}\inst{\ref{aff11}}
\and K.~McCarthy\orcid{0000-0001-6857-018X}\inst{\ref{aff11}}
\and G.~Parimbelli\orcid{0000-0002-2539-2472}\inst{\ref{aff16},\ref{aff17},\ref{aff5}}
\and F.~Passalacqua\orcid{0000-0002-8606-4093}\inst{\ref{aff18},\ref{aff19}}
\and W.~J.~Percival\orcid{0000-0002-0644-5727}\inst{\ref{aff20},\ref{aff21},\ref{aff22}}
\and C.~Scarlata\orcid{0000-0002-9136-8876}\inst{\ref{aff23}}
\and E.~Sefusatti\orcid{0000-0003-0473-1567}\inst{\ref{aff9},\ref{aff10},\ref{aff7}}
\and Y.~Wang\orcid{0000-0002-4749-2984}\inst{\ref{aff24}}
\and M.~Bonici\orcid{0000-0002-8430-126X}\inst{\ref{aff20},\ref{aff14}}
\and F.~Oppizzi\orcid{0000-0003-3904-8370}\inst{\ref{aff2}}
\and N.~Aghanim\orcid{0000-0002-6688-8992}\inst{\ref{aff25}}
\and B.~Altieri\orcid{0000-0003-3936-0284}\inst{\ref{aff26}}
\and A.~Amara\inst{\ref{aff27}}
\and S.~Andreon\orcid{0000-0002-2041-8784}\inst{\ref{aff1}}
\and N.~Auricchio\orcid{0000-0003-4444-8651}\inst{\ref{aff28}}
\and C.~Baccigalupi\orcid{0000-0002-8211-1630}\inst{\ref{aff10},\ref{aff9},\ref{aff7},\ref{aff5}}
\and M.~Baldi\orcid{0000-0003-4145-1943}\inst{\ref{aff29},\ref{aff28},\ref{aff30}}
\and A.~Balestra\orcid{0000-0002-6967-261X}\inst{\ref{aff31}}
\and S.~Bardelli\orcid{0000-0002-8900-0298}\inst{\ref{aff28}}
\and P.~Battaglia\orcid{0000-0002-7337-5909}\inst{\ref{aff28}}
\and A.~Biviano\orcid{0000-0002-0857-0732}\inst{\ref{aff9},\ref{aff10}}
\and A.~Bonchi\orcid{0000-0002-2667-5482}\inst{\ref{aff32}}
\and D.~Bonino\orcid{0000-0002-3336-9977}\inst{\ref{aff33}}
\and M.~Brescia\orcid{0000-0001-9506-5680}\inst{\ref{aff34},\ref{aff35}}
\and J.~Brinchmann\orcid{0000-0003-4359-8797}\inst{\ref{aff36},\ref{aff37}}
\and S.~Camera\orcid{0000-0003-3399-3574}\inst{\ref{aff38},\ref{aff39},\ref{aff33}}
\and G.~Ca\~nas-Herrera\orcid{0000-0003-2796-2149}\inst{\ref{aff40},\ref{aff41},\ref{aff42}}
\and V.~Capobianco\orcid{0000-0002-3309-7692}\inst{\ref{aff33}}
\and V.~F.~Cardone\inst{\ref{aff43},\ref{aff44}}
\and J.~Carretero\orcid{0000-0002-3130-0204}\inst{\ref{aff45},\ref{aff46}}
\and S.~Casas\orcid{0000-0002-4751-5138}\inst{\ref{aff47}}
\and M.~Castellano\orcid{0000-0001-9875-8263}\inst{\ref{aff43}}
\and G.~Castignani\orcid{0000-0001-6831-0687}\inst{\ref{aff28}}
\and S.~Cavuoti\orcid{0000-0002-3787-4196}\inst{\ref{aff35},\ref{aff48}}
\and K.~C.~Chambers\orcid{0000-0001-6965-7789}\inst{\ref{aff49}}
\and A.~Cimatti\inst{\ref{aff50}}
\and C.~Colodro-Conde\inst{\ref{aff51}}
\and G.~Congedo\orcid{0000-0003-2508-0046}\inst{\ref{aff52}}
\and C.~J.~Conselice\orcid{0000-0003-1949-7638}\inst{\ref{aff53}}
\and L.~Conversi\orcid{0000-0002-6710-8476}\inst{\ref{aff54},\ref{aff26}}
\and Y.~Copin\orcid{0000-0002-5317-7518}\inst{\ref{aff55}}
\and F.~Courbin\orcid{0000-0003-0758-6510}\inst{\ref{aff56},\ref{aff57}}
\and H.~M.~Courtois\orcid{0000-0003-0509-1776}\inst{\ref{aff58}}
\and M.~Crocce\orcid{0000-0002-9745-6228}\inst{\ref{aff16},\ref{aff59}}
\and A.~Da~Silva\orcid{0000-0002-6385-1609}\inst{\ref{aff60},\ref{aff61}}
\and H.~Degaudenzi\orcid{0000-0002-5887-6799}\inst{\ref{aff62}}
\and G.~De~Lucia\orcid{0000-0002-6220-9104}\inst{\ref{aff9}}
\and A.~M.~Di~Giorgio\orcid{0000-0002-4767-2360}\inst{\ref{aff63}}
\and H.~Dole\orcid{0000-0002-9767-3839}\inst{\ref{aff25}}
\and M.~Douspis\orcid{0000-0003-4203-3954}\inst{\ref{aff25}}
\and F.~Dubath\orcid{0000-0002-6533-2810}\inst{\ref{aff62}}
\and C.~A.~J.~Duncan\orcid{0009-0003-3573-0791}\inst{\ref{aff53}}
\and X.~Dupac\inst{\ref{aff26}}
\and S.~Dusini\orcid{0000-0002-1128-0664}\inst{\ref{aff19}}
\and S.~Escoffier\orcid{0000-0002-2847-7498}\inst{\ref{aff64}}
\and M.~Farina\orcid{0000-0002-3089-7846}\inst{\ref{aff63}}
\and R.~Farinelli\inst{\ref{aff28}}
\and F.~Faustini\orcid{0000-0001-6274-5145}\inst{\ref{aff43},\ref{aff32}}
\and S.~Ferriol\inst{\ref{aff55}}
\and F.~Finelli\orcid{0000-0002-6694-3269}\inst{\ref{aff28},\ref{aff65}}
\and S.~Fotopoulou\orcid{0000-0002-9686-254X}\inst{\ref{aff66}}
\and N.~Fourmanoit\orcid{0009-0005-6816-6925}\inst{\ref{aff64}}
\and M.~Frailis\orcid{0000-0002-7400-2135}\inst{\ref{aff9}}
\and E.~Franceschi\orcid{0000-0002-0585-6591}\inst{\ref{aff28}}
\and M.~Fumana\orcid{0000-0001-6787-5950}\inst{\ref{aff14}}
\and S.~Galeotta\orcid{0000-0002-3748-5115}\inst{\ref{aff9}}
\and K.~George\orcid{0000-0002-1734-8455}\inst{\ref{aff67}}
\and W.~Gillard\orcid{0000-0003-4744-9748}\inst{\ref{aff64}}
\and B.~Gillis\orcid{0000-0002-4478-1270}\inst{\ref{aff52}}
\and C.~Giocoli\orcid{0000-0002-9590-7961}\inst{\ref{aff28},\ref{aff30}}
\and J.~Gracia-Carpio\inst{\ref{aff68}}
\and A.~Grazian\orcid{0000-0002-5688-0663}\inst{\ref{aff31}}
\and F.~Grupp\inst{\ref{aff68},\ref{aff67}}
\and L.~Guzzo\orcid{0000-0001-8264-5192}\inst{\ref{aff69},\ref{aff1},\ref{aff70}}
\and S.~V.~H.~Haugan\orcid{0000-0001-9648-7260}\inst{\ref{aff71}}
\and W.~Holmes\inst{\ref{aff11}}
\and F.~Hormuth\inst{\ref{aff72}}
\and A.~Hornstrup\orcid{0000-0002-3363-0936}\inst{\ref{aff73},\ref{aff74}}
\and P.~Hudelot\inst{\ref{aff15}}
\and K.~Jahnke\orcid{0000-0003-3804-2137}\inst{\ref{aff75}}
\and M.~Jhabvala\inst{\ref{aff76}}
\and B.~Joachimi\orcid{0000-0001-7494-1303}\inst{\ref{aff77}}
\and E.~Keih\"anen\orcid{0000-0003-1804-7715}\inst{\ref{aff78}}
\and S.~Kermiche\orcid{0000-0002-0302-5735}\inst{\ref{aff64}}
\and A.~Kiessling\orcid{0000-0002-2590-1273}\inst{\ref{aff11}}
\and M.~Kilbinger\orcid{0000-0001-9513-7138}\inst{\ref{aff79}}
\and B.~Kubik\orcid{0009-0006-5823-4880}\inst{\ref{aff55}}
\and M.~K\"ummel\orcid{0000-0003-2791-2117}\inst{\ref{aff67}}
\and M.~Kunz\orcid{0000-0002-3052-7394}\inst{\ref{aff80}}
\and H.~Kurki-Suonio\orcid{0000-0002-4618-3063}\inst{\ref{aff81},\ref{aff82}}
\and A.~M.~C.~Le~Brun\orcid{0000-0002-0936-4594}\inst{\ref{aff83}}
\and P.~Liebing\inst{\ref{aff84}}
\and S.~Ligori\orcid{0000-0003-4172-4606}\inst{\ref{aff33}}
\and P.~B.~Lilje\orcid{0000-0003-4324-7794}\inst{\ref{aff71}}
\and V.~Lindholm\orcid{0000-0003-2317-5471}\inst{\ref{aff81},\ref{aff82}}
\and I.~Lloro\orcid{0000-0001-5966-1434}\inst{\ref{aff85}}
\and G.~Mainetti\orcid{0000-0003-2384-2377}\inst{\ref{aff86}}
\and D.~Maino\inst{\ref{aff69},\ref{aff14},\ref{aff70}}
\and E.~Maiorano\orcid{0000-0003-2593-4355}\inst{\ref{aff28}}
\and O.~Mansutti\orcid{0000-0001-5758-4658}\inst{\ref{aff9}}
\and S.~Marcin\inst{\ref{aff87}}
\and O.~Marggraf\orcid{0000-0001-7242-3852}\inst{\ref{aff88}}
\and M.~Martinelli\orcid{0000-0002-6943-7732}\inst{\ref{aff43},\ref{aff44}}
\and N.~Martinet\orcid{0000-0003-2786-7790}\inst{\ref{aff4}}
\and F.~Marulli\orcid{0000-0002-8850-0303}\inst{\ref{aff89},\ref{aff28},\ref{aff30}}
\and R.~Massey\orcid{0000-0002-6085-3780}\inst{\ref{aff90}}
\and S.~Maurogordato\inst{\ref{aff91}}
\and E.~Medinaceli\orcid{0000-0002-4040-7783}\inst{\ref{aff28}}
\and S.~Mei\orcid{0000-0002-2849-559X}\inst{\ref{aff92},\ref{aff93}}
\and M.~Melchior\inst{\ref{aff94}}
\and Y.~Mellier\inst{\ref{aff95},\ref{aff15}}
\and M.~Meneghetti\orcid{0000-0003-1225-7084}\inst{\ref{aff28},\ref{aff30}}
\and E.~Merlin\orcid{0000-0001-6870-8900}\inst{\ref{aff43}}
\and G.~Meylan\inst{\ref{aff96}}
\and A.~Mora\orcid{0000-0002-1922-8529}\inst{\ref{aff97}}
\and M.~Moresco\orcid{0000-0002-7616-7136}\inst{\ref{aff89},\ref{aff28}}
\and L.~Moscardini\orcid{0000-0002-3473-6716}\inst{\ref{aff89},\ref{aff28},\ref{aff30}}
\and R.~Nakajima\orcid{0009-0009-1213-7040}\inst{\ref{aff88}}
\and C.~Neissner\orcid{0000-0001-8524-4968}\inst{\ref{aff98},\ref{aff46}}
\and S.-M.~Niemi\orcid{0009-0005-0247-0086}\inst{\ref{aff40}}
\and J.~W.~Nightingale\orcid{0000-0002-8987-7401}\inst{\ref{aff99}}
\and C.~Padilla\orcid{0000-0001-7951-0166}\inst{\ref{aff98}}
\and S.~Paltani\orcid{0000-0002-8108-9179}\inst{\ref{aff62}}
\and F.~Pasian\orcid{0000-0002-4869-3227}\inst{\ref{aff9}}
\and K.~Pedersen\inst{\ref{aff100}}
\and V.~Pettorino\inst{\ref{aff40}}
\and S.~Pires\orcid{0000-0002-0249-2104}\inst{\ref{aff79}}
\and G.~Polenta\orcid{0000-0003-4067-9196}\inst{\ref{aff32}}
\and M.~Poncet\inst{\ref{aff101}}
\and L.~A.~Popa\inst{\ref{aff102}}
\and L.~Pozzetti\orcid{0000-0001-7085-0412}\inst{\ref{aff28}}
\and F.~Raison\orcid{0000-0002-7819-6918}\inst{\ref{aff68}}
\and R.~Rebolo\orcid{0000-0003-3767-7085}\inst{\ref{aff51},\ref{aff103},\ref{aff104}}
\and A.~Renzi\orcid{0000-0001-9856-1970}\inst{\ref{aff18},\ref{aff19}}
\and J.~Rhodes\orcid{0000-0002-4485-8549}\inst{\ref{aff11}}
\and G.~Riccio\inst{\ref{aff35}}
\and E.~Romelli\orcid{0000-0003-3069-9222}\inst{\ref{aff9}}
\and M.~Roncarelli\orcid{0000-0001-9587-7822}\inst{\ref{aff28}}
\and E.~Rossetti\orcid{0000-0003-0238-4047}\inst{\ref{aff29}}
\and R.~Saglia\orcid{0000-0003-0378-7032}\inst{\ref{aff67},\ref{aff68}}
\and Z.~Sakr\orcid{0000-0002-4823-3757}\inst{\ref{aff105},\ref{aff106},\ref{aff107}}
\and D.~Sapone\orcid{0000-0001-7089-4503}\inst{\ref{aff108}}
\and B.~Sartoris\orcid{0000-0003-1337-5269}\inst{\ref{aff67},\ref{aff9}}
\and J.~A.~Schewtschenko\orcid{0000-0002-4913-6393}\inst{\ref{aff52}}
\and P.~Schneider\orcid{0000-0001-8561-2679}\inst{\ref{aff88}}
\and T.~Schrabback\orcid{0000-0002-6987-7834}\inst{\ref{aff109}}
\and M.~Scodeggio\inst{\ref{aff14}}
\and A.~Secroun\orcid{0000-0003-0505-3710}\inst{\ref{aff64}}
\and G.~Seidel\orcid{0000-0003-2907-353X}\inst{\ref{aff75}}
\and M.~Seiffert\orcid{0000-0002-7536-9393}\inst{\ref{aff11}}
\and S.~Serrano\orcid{0000-0002-0211-2861}\inst{\ref{aff59},\ref{aff110},\ref{aff16}}
\and P.~Simon\inst{\ref{aff88}}
\and C.~Sirignano\orcid{0000-0002-0995-7146}\inst{\ref{aff18},\ref{aff19}}
\and G.~Sirri\orcid{0000-0003-2626-2853}\inst{\ref{aff30}}
\and L.~Stanco\orcid{0000-0002-9706-5104}\inst{\ref{aff19}}
\and J.~Steinwagner\orcid{0000-0001-7443-1047}\inst{\ref{aff68}}
\and C.~Surace\orcid{0000-0003-2592-0113}\inst{\ref{aff4}}
\and P.~Tallada-Cresp\'{i}\orcid{0000-0002-1336-8328}\inst{\ref{aff45},\ref{aff46}}
\and D.~Tavagnacco\orcid{0000-0001-7475-9894}\inst{\ref{aff9}}
\and A.~N.~Taylor\inst{\ref{aff52}}
\and I.~Tereno\orcid{0000-0002-4537-6218}\inst{\ref{aff60},\ref{aff111}}
\and N.~Tessore\orcid{0000-0002-9696-7931}\inst{\ref{aff77}}
\and S.~Toft\orcid{0000-0003-3631-7176}\inst{\ref{aff112},\ref{aff113}}
\and R.~Toledo-Moreo\orcid{0000-0002-2997-4859}\inst{\ref{aff114}}
\and F.~Torradeflot\orcid{0000-0003-1160-1517}\inst{\ref{aff46},\ref{aff45}}
\and I.~Tutusaus\orcid{0000-0002-3199-0399}\inst{\ref{aff106}}
\and L.~Valenziano\orcid{0000-0002-1170-0104}\inst{\ref{aff28},\ref{aff65}}
\and J.~Valiviita\orcid{0000-0001-6225-3693}\inst{\ref{aff81},\ref{aff82}}
\and T.~Vassallo\orcid{0000-0001-6512-6358}\inst{\ref{aff67},\ref{aff9}}
\and G.~Verdoes~Kleijn\orcid{0000-0001-5803-2580}\inst{\ref{aff115}}
\and D.~Vibert\orcid{0009-0008-0607-631X}\inst{\ref{aff4}}
\and J.~Weller\orcid{0000-0002-8282-2010}\inst{\ref{aff67},\ref{aff68}}
\and G.~Zamorani\orcid{0000-0002-2318-301X}\inst{\ref{aff28}}
\and F.~M.~Zerbi\inst{\ref{aff1}}
\and E.~Zucca\orcid{0000-0002-5845-8132}\inst{\ref{aff28}}
\and V.~Allevato\orcid{0000-0001-7232-5152}\inst{\ref{aff35}}
\and M.~Ballardini\orcid{0000-0003-4481-3559}\inst{\ref{aff116},\ref{aff117},\ref{aff28}}
\and M.~Bolzonella\orcid{0000-0003-3278-4607}\inst{\ref{aff28}}
\and E.~Bozzo\orcid{0000-0002-8201-1525}\inst{\ref{aff62}}
\and C.~Burigana\orcid{0000-0002-3005-5796}\inst{\ref{aff118},\ref{aff65}}
\and R.~Cabanac\orcid{0000-0001-6679-2600}\inst{\ref{aff106}}
\and A.~Cappi\inst{\ref{aff28},\ref{aff91}}
\and D.~Di~Ferdinando\inst{\ref{aff30}}
\and J.~A.~Escartin~Vigo\inst{\ref{aff68}}
\and L.~Gabarra\orcid{0000-0002-8486-8856}\inst{\ref{aff119}}
\and W.~G.~Hartley\inst{\ref{aff62}}
\and J.~Mart\'{i}n-Fleitas\orcid{0000-0002-8594-569X}\inst{\ref{aff97}}
\and S.~Matthew\orcid{0000-0001-8448-1697}\inst{\ref{aff52}}
\and N.~Mauri\orcid{0000-0001-8196-1548}\inst{\ref{aff50},\ref{aff30}}
\and R.~B.~Metcalf\orcid{0000-0003-3167-2574}\inst{\ref{aff89},\ref{aff28}}
\and A.~Pezzotta\orcid{0000-0003-0726-2268}\inst{\ref{aff120},\ref{aff68}}
\and M.~P\"ontinen\orcid{0000-0001-5442-2530}\inst{\ref{aff81}}
\and C.~Porciani\orcid{0000-0002-7797-2508}\inst{\ref{aff88}}
\and V.~Scottez\inst{\ref{aff95},\ref{aff121}}
\and M.~Sereno\orcid{0000-0003-0302-0325}\inst{\ref{aff28},\ref{aff30}}
\and M.~Tenti\orcid{0000-0002-4254-5901}\inst{\ref{aff30}}
\and M.~Viel\orcid{0000-0002-2642-5707}\inst{\ref{aff10},\ref{aff9},\ref{aff5},\ref{aff7},\ref{aff6}}
\and M.~Wiesmann\orcid{0009-0000-8199-5860}\inst{\ref{aff71}}
\and Y.~Akrami\orcid{0000-0002-2407-7956}\inst{\ref{aff122},\ref{aff123}}
\and S.~Alvi\orcid{0000-0001-5779-8568}\inst{\ref{aff116}}
\and I.~T.~Andika\orcid{0000-0001-6102-9526}\inst{\ref{aff124},\ref{aff125}}
\and M.~Archidiacono\orcid{0000-0003-4952-9012}\inst{\ref{aff69},\ref{aff70}}
\and F.~Atrio-Barandela\orcid{0000-0002-2130-2513}\inst{\ref{aff126}}
\and S.~Avila\orcid{0000-0001-5043-3662}\inst{\ref{aff45}}
\and A.~Balaguera-Antolinez\orcid{0000-0001-5028-3035}\inst{\ref{aff51},\ref{aff127}}
\and C.~Benoist\inst{\ref{aff91}}
\and D.~Bertacca\orcid{0000-0002-2490-7139}\inst{\ref{aff18},\ref{aff31},\ref{aff19}}
\and M.~Bethermin\orcid{0000-0002-3915-2015}\inst{\ref{aff128}}
\and L.~Blot\orcid{0000-0002-9622-7167}\inst{\ref{aff129},\ref{aff83}}
\and H.~B\"ohringer\orcid{0000-0001-8241-4204}\inst{\ref{aff68},\ref{aff130},\ref{aff131}}
\and S.~Borgani\orcid{0000-0001-6151-6439}\inst{\ref{aff8},\ref{aff10},\ref{aff9},\ref{aff7},\ref{aff6}}
\and M.~L.~Brown\orcid{0000-0002-0370-8077}\inst{\ref{aff53}}
\and A.~Calabro\orcid{0000-0003-2536-1614}\inst{\ref{aff43}}
\and B.~Camacho~Quevedo\orcid{0000-0002-8789-4232}\inst{\ref{aff59},\ref{aff16}}
\and F.~Caro\inst{\ref{aff43}}
\and C.~S.~Carvalho\inst{\ref{aff111}}
\and T.~Castro\orcid{0000-0002-6292-3228}\inst{\ref{aff9},\ref{aff7},\ref{aff10},\ref{aff6}}
\and F.~Cogato\orcid{0000-0003-4632-6113}\inst{\ref{aff89},\ref{aff28}}
\and A.~R.~Cooray\orcid{0000-0002-3892-0190}\inst{\ref{aff132}}
\and O.~Cucciati\orcid{0000-0002-9336-7551}\inst{\ref{aff28}}
\and S.~Davini\orcid{0000-0003-3269-1718}\inst{\ref{aff2}}
\and F.~De~Paolis\orcid{0000-0001-6460-7563}\inst{\ref{aff133},\ref{aff134},\ref{aff135}}
\and G.~Desprez\orcid{0000-0001-8325-1742}\inst{\ref{aff115}}
\and A.~D\'iaz-S\'anchez\orcid{0000-0003-0748-4768}\inst{\ref{aff136}}
\and J.~J.~Diaz\inst{\ref{aff51}}
\and S.~Di~Domizio\orcid{0000-0003-2863-5895}\inst{\ref{aff3},\ref{aff2}}
\and J.~M.~Diego\orcid{0000-0001-9065-3926}\inst{\ref{aff137}}
\and P.~Dimauro\orcid{0000-0001-7399-2854}\inst{\ref{aff43},\ref{aff138}}
\and A.~Enia\orcid{0000-0002-0200-2857}\inst{\ref{aff29},\ref{aff28}}
\and Y.~Fang\inst{\ref{aff67}}
\and A.~G.~Ferrari\orcid{0009-0005-5266-4110}\inst{\ref{aff30}}
\and A.~Finoguenov\orcid{0000-0002-4606-5403}\inst{\ref{aff81}}
\and A.~Fontana\orcid{0000-0003-3820-2823}\inst{\ref{aff43}}
\and A.~Franco\orcid{0000-0002-4761-366X}\inst{\ref{aff134},\ref{aff133},\ref{aff135}}
\and K.~Ganga\orcid{0000-0001-8159-8208}\inst{\ref{aff92}}
\and J.~Garc\'ia-Bellido\orcid{0000-0002-9370-8360}\inst{\ref{aff122}}
\and T.~Gasparetto\orcid{0000-0002-7913-4866}\inst{\ref{aff9}}
\and V.~Gautard\inst{\ref{aff139}}
\and E.~Gaztanaga\orcid{0000-0001-9632-0815}\inst{\ref{aff16},\ref{aff59},\ref{aff140}}
\and F.~Giacomini\orcid{0000-0002-3129-2814}\inst{\ref{aff30}}
\and F.~Gianotti\orcid{0000-0003-4666-119X}\inst{\ref{aff28}}
\and G.~Gozaliasl\orcid{0000-0002-0236-919X}\inst{\ref{aff141},\ref{aff81}}
\and M.~Guidi\orcid{0000-0001-9408-1101}\inst{\ref{aff29},\ref{aff28}}
\and C.~M.~Gutierrez\orcid{0000-0001-7854-783X}\inst{\ref{aff142}}
\and A.~Hall\orcid{0000-0002-3139-8651}\inst{\ref{aff52}}
\and S.~Hemmati\orcid{0000-0003-2226-5395}\inst{\ref{aff143}}
\and C.~Hern\'andez-Monteagudo\orcid{0000-0001-5471-9166}\inst{\ref{aff104},\ref{aff51}}
\and H.~Hildebrandt\orcid{0000-0002-9814-3338}\inst{\ref{aff144}}
\and J.~Hjorth\orcid{0000-0002-4571-2306}\inst{\ref{aff100}}
\and S.~Joudaki\orcid{0000-0001-8820-673X}\inst{\ref{aff45}}
\and J.~J.~E.~Kajava\orcid{0000-0002-3010-8333}\inst{\ref{aff145},\ref{aff146}}
\and Y.~Kang\orcid{0009-0000-8588-7250}\inst{\ref{aff62}}
\and V.~Kansal\orcid{0000-0002-4008-6078}\inst{\ref{aff147},\ref{aff148}}
\and D.~Karagiannis\orcid{0000-0002-4927-0816}\inst{\ref{aff116},\ref{aff149}}
\and K.~Kiiveri\inst{\ref{aff78}}
\and C.~C.~Kirkpatrick\inst{\ref{aff78}}
\and S.~Kruk\orcid{0000-0001-8010-8879}\inst{\ref{aff26}}
\and V.~Le~Brun\orcid{0000-0002-5027-1939}\inst{\ref{aff4}}
\and J.~Le~Graet\orcid{0000-0001-6523-7971}\inst{\ref{aff64}}
\and L.~Legrand\orcid{0000-0003-0610-5252}\inst{\ref{aff150},\ref{aff151}}
\and M.~Lembo\orcid{0000-0002-5271-5070}\inst{\ref{aff116},\ref{aff117}}
\and F.~Lepori\orcid{0009-0000-5061-7138}\inst{\ref{aff152}}
\and G.~Leroy\orcid{0009-0004-2523-4425}\inst{\ref{aff153},\ref{aff90}}
\and G.~F.~Lesci\orcid{0000-0002-4607-2830}\inst{\ref{aff89},\ref{aff28}}
\and L.~Leuzzi\orcid{0009-0006-4479-7017}\inst{\ref{aff89},\ref{aff28}}
\and T.~I.~Liaudat\orcid{0000-0002-9104-314X}\inst{\ref{aff154}}
\and A.~Loureiro\orcid{0000-0002-4371-0876}\inst{\ref{aff155},\ref{aff156}}
\and J.~Macias-Perez\orcid{0000-0002-5385-2763}\inst{\ref{aff157}}
\and M.~Magliocchetti\orcid{0000-0001-9158-4838}\inst{\ref{aff63}}
\and F.~Mannucci\orcid{0000-0002-4803-2381}\inst{\ref{aff158}}
\and R.~Maoli\orcid{0000-0002-6065-3025}\inst{\ref{aff159},\ref{aff43}}
\and C.~J.~A.~P.~Martins\orcid{0000-0002-4886-9261}\inst{\ref{aff160},\ref{aff36}}
\and L.~Maurin\orcid{0000-0002-8406-0857}\inst{\ref{aff25}}
\and M.~Miluzio\inst{\ref{aff26},\ref{aff161}}
\and C.~Moretti\orcid{0000-0003-3314-8936}\inst{\ref{aff5},\ref{aff6},\ref{aff9},\ref{aff10},\ref{aff7}}
\and G.~Morgante\inst{\ref{aff28}}
\and S.~Nadathur\orcid{0000-0001-9070-3102}\inst{\ref{aff140}}
\and K.~Naidoo\orcid{0000-0002-9182-1802}\inst{\ref{aff140}}
\and A.~Navarro-Alsina\orcid{0000-0002-3173-2592}\inst{\ref{aff88}}
\and K.~Paterson\orcid{0000-0001-8340-3486}\inst{\ref{aff75}}
\and L.~Patrizii\inst{\ref{aff30}}
\and A.~Pisani\orcid{0000-0002-6146-4437}\inst{\ref{aff64},\ref{aff162}}
\and D.~Potter\orcid{0000-0002-0757-5195}\inst{\ref{aff152}}
\and S.~Quai\orcid{0000-0002-0449-8163}\inst{\ref{aff89},\ref{aff28}}
\and M.~Radovich\orcid{0000-0002-3585-866X}\inst{\ref{aff31}}
\and P.-F.~Rocci\inst{\ref{aff25}}
\and S.~Sacquegna\orcid{0000-0002-8433-6630}\inst{\ref{aff133},\ref{aff134},\ref{aff135}}
\and M.~Sahl\'en\orcid{0000-0003-0973-4804}\inst{\ref{aff163}}
\and D.~B.~Sanders\orcid{0000-0002-1233-9998}\inst{\ref{aff49}}
\and A.~Schneider\orcid{0000-0001-7055-8104}\inst{\ref{aff152}}
\and D.~Sciotti\orcid{0009-0008-4519-2620}\inst{\ref{aff43},\ref{aff44}}
\and E.~Sellentin\inst{\ref{aff164},\ref{aff42}}
\and L.~C.~Smith\orcid{0000-0002-3259-2771}\inst{\ref{aff165}}
\and J.~G.~Sorce\orcid{0000-0002-2307-2432}\inst{\ref{aff166},\ref{aff25}}
\and K.~Tanidis\orcid{0000-0001-9843-5130}\inst{\ref{aff119}}
\and C.~Tao\orcid{0000-0001-7961-8177}\inst{\ref{aff64}}
\and G.~Testera\inst{\ref{aff2}}
\and R.~Teyssier\orcid{0000-0001-7689-0933}\inst{\ref{aff162}}
\and S.~Tosi\orcid{0000-0002-7275-9193}\inst{\ref{aff3},\ref{aff2},\ref{aff1}}
\and A.~Troja\orcid{0000-0003-0239-4595}\inst{\ref{aff18},\ref{aff19}}
\and M.~Tucci\inst{\ref{aff62}}
\and C.~Valieri\inst{\ref{aff30}}
\and A.~Venhola\orcid{0000-0001-6071-4564}\inst{\ref{aff167}}
\and D.~Vergani\orcid{0000-0003-0898-2216}\inst{\ref{aff28}}
\and G.~Verza\orcid{0000-0002-1886-8348}\inst{\ref{aff168}}
\and N.~A.~Walton\orcid{0000-0003-3983-8778}\inst{\ref{aff165}}}
                                                                                   
\institute{INAF-Osservatorio Astronomico di Brera, Via Brera 28, 20122 Milano, Italy\label{aff1}
\and
INFN-Sezione di Genova, Via Dodecaneso 33, 16146, Genova, Italy\label{aff2}
\and
Dipartimento di Fisica, Universit\`a di Genova, Via Dodecaneso 33, 16146, Genova, Italy\label{aff3}
\and
Aix-Marseille Universit\'e, CNRS, CNES, LAM, Marseille, France\label{aff4}
\and
SISSA, International School for Advanced Studies, Via Bonomea 265, 34136 Trieste TS, Italy\label{aff5}
\and
ICSC - Centro Nazionale di Ricerca in High Performance Computing, Big Data e Quantum Computing, Via Magnanelli 2, Bologna, Italy\label{aff6}
\and
INFN, Sezione di Trieste, Via Valerio 2, 34127 Trieste TS, Italy\label{aff7}
\and
Dipartimento di Fisica - Sezione di Astronomia, Universit\`a di Trieste, Via Tiepolo 11, 34131 Trieste, Italy\label{aff8}
\and
INAF-Osservatorio Astronomico di Trieste, Via G. B. Tiepolo 11, 34143 Trieste, Italy\label{aff9}
\and
IFPU, Institute for Fundamental Physics of the Universe, via Beirut 2, 34151 Trieste, Italy\label{aff10}
\and
Jet Propulsion Laboratory, California Institute of Technology, 4800 Oak Grove Drive, Pasadena, CA, 91109, USA\label{aff11}
\and
Johns Hopkins University 3400 North Charles Street Baltimore, MD 21218, USA\label{aff12}
\and
California Institute of Technology, 1200 E California Blvd, Pasadena, CA 91125, USA\label{aff13}
\and
INAF-IASF Milano, Via Alfonso Corti 12, 20133 Milano, Italy\label{aff14}
\and
Institut d'Astrophysique de Paris, UMR 7095, CNRS, and Sorbonne Universit\'e, 98 bis boulevard Arago, 75014 Paris, France\label{aff15}
\and
Institute of Space Sciences (ICE, CSIC), Campus UAB, Carrer de Can Magrans, s/n, 08193 Barcelona, Spain\label{aff16}
\and
Dipartimento di Fisica, Universit\`a degli studi di Genova, and INFN-Sezione di Genova, via Dodecaneso 33, 16146, Genova, Italy\label{aff17}
\and
Dipartimento di Fisica e Astronomia "G. Galilei", Universit\`a di Padova, Via Marzolo 8, 35131 Padova, Italy\label{aff18}
\and
INFN-Padova, Via Marzolo 8, 35131 Padova, Italy\label{aff19}
\and
Waterloo Centre for Astrophysics, University of Waterloo, Waterloo, Ontario N2L 3G1, Canada\label{aff20}
\and
Department of Physics and Astronomy, University of Waterloo, Waterloo, Ontario N2L 3G1, Canada\label{aff21}
\and
Perimeter Institute for Theoretical Physics, Waterloo, Ontario N2L 2Y5, Canada\label{aff22}
\and
Minnesota Institute for Astrophysics, University of Minnesota, 116 Church St SE, Minneapolis, MN 55455, USA\label{aff23}
\and
Infrared Processing and Analysis Center, California Institute of Technology, Pasadena, CA 91125, USA\label{aff24}
\and
Universit\'e Paris-Saclay, CNRS, Institut d'astrophysique spatiale, 91405, Orsay, France\label{aff25}
\and
ESAC/ESA, Camino Bajo del Castillo, s/n., Urb. Villafranca del Castillo, 28692 Villanueva de la Ca\~nada, Madrid, Spain\label{aff26}
\and
School of Mathematics and Physics, University of Surrey, Guildford, Surrey, GU2 7XH, UK\label{aff27}
\and
INAF-Osservatorio di Astrofisica e Scienza dello Spazio di Bologna, Via Piero Gobetti 93/3, 40129 Bologna, Italy\label{aff28}
\and
Dipartimento di Fisica e Astronomia, Universit\`a di Bologna, Via Gobetti 93/2, 40129 Bologna, Italy\label{aff29}
\and
INFN-Sezione di Bologna, Viale Berti Pichat 6/2, 40127 Bologna, Italy\label{aff30}
\and
INAF-Osservatorio Astronomico di Padova, Via dell'Osservatorio 5, 35122 Padova, Italy\label{aff31}
\and
Space Science Data Center, Italian Space Agency, via del Politecnico snc, 00133 Roma, Italy\label{aff32}
\and
INAF-Osservatorio Astrofisico di Torino, Via Osservatorio 20, 10025 Pino Torinese (TO), Italy\label{aff33}
\and
Department of Physics "E. Pancini", University Federico II, Via Cinthia 6, 80126, Napoli, Italy\label{aff34}
\and
INAF-Osservatorio Astronomico di Capodimonte, Via Moiariello 16, 80131 Napoli, Italy\label{aff35}
\and
Instituto de Astrof\'isica e Ci\^encias do Espa\c{c}o, Universidade do Porto, CAUP, Rua das Estrelas, PT4150-762 Porto, Portugal\label{aff36}
\and
Faculdade de Ci\^encias da Universidade do Porto, Rua do Campo de Alegre, 4150-007 Porto, Portugal\label{aff37}
\and
Dipartimento di Fisica, Universit\`a degli Studi di Torino, Via P. Giuria 1, 10125 Torino, Italy\label{aff38}
\and
INFN-Sezione di Torino, Via P. Giuria 1, 10125 Torino, Italy\label{aff39}
\and
European Space Agency/ESTEC, Keplerlaan 1, 2201 AZ Noordwijk, The Netherlands\label{aff40}
\and
Institute Lorentz, Leiden University, Niels Bohrweg 2, 2333 CA Leiden, The Netherlands\label{aff41}
\and
Leiden Observatory, Leiden University, Einsteinweg 55, 2333 CC Leiden, The Netherlands\label{aff42}
\and
INAF-Osservatorio Astronomico di Roma, Via Frascati 33, 00078 Monteporzio Catone, Italy\label{aff43}
\and
INFN-Sezione di Roma, Piazzale Aldo Moro, 2 - c/o Dipartimento di Fisica, Edificio G. Marconi, 00185 Roma, Italy\label{aff44}
\and
Centro de Investigaciones Energ\'eticas, Medioambientales y Tecnol\'ogicas (CIEMAT), Avenida Complutense 40, 28040 Madrid, Spain\label{aff45}
\and
Port d'Informaci\'{o} Cient\'{i}fica, Campus UAB, C. Albareda s/n, 08193 Bellaterra (Barcelona), Spain\label{aff46}
\and
Institute for Theoretical Particle Physics and Cosmology (TTK), RWTH Aachen University, 52056 Aachen, Germany\label{aff47}
\and
INFN section of Naples, Via Cinthia 6, 80126, Napoli, Italy\label{aff48}
\and
Institute for Astronomy, University of Hawaii, 2680 Woodlawn Drive, Honolulu, HI 96822, USA\label{aff49}
\and
Dipartimento di Fisica e Astronomia "Augusto Righi" - Alma Mater Studiorum Universit\`a di Bologna, Viale Berti Pichat 6/2, 40127 Bologna, Italy\label{aff50}
\and
Instituto de Astrof\'{\i}sica de Canarias, V\'{\i}a L\'actea, 38205 La Laguna, Tenerife, Spain\label{aff51}
\and
Institute for Astronomy, University of Edinburgh, Royal Observatory, Blackford Hill, Edinburgh EH9 3HJ, UK\label{aff52}
\and
Jodrell Bank Centre for Astrophysics, Department of Physics and Astronomy, University of Manchester, Oxford Road, Manchester M13 9PL, UK\label{aff53}
\and
European Space Agency/ESRIN, Largo Galileo Galilei 1, 00044 Frascati, Roma, Italy\label{aff54}
\and
Universit\'e Claude Bernard Lyon 1, CNRS/IN2P3, IP2I Lyon, UMR 5822, Villeurbanne, F-69100, France\label{aff55}
\and
Institut de Ci\`{e}ncies del Cosmos (ICCUB), Universitat de Barcelona (IEEC-UB), Mart\'{i} i Franqu\`{e}s 1, 08028 Barcelona, Spain\label{aff56}
\and
Instituci\'o Catalana de Recerca i Estudis Avan\c{c}ats (ICREA), Passeig de Llu\'{\i}s Companys 23, 08010 Barcelona, Spain\label{aff57}
\and
UCB Lyon 1, CNRS/IN2P3, IUF, IP2I Lyon, 4 rue Enrico Fermi, 69622 Villeurbanne, France\label{aff58}
\and
Institut d'Estudis Espacials de Catalunya (IEEC),  Edifici RDIT, Campus UPC, 08860 Castelldefels, Barcelona, Spain\label{aff59}
\and
Departamento de F\'isica, Faculdade de Ci\^encias, Universidade de Lisboa, Edif\'icio C8, Campo Grande, PT1749-016 Lisboa, Portugal\label{aff60}
\and
Instituto de Astrof\'isica e Ci\^encias do Espa\c{c}o, Faculdade de Ci\^encias, Universidade de Lisboa, Campo Grande, 1749-016 Lisboa, Portugal\label{aff61}
\and
Department of Astronomy, University of Geneva, ch. d'Ecogia 16, 1290 Versoix, Switzerland\label{aff62}
\and
INAF-Istituto di Astrofisica e Planetologia Spaziali, via del Fosso del Cavaliere, 100, 00100 Roma, Italy\label{aff63}
\and
Aix-Marseille Universit\'e, CNRS/IN2P3, CPPM, Marseille, France\label{aff64}
\and
INFN-Bologna, Via Irnerio 46, 40126 Bologna, Italy\label{aff65}
\and
School of Physics, HH Wills Physics Laboratory, University of Bristol, Tyndall Avenue, Bristol, BS8 1TL, UK\label{aff66}
\and
Universit\"ats-Sternwarte M\"unchen, Fakult\"at f\"ur Physik, Ludwig-Maximilians-Universit\"at M\"unchen, Scheinerstrasse 1, 81679 M\"unchen, Germany\label{aff67}
\and
Max Planck Institute for Extraterrestrial Physics, Giessenbachstr. 1, 85748 Garching, Germany\label{aff68}
\and
Dipartimento di Fisica "Aldo Pontremoli", Universit\`a degli Studi di Milano, Via Celoria 16, 20133 Milano, Italy\label{aff69}
\and
INFN-Sezione di Milano, Via Celoria 16, 20133 Milano, Italy\label{aff70}
\and
Institute of Theoretical Astrophysics, University of Oslo, P.O. Box 1029 Blindern, 0315 Oslo, Norway\label{aff71}
\and
Felix Hormuth Engineering, Goethestr. 17, 69181 Leimen, Germany\label{aff72}
\and
Technical University of Denmark, Elektrovej 327, 2800 Kgs. Lyngby, Denmark\label{aff73}
\and
Cosmic Dawn Center (DAWN), Denmark\label{aff74}
\and
Max-Planck-Institut f\"ur Astronomie, K\"onigstuhl 17, 69117 Heidelberg, Germany\label{aff75}
\and
NASA Goddard Space Flight Center, Greenbelt, MD 20771, USA\label{aff76}
\and
Department of Physics and Astronomy, University College London, Gower Street, London WC1E 6BT, UK\label{aff77}
\and
Department of Physics and Helsinki Institute of Physics, Gustaf H\"allstr\"omin katu 2, 00014 University of Helsinki, Finland\label{aff78}
\and
Universit\'e Paris-Saclay, Universit\'e Paris Cit\'e, CEA, CNRS, AIM, 91191, Gif-sur-Yvette, France\label{aff79}
\and
Universit\'e de Gen\`eve, D\'epartement de Physique Th\'eorique and Centre for Astroparticle Physics, 24 quai Ernest-Ansermet, CH-1211 Gen\`eve 4, Switzerland\label{aff80}
\and
Department of Physics, P.O. Box 64, 00014 University of Helsinki, Finland\label{aff81}
\and
Helsinki Institute of Physics, Gustaf H{\"a}llstr{\"o}min katu 2, University of Helsinki, Helsinki, Finland\label{aff82}
\and
Laboratoire d'etude de l'Univers et des phenomenes eXtremes, Observatoire de Paris, Universit\'e PSL, Sorbonne Universit\'e, CNRS, 92190 Meudon, France\label{aff83}
\and
Mullard Space Science Laboratory, University College London, Holmbury St Mary, Dorking, Surrey RH5 6NT, UK\label{aff84}
\and
NOVA optical infrared instrumentation group at ASTRON, Oude Hoogeveensedijk 4, 7991PD, Dwingeloo, The Netherlands\label{aff85}
\and
Centre de Calcul de l'IN2P3/CNRS, 21 avenue Pierre de Coubertin 69627 Villeurbanne Cedex, France\label{aff86}
\and
University of Applied Sciences and Arts of Northwestern Switzerland, School of Computer Science, 5210 Windisch, Switzerland\label{aff87}
\and
Universit\"at Bonn, Argelander-Institut f\"ur Astronomie, Auf dem H\"ugel 71, 53121 Bonn, Germany\label{aff88}
\and
Dipartimento di Fisica e Astronomia "Augusto Righi" - Alma Mater Studiorum Universit\`a di Bologna, via Piero Gobetti 93/2, 40129 Bologna, Italy\label{aff89}
\and
Department of Physics, Institute for Computational Cosmology, Durham University, South Road, Durham, DH1 3LE, UK\label{aff90}
\and
Universit\'e C\^{o}te d'Azur, Observatoire de la C\^{o}te d'Azur, CNRS, Laboratoire Lagrange, Bd de l'Observatoire, CS 34229, 06304 Nice cedex 4, France\label{aff91}
\and
Universit\'e Paris Cit\'e, CNRS, Astroparticule et Cosmologie, 75013 Paris, France\label{aff92}
\and
CNRS-UCB International Research Laboratory, Centre Pierre Bin\'etruy, IRL2007, CPB-IN2P3, Berkeley, USA\label{aff93}
\and
University of Applied Sciences and Arts of Northwestern Switzerland, School of Engineering, 5210 Windisch, Switzerland\label{aff94}
\and
Institut d'Astrophysique de Paris, 98bis Boulevard Arago, 75014, Paris, France\label{aff95}
\and
Institute of Physics, Laboratory of Astrophysics, Ecole Polytechnique F\'ed\'erale de Lausanne (EPFL), Observatoire de Sauverny, 1290 Versoix, Switzerland\label{aff96}
\and
Aurora Technology for European Space Agency (ESA), Camino bajo del Castillo, s/n, Urbanizacion Villafranca del Castillo, Villanueva de la Ca\~nada, 28692 Madrid, Spain\label{aff97}
\and
Institut de F\'{i}sica d'Altes Energies (IFAE), The Barcelona Institute of Science and Technology, Campus UAB, 08193 Bellaterra (Barcelona), Spain\label{aff98}
\and
School of Mathematics, Statistics and Physics, Newcastle University, Herschel Building, Newcastle-upon-Tyne, NE1 7RU, UK\label{aff99}
\and
DARK, Niels Bohr Institute, University of Copenhagen, Jagtvej 155, 2200 Copenhagen, Denmark\label{aff100}
\and
Centre National d'Etudes Spatiales -- Centre spatial de Toulouse, 18 avenue Edouard Belin, 31401 Toulouse Cedex 9, France\label{aff101}
\and
Institute of Space Science, Str. Atomistilor, nr. 409 M\u{a}gurele, Ilfov, 077125, Romania\label{aff102}
\and
Consejo Superior de Investigaciones Cientificas, Calle Serrano 117, 28006 Madrid, Spain\label{aff103}
\and
Universidad de La Laguna, Departamento de Astrof\'{\i}sica, 38206 La Laguna, Tenerife, Spain\label{aff104}
\and
Institut f\"ur Theoretische Physik, University of Heidelberg, Philosophenweg 16, 69120 Heidelberg, Germany\label{aff105}
\and
Institut de Recherche en Astrophysique et Plan\'etologie (IRAP), Universit\'e de Toulouse, CNRS, UPS, CNES, 14 Av. Edouard Belin, 31400 Toulouse, France\label{aff106}
\and
Universit\'e St Joseph; Faculty of Sciences, Beirut, Lebanon\label{aff107}
\and
Departamento de F\'isica, FCFM, Universidad de Chile, Blanco Encalada 2008, Santiago, Chile\label{aff108}
\and
Universit\"at Innsbruck, Institut f\"ur Astro- und Teilchenphysik, Technikerstr. 25/8, 6020 Innsbruck, Austria\label{aff109}
\and
Satlantis, University Science Park, Sede Bld 48940, Leioa-Bilbao, Spain\label{aff110}
\and
Instituto de Astrof\'isica e Ci\^encias do Espa\c{c}o, Faculdade de Ci\^encias, Universidade de Lisboa, Tapada da Ajuda, 1349-018 Lisboa, Portugal\label{aff111}
\and
Cosmic Dawn Center (DAWN)\label{aff112}
\and
Niels Bohr Institute, University of Copenhagen, Jagtvej 128, 2200 Copenhagen, Denmark\label{aff113}
\and
Universidad Polit\'ecnica de Cartagena, Departamento de Electr\'onica y Tecnolog\'ia de Computadoras,  Plaza del Hospital 1, 30202 Cartagena, Spain\label{aff114}
\and
Kapteyn Astronomical Institute, University of Groningen, PO Box 800, 9700 AV Groningen, The Netherlands\label{aff115}
\and
Dipartimento di Fisica e Scienze della Terra, Universit\`a degli Studi di Ferrara, Via Giuseppe Saragat 1, 44122 Ferrara, Italy\label{aff116}
\and
Istituto Nazionale di Fisica Nucleare, Sezione di Ferrara, Via Giuseppe Saragat 1, 44122 Ferrara, Italy\label{aff117}
\and
INAF, Istituto di Radioastronomia, Via Piero Gobetti 101, 40129 Bologna, Italy\label{aff118}
\and
Department of Physics, Oxford University, Keble Road, Oxford OX1 3RH, UK\label{aff119}
\and
INAF - Osservatorio Astronomico di Brera, via Emilio Bianchi 46, 23807 Merate, Italy\label{aff120}
\and
ICL, Junia, Universit\'e Catholique de Lille, LITL, 59000 Lille, France\label{aff121}
\and
Instituto de F\'isica Te\'orica UAM-CSIC, Campus de Cantoblanco, 28049 Madrid, Spain\label{aff122}
\and
CERCA/ISO, Department of Physics, Case Western Reserve University, 10900 Euclid Avenue, Cleveland, OH 44106, USA\label{aff123}
\and
Technical University of Munich, TUM School of Natural Sciences, Physics Department, James-Franck-Str.~1, 85748 Garching, Germany\label{aff124}
\and
Max-Planck-Institut f\"ur Astrophysik, Karl-Schwarzschild-Str.~1, 85748 Garching, Germany\label{aff125}
\and
Departamento de F{\'\i}sica Fundamental. Universidad de Salamanca. Plaza de la Merced s/n. 37008 Salamanca, Spain\label{aff126}
\and
Instituto de Astrof\'isica de Canarias (IAC); Departamento de Astrof\'isica, Universidad de La Laguna (ULL), 38200, La Laguna, Tenerife, Spain\label{aff127}
\and
Universit\'e de Strasbourg, CNRS, Observatoire astronomique de Strasbourg, UMR 7550, 67000 Strasbourg, France\label{aff128}
\and
Center for Data-Driven Discovery, Kavli IPMU (WPI), UTIAS, The University of Tokyo, Kashiwa, Chiba 277-8583, Japan\label{aff129}
\and
Ludwig-Maximilians-University, Schellingstrasse 4, 80799 Munich, Germany\label{aff130}
\and
Max-Planck-Institut f\"ur Physik, Boltzmannstr. 8, 85748 Garching, Germany\label{aff131}
\and
Department of Physics \& Astronomy, University of California Irvine, Irvine CA 92697, USA\label{aff132}
\and
Department of Mathematics and Physics E. De Giorgi, University of Salento, Via per Arnesano, CP-I93, 73100, Lecce, Italy\label{aff133}
\and
INFN, Sezione di Lecce, Via per Arnesano, CP-193, 73100, Lecce, Italy\label{aff134}
\and
INAF-Sezione di Lecce, c/o Dipartimento Matematica e Fisica, Via per Arnesano, 73100, Lecce, Italy\label{aff135}
\and
Departamento F\'isica Aplicada, Universidad Polit\'ecnica de Cartagena, Campus Muralla del Mar, 30202 Cartagena, Murcia, Spain\label{aff136}
\and
Instituto de F\'isica de Cantabria, Edificio Juan Jord\'a, Avenida de los Castros, 39005 Santander, Spain\label{aff137}
\and
Observatorio Nacional, Rua General Jose Cristino, 77-Bairro Imperial de Sao Cristovao, Rio de Janeiro, 20921-400, Brazil\label{aff138}
\and
CEA Saclay, DFR/IRFU, Service d'Astrophysique, Bat. 709, 91191 Gif-sur-Yvette, France\label{aff139}
\and
Institute of Cosmology and Gravitation, University of Portsmouth, Portsmouth PO1 3FX, UK\label{aff140}
\and
Department of Computer Science, Aalto University, PO Box 15400, Espoo, FI-00 076, Finland\label{aff141}
\and
Instituto de Astrof\'\i sica de Canarias, c/ Via Lactea s/n, La Laguna 38200, Spain. Departamento de Astrof\'\i sica de la Universidad de La Laguna, Avda. Francisco Sanchez, La Laguna, 38200, Spain\label{aff142}
\and
Caltech/IPAC, 1200 E. California Blvd., Pasadena, CA 91125, USA\label{aff143}
\and
Ruhr University Bochum, Faculty of Physics and Astronomy, Astronomical Institute (AIRUB), German Centre for Cosmological Lensing (GCCL), 44780 Bochum, Germany\label{aff144}
\and
Department of Physics and Astronomy, Vesilinnantie 5, 20014 University of Turku, Finland\label{aff145}
\and
Serco for European Space Agency (ESA), Camino bajo del Castillo, s/n, Urbanizacion Villafranca del Castillo, Villanueva de la Ca\~nada, 28692 Madrid, Spain\label{aff146}
\and
ARC Centre of Excellence for Dark Matter Particle Physics, Melbourne, Australia\label{aff147}
\and
Centre for Astrophysics \& Supercomputing, Swinburne University of Technology,  Hawthorn, Victoria 3122, Australia\label{aff148}
\and
Department of Physics and Astronomy, University of the Western Cape, Bellville, Cape Town, 7535, South Africa\label{aff149}
\and
DAMTP, Centre for Mathematical Sciences, Wilberforce Road, Cambridge CB3 0WA, UK\label{aff150}
\and
Kavli Institute for Cosmology Cambridge, Madingley Road, Cambridge, CB3 0HA, UK\label{aff151}
\and
Department of Astrophysics, University of Zurich, Winterthurerstrasse 190, 8057 Zurich, Switzerland\label{aff152}
\and
Department of Physics, Centre for Extragalactic Astronomy, Durham University, South Road, Durham, DH1 3LE, UK\label{aff153}
\and
IRFU, CEA, Universit\'e Paris-Saclay 91191 Gif-sur-Yvette Cedex, France\label{aff154}
\and
Oskar Klein Centre for Cosmoparticle Physics, Department of Physics, Stockholm University, Stockholm, SE-106 91, Sweden\label{aff155}
\and
Astrophysics Group, Blackett Laboratory, Imperial College London, London SW7 2AZ, UK\label{aff156}
\and
Univ. Grenoble Alpes, CNRS, Grenoble INP, LPSC-IN2P3, 53, Avenue des Martyrs, 38000, Grenoble, France\label{aff157}
\and
INAF-Osservatorio Astrofisico di Arcetri, Largo E. Fermi 5, 50125, Firenze, Italy\label{aff158}
\and
Dipartimento di Fisica, Sapienza Universit\`a di Roma, Piazzale Aldo Moro 2, 00185 Roma, Italy\label{aff159}
\and
Centro de Astrof\'{\i}sica da Universidade do Porto, Rua das Estrelas, 4150-762 Porto, Portugal\label{aff160}
\and
HE Space for European Space Agency (ESA), Camino bajo del Castillo, s/n, Urbanizacion Villafranca del Castillo, Villanueva de la Ca\~nada, 28692 Madrid, Spain\label{aff161}
\and
Department of Astrophysical Sciences, Peyton Hall, Princeton University, Princeton, NJ 08544, USA\label{aff162}
\and
Theoretical astrophysics, Department of Physics and Astronomy, Uppsala University, Box 515, 751 20 Uppsala, Sweden\label{aff163}
\and
Mathematical Institute, University of Leiden, Einsteinweg 55, 2333 CA Leiden, The Netherlands\label{aff164}
\and
Institute of Astronomy, University of Cambridge, Madingley Road, Cambridge CB3 0HA, UK\label{aff165}
\and
Univ. Lille, CNRS, Centrale Lille, UMR 9189 CRIStAL, 59000 Lille, France\label{aff166}
\and
Space physics and astronomy research unit, University of Oulu, Pentti Kaiteran katu 1, FI-90014 Oulu, Finland\label{aff167}
\and
Center for Computational Astrophysics, Flatiron Institute, 162 5th Avenue, 10010, New York, NY, USA\label{aff168}}

\abstract
{The \Euclid galaxy survey is designed to measure the spectroscopic redshift of emission-line galaxies (ELGs) by identifying the \ha\, emission line in their slitless spectra. The efficacy of this approach crucially depends on the signal-to-noise ratio (S/N) of the line, as sometimes noise fluctuations in the spectrum continuum can be misidentified as \ha. In addition, other genuine strong emission lines can be mistaken for \ha, depending on the redshift of the source. Both effects lead to ambiguities in the redshift measurement that can result in catastrophic redshift errors and the inclusion of `interloper' galaxies in the sample.
}
{This paper forecasts the impact on the galaxy clustering analysis of the expected redshift errors in the \Euclid spectroscopic sample. Specifically, it investigates the effect of the redshift interloper contamination on the galaxy two-point correlation function (2PCF) and, in turn, on the inferred growth rate of structure \(f \sigma_8\) and Alcock--Paczynski (AP) parameters \(\alpha_\parallel\) and \(\alpha_\perp\).
}
{
This work is based on the analysis of \num{1000} synthetic spectroscopic catalogues, the EuclidLargeMocks, which mimic the area and selection function of the \Euclid Data Release 1 (DR1) sample. We estimated the 2PCF of contaminated catalogues and separated the different contributions, particularly those coming from galaxies with correctly measured redshift and from contaminants. We explored different models of increasing complexity to describe the measured 2PCF at a fixed cosmology, with the aim of identifying the most efficient model to reproduce the data. Finally, we performed a cosmological inference and evaluated the systematic error on the inferred \(f \sigma_8\), \(\alpha_\parallel\) , and \(\alpha_\perp\) values associated with different models.
}
{
Our results demonstrate that a minimal modelling approach, which only accounts for an attenuation of the clustering signal regardless of the type of contaminants, is sufficient to recover the correct values of \(f \sigma_8\) ,  \(\alpha_\parallel\), and \(\alpha_\perp\) at DR1. The accuracy and precision of the estimated AP parameters are largely insensitive to the presence of interlopers. The adoption of a minimal modelling induces a 1\%--3\% systematic error on the growth rate of structure estimation, depending on the considered redshift. However, this error remains smaller than the statistical error expected for the \Euclid DR1 analysis.
}
{}


   \titlerunning{
   The impact of redshift interlopers on the two-point correlation function analysis
   }
   \authorrunning{
   Euclid Collaboration: I.~Risso et al.
   }
   
    \maketitle
    \nolinenumbers

\section{\label{sc:Intro}Introduction}

Galaxy surveys aim to map the large-scale structure of the Universe using galaxies as tracers of the underlying matter distribution to infer the cosmological model. One of the largest surveys is being conducted by the \Euclid space mission \citep{EuclidSkyOverview}, which was launched by the European Space Agency (ESA) on the 1 July 2023. Its primary goal is to probe the expansion history of the Universe and the evolution of cosmic structures over the past ten billion years and, in turn, indirectly probe  the nature of its two dominant components: dark matter and dark energy. The \Euclid satellite uses slitless spectroscopy and the Near-Infrared Spectrometer and Photometer (NISP, \citealp{EuclidSkyNISP}) to measure the redshift of tens of millions of galaxies and create one of the largest and most detailed three-dimensional maps of the Universe. The redshift of the observed galaxies is primarily determined by the position of the strongest emission lines in their spectra, in particular the \ha\, line. Since the measured redshift of galaxies is used to estimate their radial distance from us, systematic errors in the redshift determination can introduce contaminants in the spectroscopic sample and ultimately alter the observed galaxy spatial distribution.

The slitless spectroscopy used in \Euclid implies that the observed spectra will generally have a lower resolution and that there will be more contamination from adjacent objects than when using slit or fibre spectroscopy. This leads to larger redshift measurement uncertainties but also to systematically wrong redshift determinations.  The redshift error can be several orders of magnitude larger than the statistical uncertainty targeted by the experiment, which is of $\Delta z \sim 0.001$ \citep{EuclidSkyOverview}.
\Euclid, like the upcoming NASA Nancy Grace Roman Space Telescope satellite\footnote{\url{https://roman.gsfc.nasa.gov/}}, has a medium-low spectral resolution of $R = \lambda / \Delta\lambda < 1000$ and a limited bandwidth, which leaves room for ambiguity in emission-line identification. Moreover, in order to measure the redshift of millions of galaxies, emission lines are detected at a S/N that is typically lower than the threshold adopted in targeted ground-based spectroscopic surveys (such as DESI, \citealt{DESI2019}),
just sufficient to determine the redshift using a single prominent emission line. As a result, a non-negligible fraction of the objects in the \Euclid spectroscopic catalogue will be interloper galaxies, that is, galaxies whose estimated redshift has a catastrophic error. In that case, the detected line is not the expected one but is either another emission line or a notably prominent noise spike. This can affect the clustering statistics and, in turn, the cosmological parameters  obtained from them.

The impact of redshift interlopers has been studied in previous works. \citet{Pullen_2016} introduced the formalism to model the galaxy power spectrum in the presence of interlopers. \citet{Addison_2019}  adopt the same formalism to forecast the impact of interloper galaxies on the baryon acoustic oscillations (BAO) and redshift-space distortion (RSD) analysis of future spectroscopic surveys targeting emission-line galaxies (ELGs). 
\citet{Foroozan_2022} and \citet{Nguyen_2024} present two-point correlation function (2PCF) models in the presence of small displacement interlopers and assessed the performance of their methods to recover unbiased estimates of the BAO parameters. 
\citet{LymanInter} present an analysis of contamination in Lyman-break galaxy samples at high-redshift by studying the spatial correlation with intermediate-redshift galaxies.
Furthermore, methods to mitigate the impact of interlopers have been studied in recent years using simulations 
\citep{Farrow_2021,Blanchard-EP7,Peng_2023}.

Within the context of the \Euclid preparation, 
Euclid Collaboration: Monaco et al. (in prep.) describe the strategy to identify all potential sources of data systematics in the pipeline for the spectroscopic data analysis.  
This paper focuses on assessing the impact of redshift errors on 2PCF measurements and configuration-space galaxy clustering analysis at \Euclid DR1. The counterpart to this study in Fourier space is described in the companion paper, Euclid Collaboration: Lee et al. (in prep.).

Assessing the impact of redshift errors in \Euclid requires the consideration of realistic types and fractions of redshift interlopers. The Euclid Consortium has released a suite of \num{1000} mock catalogues, named EuclidLargeMocks
\citep{EP-Monaco1},
which currently offers the best balance between robust statistical power and a realistic modelling of selection effects. This suite effectively mimics the anticipated types and proportions of interlopers in the Euclid Wide Survey (EWS), and we adopt it for this analysis. Using these catalogues, we study  how the assumption of an incomplete model for the measured 2PCF in the presence of interlopers affects the cosmological parameter estimates. We focus on the growth rate of structures and on the Alcock--Paczynski (AP) parameters, and conduct a separate analysis for each case. Although both analyses are based on the same set of measurements, they rely on fundamentally different theoretical models for the 2PCF and target distinct ranges of scales. For this reason, we chose to separate the analyses and present the methodology and results in distinct sections. 

The paper is structured as follows. In Sect.~\ref{sc:Interlopers}, we introduce the \Euclid mission and the types of interloper galaxies that we expect to find in the spectroscopic catalogue. We provide a quantitative assessment of these contaminants and of their effect on the galaxy clustering 2PCF. In Sect.~\ref{sc:Estimator}, we present the estimator of the 2PCF and the predicted 2PCF in the presence of interlopers. In Sect.~\ref{sec:catalogues}, we describe the mock catalogues and 2PCF measurements.
In Sect.~\ref{sc:2PCFmeas}, we evaluate the amplitude and relevance of the interloper galaxy contributions to the measured 2PCF.
In Sect.~\ref{sc:CosmoFit}, we perform a Monte Carlo Markov chain (MCMC) analysis of the full shape of the 2PCF and study how constraints on the growth rate, $f\sigma_8$ , change when using different theoretical models of varying complexity to describe the contaminated signal. In Sect.~\ref{sc:BAOFit}, we focus on the modelling of the BAO signal and study the bias on the derived AP parameters induced by adopting an inadequate model that does not account for the interloper presence. In Sect.~\ref{sc:Conclusions} we conclude with a comprehensive discussion of the results and we draw our final conclusions.

\section{\label{sc:Interlopers}Interloper galaxies in the \Euclid mission}

The \Euclid mission anchors the determination of galaxy redshifts to the detection of the \ha\, line, the most intense emission line expected in the optical and near-infrared rest-frame wavelength of an ELG spectrum. This detection is carried out by the NISP instrument, designed to cover during the EWS a wavelength range 1206--1892\,\si{nm} that enables the detection of the \ha\, line in the redshift range $0.84 \leq z \leq 1.88$ \citep{EuclidSkyOverview}. To maximize the number of observed galaxies in a given exposure time, NISP performs slitless spectroscopy, thus capturing the spectra of all objects entering the telescope field of view. However, this strategy results in a medium-low spectral resolution ($R>480$, \citealp{EuclidSkyNISP}). As a result, the \ha\, line and the $\ion{N}{ii}\,\lambda\lambda$6549,6584 doublet are blended into a single emission feature and cannot be separated at the detection threshold in signal-to-noise ratio (S/N) adopted to select the \Euclid spectroscopic sample \citep{Scaramella-EP1}. Moreover, the limited wavelength range and S/N of the spectra generally prevent the detection of multiple emission lines.
This leads to the presence of interlopers in the catalogues, since in most of the cases we have to rely on a single-line detection to assign a redshift value. When the measured spectrum has only one significant emission line, a prior on this line being \ha\, is used \citep{Q1-TP007}, since this is the most prominent expected emission line. With no additional spectral features, this guess can result in an interloper detection.

The relation between the true and measured redshifts for any type of galaxy, including interloper ones, can be derived from the redshift definition as
\begin{equation}
\label{eq:z-def-with-lambdas}
    1 + z = \lambda_{\rm obs} / \lambda_{\rm rest} \, ,
\end{equation}
where $\lambda_{\rm obs}$ is the observed wavelength of the line and $\lambda_{\rm rest}$ is the expected rest-frame wavelength. When the observed wavelength of a feature is interpreted as the rest-frame wavelength of the incorrect line at a incorrect redshift, the relation becomes\footnote{For comparison to \cite{Pullen_2016}: $z_{\rm true} \equiv {z_{\rm Int}}$, $z_{\rm meas} \equiv {z_{\rm SELG}}$, and $\lambda_{\rm wrong} \equiv {\lambda_{\rm SEL}}$.}
\begin{equation}
\label{eq:zmeas-vs-ztrue}
    \frac{1 + z_{\rm true}}{1 + z_{\rm meas}} = \frac{\lambda_{\rm wrong}}{\lambda_{\rm true}}\, ,
\end{equation}
where $\lambda_{\rm wrong}$ and $\lambda_{\rm true}$ are respectively the incorrect and true wavelengths. 
In case of an interloper detection, the ratio in Eq.~\eqref{eq:zmeas-vs-ztrue} significantly deviates from unity.

\subsection{Classification of interlopers}
\label{sec:classification_of_interlopers}

There are two possible ways of incorrectly identifying the \ha\, line, leading to two distinct types of interlopers:
\begin{itemize}
\item `Line interlopers' are galaxies with detected genuine emission lines incorrectly classified as \ha. Apart from \ha\,, some other emission lines are sufficiently intense to be detected (see 
Euclid Collaboration: Granett et al., in prep., 
and Sect.~\ref{sec:foreseen-interlopers}). Those lines enter the NISP wavelength range one by one in different redshift intervals, leading to possible line misidentifications. The systematic error in the redshift estimate given by Eq.~\eqref{eq:zmeas-vs-ztrue} is deterministic and it depends on the ratio between the \ha \,wavelength and the one of the misidentified line.

\item `Noise interlopers' consists of objects from the parent sample that happen to enter the spectroscopic catalogue because of the presence an intense noise fluctuation resembling an emission line in their low-S/N spectrum.
Typically, they correspond to galaxies whose spectral features are either weak or located outside the wavelength range of the instrument. Stars can also be mistaken for galaxies when their spectra have a low S/N. All these objects have featureless spectra and high noise. This misidentification results in a redshift estimate that is catastrophically different from the true redshift. Unlike line interlopers, however, there is no one-to-one relationship between the true and measured redshifts in this case, since the detection is based on random spikes in the noisy spectra. 
\end{itemize}

\subsection{Foreseen \Euclid interloper galaxies}
\label{sec:foreseen-interlopers}

To characterize the population of interloper galaxies expected in the \Euclid spectroscopic sample, we make use of the \Euclid redshift error baseline model derived from end-to-end simulations by  
Euclid Collaboration: Granett et al. (in prep.). These simulations rely on statistical tools that bypass the complexity of the Euclid spectroscopic data reduction pipeline, producing realistic, though approximate, data products in significantly less computational time.
The redshift error model was calibrated using a set of simulated NISP spectra with noise characteristics mimicking those expected in the EWS. The spectra were constructed from the EL-COSMOS catalogue \citep{cosmos2020} with the Fastspec simulator 
(Euclid Collaboration: Granett et al., in prep.)
and analysed by the OU-SPE\footnote{Organisation Unit SPEctral extraction and redshift data.} processing function of the \Euclid Science Ground Segment to measure spectral features and redshift.
  
From these simulations, two emission lines were identified as primary sources of redshift errors from line misidentification: \ion{O}{iii}\,$\lambda$5008 and \ion{S}{iii}$\,\lambda$9531 (hereafter noted \ion{O}{iii} and \ion{S}{iii}). The \ion{O}{iii} line is the brightest line in the 
$\mathrm{H}\,\beta\,\lambda$4863-\ion{O}{iii}\,$\lambda\lambda$4959,5008 complex.
Its visibility range is about $1.5 < z_{\rm true} < 2.7$. Within the range $1.5 < z_{\rm true} < 1.8$, line misidentification is less likely since both \ha\, and the \ion{O}{iii} lines are potentially detectable. Line misidentification can increase at $z_{\rm true} > 1.8$, where the \ha\, line cannot be observed any more: in this range, all the prominent \ion{O}{iii} lines can be mistaken for \ha. Moving to higher redshifts, we expect the misidentification probability to decrease in general simply because the number of observable sources decreases with the redshift. Given the smaller emission wavelength of the \ion{O}{iii} line with respect to the \ha\, line, \ion{O}{iii} interlopers correspond to sources which are in reality further away compared to their estimated distance. The \ion{S}{iii} line is detectable in NISP over the redshift range $0.3 < z_{\rm true} < 0.94$ and there is only a small redshift interval where both \ion{S}{iii} and \ha~ can be detected simultaneously. Since the \ion{S}{iii} emission wavelength is larger than the \ha\, one, \ion{S}{iii} interlopers are systematically positioned further away than their actual distance. More details on the emission lines of interest for this study can be found in \cite{Gabarra-EP31}.

In addition to line interlopers, we expect to observe noise interlopers. Given the diverse nature of possible noise interlopers and the inherent random process of detecting a noise line mistaken for \ha\,, we expect a fairly uniform distribution of these interlopers across different wavelengths and redshift. This is consistent with the fact that such noise interlopers can originate from virtually any true redshift.

\begin{figure}[htbp]
    \centering
    \includegraphics[width=0.49\textwidth]{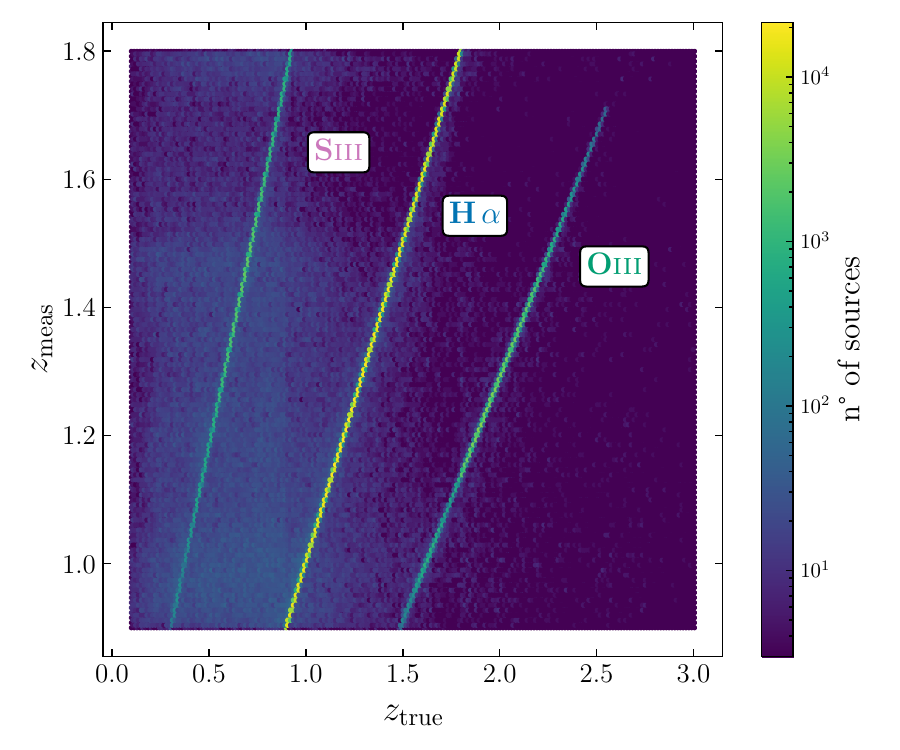}
    \caption{Representation of interloper galaxies in the ($z_{\rm true}$, $z_{\rm meas}$) plane, representative of the \Euclid spectroscopic selection. The extent of the vertical axis corresponds to the baseline observed redshift range used for the spectroscopic analysis. The plot refers to one of the EuclidLargeMocks (see 
    \citealt{EP-Monaco1},
    and Sect.~\ref{sec:elm-description}).}
    \label{fig:ztrue-vs-zmeas-2D}
\end{figure}

Figure \ref{fig:ztrue-vs-zmeas-2D} offers a graphical representation of the \Euclid interlopers' properties just described. It shows the expected distribution of the measured redshifts $z_{\rm meas}$ of galaxies versus their true redshifts $z_{\rm true}$, and highlights the off-diagonal location of all types of interlopers. 
The bisector line corresponds to the `correct galaxies', i.e. those galaxies whose redshift was correctly measured within the instrumental uncertainty. 
The coloured tracks correspond to line interlopers (identified by their labels). When represented in the $(z_{\rm true}, z_{\rm meas})$ plane, line interlopers lie along straight lines with slope different from one, whose characteristic value is determined by Eq.~\eqref{eq:zmeas-vs-ztrue}. The shaded blue distribution in the background consists of noise interlopers.
The lack of correlation in their detection randomizes the positions of the noise interlopers in the $(z_{\rm true}, z_{\rm meas})$ plane, forming a diffuse cloud of points.

\subsection{Impact of interlopers on the galaxy density contrast}
\label{sec:damping-shift-interlopers}

To estimate galaxy clustering properties, we first need to estimate the comoving positions of all galaxies from their measured redshift. For an interloper galaxy, the estimated comoving position $\mathbf{x}$ differs from the true one $\mathbf{y}$ 
due to the radial displacement caused by the incorrect redshift determination \citep{Pullen_2016}.
Quantitatively, we can link the interlopers observed and true positions geometrically via
\begin{align}
    (\mathbf{x}_{\parallel}, \mathbf{x}_{\perp}) = (\gamma_{\parallel}\mathbf{y}_{\parallel}, \gamma_{\perp}\mathbf{y}_{\perp})\, ,
\end{align}
where $\mathbf{x}_{\parallel}$, $\mathbf{y}_{\parallel}$, $\mathbf{x}_{\perp}$, and $\mathbf{y}_{\perp}$ are the radial and transverse components of the position vectors, and
\begin{eqnarray} 
\label{eq:gamma-def}
\gamma_{\perp} &=&\dfrac{D_\sfont{A}(z_{\rm meas})}{D_\sfont{A}\left(z_{\text {true }}\right)} \label{eq:gamma-perp} \, ,\\ 
\gamma_{\|} &=& \dfrac{\left(1+z_{\rm meas}\right)/H(z_{\rm meas})}{\left(1+z_{\text {true }}\right)/H(z_{\text{true}})} \, .\label{eq:gamma-parallel}
\end{eqnarray}
This is a geometrical dilation whose expression is analogous to AP distortions but which depends on different quantities: the ratio between the comoving transverse distances, $D_\sfont{A}$, and the Hubble parameters, $H$, at the true and observed redshifts.

\begin{table}
\caption{\aabf{Geometric distortion factors $\gamma$ for misidentified \ion{S}{iii} and \ion{O}{iii} galaxies in the four baseline \Euclid spectroscopic bins.}}
\centering
\begin{adjustbox}{max width=0.49\textwidth}
\begin{tabular}{|l|S|S|S|S|}
\hline
{} & {$\gamma_{\perp, [\ion{S}{iii}]}$} & {$\gamma_{\parallel, [\ion{S}{iii}]}$} & {$\gamma_{\perp, [\ion{O}{iii}]}$} & {$\gamma_{\parallel, [\ion{O}{iii}]}$} \\
\hline
{$z \in [0.9, 1.1]$} & 2.23 & 0.99 & 0.72 & 1.08 \\
{$z \in [1.1, 1.3]$} & 1.93 & 0.96 & 0.75 & 1.09 \\
{$z \in [1.3, 1.5]$} & 1.75 & 0.94 & 0.78 & 1.10 \\
{$z \in [1.5, 1.8]$} & 1.61 & 0.91 & 0.80 & 1.11 \\
\hline
\end{tabular}
\end{adjustbox}
\label{tab:gammatable}
\end{table}

For line interlopers, the value of the $\gamma$ parameters is well-defined, since the relation between the observed and true redshifts is deterministic (see Eq.~\ref{eq:zmeas-vs-ztrue}). 
\aabf{
The values of $\gamma_{\parallel}$ and $\gamma_{\perp}$  are larger or smaller than unity depending on whether line interlopers are located at redshifts higher or lower than that of the correct galaxies. If the interloper redshift is higher, the estimated separation between any two line interlopers is smaller than the true one, while, if the redshift is lower, the estimated separations are overestimated. The parameters  $\gamma_{\parallel}$ and $\gamma_{\perp}$ quantify this effect for the parallel and perpendicular components of the separation vectors, respectively.  Table~\ref{tab:gammatable} reports the reference values of these parameters for the \Euclid survey, assuming the cosmological model given in Table~\ref{tab:elm-cosmology}.
} Unlike line interlopers, the relation between the true and observed redshifts of noise interlopers is not unique. As a result, there is no single gamma value that can be associated with this type of interloper. Instead, this parameter will vary according to a probability distribution function, which is in principle independent of the true source position. By extending the formalism introduced by \citet{Pullen_2016}, the galaxy density contrast $\delta$ at an observed comoving position $\mathbf{x}$ in the presence of both line and noise interlopers can be approximated as
\begin{align}
\label{eq:density-contrast-with-interlopers}
    \delta(\mathbf{x}) \,=\, &(1-f_{\rm tot})\, \delta_{\rm c} (\mathbf{x}) + 
    \sum_{i} f_{i}\, \delta_{i}(\mathbf{x}_{\parallel}/\gamma_{\parallel}, \mathbf{x}_{\perp}/\gamma_{\perp}) \\
    &+ f_{\rm n}\, \int \int \mathcal{P}_{\rm n} (\gamma_{\parallel}, \gamma_{\perp}) \, \delta_{\rm n} (\mathbf{x}_{\parallel}/\gamma_{\parallel}, \mathbf{x}_{\perp}/\gamma_{\perp}) \,\mathrm{d}\gamma_{\parallel} \, \mathrm{d}\gamma_{\perp} \,, \nonumber
\end{align}
where the subscript `c' stands for `correct' galaxies (defined at the end of Sect.~\ref{sec:foreseen-interlopers}), `n' stands for noise interlopers, the sum runs over all the types `$i$' of line interlopers, $f_{\rm tot}$ is the total fraction of contaminants accounting for all types of interlopers, $\mathcal{P}_{\rm n}$ is the joint probability distribution function of $\gamma_{\parallel}$ and $\gamma_{\perp}$ for noise interlopers, and $f_{\rm n}$ is the fraction of noise interlopers. The fraction relative to each population is defined as the ratio between the number of galaxies of a certain type with respect to the total number of galaxies in the sample as
\begin{align}
    f_{\rm g} = \frac{n_{\rm g}}{n_{\rm tot}} \, .
\end{align}
By definition, we have that
\begin{align}
    f_{\mathrm{tot}} = \sum_{\mathrm{g}} f_{\rm g} =   \sum_{i} f_{i}  + f_{\mathrm{n}} \, .
\end{align}
Equation \eqref{eq:density-contrast-with-interlopers} shows that the density contrast measured in the presence of interlopers is diluted compared to the one measured on a catalogue made only of correct galaxies. The attenuation of the signal is proportional to the total contamination fraction. This can be intuitively understood by considering the extreme case of noise interlopers, which are clustered objects randomly displaced along the line of sight, thereby leading to a smoothed version of the original galaxy density contrast. 

The contributions of interloper galaxies are weighted by the fraction of each interloper in the catalogue. We expect the contaminant terms to be subdominant and one of \Euclid survey requirements is to keep the interloper contamination fraction below the 20\% threshold \citep{EuclidSkyOverview}. Yet, these terms inevitably modify the measured clustering statistics since interlopers have their own clustering properties. 
We show the effect of interlopers on the 2PCF in the presence of the expected types and fractions of interlopers for \Euclid in Sect.~\ref{sc:2PCFmeas}.

\subsection{Interloper fractions and associated clustering properties}
\label{sec:cpc}

The EWS described in \citet{Scaramella-EP1} is complemented by periodic deeper observations on a smaller area, which constitute the Euclid Deep Survey (EDS). The EDS will be used to accurately characterise the typical EWS galaxy population, as EDS fields are meant to provide a 99\% complete and 99\% pure spectroscopic sample of the depth of the EWS, thanks to a high cumulative exposure time that will be reached along the mission \citep{EuclidSkyOverview}. By the end of the survey, the EDS will span an area of $53 \deg^2$ and be observed with both the blue and red grisms \citep{EuclidSkyOverview}. The EDS will enable us to measure the spectra of observed galaxies with a higher S/N compared to the shallower exposures of the EWS. By comparing the same fields, first observed at EDS depth and then in the EWS, we can identify and characterize all interlopers included in the contaminated EWS sample, as well as their redshift distributions. No noise interlopers are expected in EDS observations given the higher depth and higher spectrum S/N. Similarly, line interlopers should not be present in the EDS, as more than one line can be detected due to the higher S/N, leading to an unambiguous identification of the \ion{O}{iii} or \ion{S}{iii} line in the spectra for instance.

\section{\label{sc:Estimator} Estimated 2PCF in the presence of interlopers}

The galaxy clustering analysis in configuration space in \Euclid will use the galaxy 2PCF, which will be estimated using the Landy--Szalay (LS) estimator \citep{LandySzalay1993}. This estimator arises from first defining a catalogue overdensity, defined as the fractional difference between the data and random catalogue counts, and taking the auto-correlation of it. The random catalogue, which comprises randomly distributed points within the survey volume, allows the mapping of the geometry and selection function of the survey.  
Schematically, the overdensity in galaxy counts at any position $\bx$ is
\begin{equation}
    \delta(\bx) = \frac{D(\bx)-R(\bx)}{R(\bx)}
\end{equation}
and leads to the auto-correlation estimator
\begin{equation}
    \xi(\br) = \frac{\mathrm{DD}(\br) -2 \mathrm{DR}(\br) + \mathrm{RR}(\br)}{\mathrm{RR}(\br)}\, ,
\end{equation}
where $D(\bx)$ and $R(\bx)$ are data and random catalogue counts at position $\bx$, and $\mathrm{DD}(\br)$, $\mathrm{DR}(\br)$, $\mathrm{RR}(\br)$ are respectively the normalized data-data, data-random and random-random pair counts as a function of the pair separation vector $\br$. The normalization of pair counts originates from the fact that the random catalogue contains a much larger number of objects than the data catalogue, such that
\begin{align}
   \mathrm{DD}(\br) &= \frac{2 \, \widehat{\mathrm{DD}} (\br)}{N_{\rm D}(N_{\rm D}-1)}\, , \\
   \mathrm{DR}(\br) &= \frac{\widehat{\mathrm{DR}}(\br)}{N_{\rm D} N_{\rm R}}\, , \\
   \mathrm{RR}(\br) &= \frac{2 \, \widehat{\mathrm{RR}}(\br)}{N_{\rm R}(N_{\rm R}-1)}\, ,
\end{align}
where $\widehat{\mathrm{DD}}(\br)$, $\widehat{\mathrm{DR}}(\br)$, $\widehat{\mathrm{RR}}(\br)$ are raw counts, and $N_{\rm D}$ and $N_{\rm R}$ are the total number of objects in the data and random catalogues respectively. 

Similarly, by defining the overdensity of two populations $\delta_1 = (D_1 - R_1)/R_1$ and $\delta_2 = (D_2 - R_2)/R_2$, where now $D_1$ ($D_2$) and $R_1$ ($R_2$) stand for the data and random catalogue counts of the population $1$ ($2$), we obtain the 2-point cross-correlation function estimator
\begin{equation}
    \xi_{12}(\br) = \frac{\mathrm{D}_1 \mathrm{D}_2 (\br) - \mathrm{D}_1 \mathrm{R}_2 (\br) - \mathrm{R}_1 \mathrm{D}_2 (\br) + \mathrm{R}_1 \mathrm{R}_2(\br)}{\mathrm{R}_1 \mathrm{R}_2(\br)}\, ,
\end{equation}
where $\mathrm{D}_1 \mathrm{D}_2$, $\mathrm{D}_1 \mathrm{R}_2$, $\mathrm{R}_1 \mathrm{D}_2$, and $\mathrm{R}_1 \mathrm{R}_2$ are the data 1-data 2, data 1-random 2, random 1-data 2 and random 1-random 2 normalised pair counts, respectively.

We now consider the case of the measured 2PCF $\xi_{\rm m}$\footnote{The subscript `m' stands for `measured', since this is the only 2PCF that we are actually going to measure in the survey.} obtained by applying the auto-correlation estimator on a data catalogue containing redshift interlopers. We can decompose the contaminated data and random catalogue counts in three different components according to the three classes of redshifts by writing
\begin{align}
\label{eq:contam-cat-sum}
    D_{\rm m}(\bx) &= D_{\rm c}(\bx) + D_{\ell}(\bx) + D_{\rm n}(\bx)\, , \\
\label{eq:contam-random-sum}
    R_{\rm m}(\bx) &= R_{\rm c}(\bx) + R_{\ell}(\bx) + R_{\rm n}(\bx) \, ,
\end{align}
where we considered only one type of line interlopers for simplicity although the generalization to more than one is trivial. The subscripts $\rm m$, $\rm c$, $\rm \ell$, and $\rm n$ refer to measured (i.e. all observed objects), correct, line interloper, and noise interloper populations, respectively. In the random catalogues associated with correct, line interloper, and noise interloper populations,  
the radial distributions follow respectively those of correct, line interloper, and noise interloper galaxies. 
\aabf{Here, we consider a simplified case where the only systematic in the data is redshift error, with no angular mask applied. This matches the configuration of the mock catalogues used in this work and it is equivalent to assuming that radial and angular systematics can be treated independently. In this context, the angular distribution of the random points is taken to be uniform across the survey area. It is worth noting that the selection function of the \Euclid spectroscopic catalogue, based on a forward-modelling approach, does not rely on this assumption. The validity of this assumption needs to be verified with the real data. The impact of a realistic angular mask on clustering statistics is investigated in Monaco et al. (in prep.) and it will be the subject of dedicated \Euclid papers prepared in light of the first real data.}

If we now define the overdensity associated with the total contaminated catalogue $\delta_{\rm m} = (D_{\rm m} - R_{\rm m})/R_{\rm m}$, the expression for the associated auto-correlation estimator is
\begin{align}
\label{eq:contam-2pcf-RR}
    \xi_{\rm m}(\br) &= (1-f_{\rm tot})^2 \, \frac{\rm R_{\rm c} R_{\rm c} (\br)}{\rm R_{\rm{m}} R_{\rm{m}} (\br)} \xi_{\rm cc}(\br) +  f_{\ell}^2 \frac{\rm R_{\ell} R_{\ell}(\br)}{\rm R_{\rm m} R_{\rm m}(\br)} \xi_{\ell\ell}(\br) \\ \nonumber
    &+ f_{\rm n}^2 \frac{\rm R_{\rm n} R_{\rm n}(\br)}{\rm R_{\rm m} R_{\rm m}(\br)} \xi_{\rm nn}(\br)
    + 2 f_{\ell}(1-f_{\rm m}) \frac{\rm R_{\rm c} R_{\ell}(\br)}{\rm R_{\rm m} R_{\rm m}(\br)} \xi_{\rm c\ell}(\br) \\ \nonumber
    &+ 2 f_{\rm n}(1-f_{\rm tot}) \frac{\rm R_{\rm c} R_{\rm n}(\br)}{\rm R_{\rm m} R_{\rm m}(\br)} \xi_{\rm cn}(\br) + 2 f_{\ell} f_{\rm n}\frac{\rm R_{\ell} R_{\rm n}(\br)}{\rm R_{\rm m} R_{\rm m}(\br)} \xi_{\rm \ell n}(\br)\, ,
\end{align}
where we identified $\xi_{\rm cc}$, $\xi_{\rm \ell\ell}$, $\xi_{\rm nn}$ as the auto-correlation function of the correct, line interloper, and noise interloper populations respectively, and  $\xi_{\rm c\ell}$, $\xi_{\rm cn}$, $\xi_{\rm \ell n}$ as the correct-line interloper, correct-noise interloper, line-noise interlopers cross-correlation functions respectively. It is worth emphasising that, except for $\xi_{\rm cc}$, all correlation functions in the right-hand side of Eq.~\eqref{eq:contam-2pcf-RR} are the observed 2PCF and not the intrinsic ones, since they quantify the spatial correlation of misplaced objects.

In the right-hand side of Eq.~\eqref{eq:contam-2pcf-RR}, the random-random pair counts $\mathrm{R}_i\mathrm{R}_j$, where $i, j \in {\rm \{m,c,\ell,n  \}}$, correspond to the (normalized) random-random cross pairs associated with the different populations. They form ratios that factorize the different terms and, in turn, add an additional scale dependence to $\xi_{\rm m}(\br)$. In those ratios, the pair counts in the numerator and denominator differ only in the radial distribution of the associated random catalogues. Under the hypothesis of a mild difference in the observed radial distribution of the different sub-populations, the ratios of random-random pairs should tend to unity and Eq.~\eqref{eq:contam-2pcf-RR} be approximated by
\begin{align}
\label{eq:contam-2pcf-const}
    \xi_{\rm m}(\br) =& \, (1-f_{\rm tot})^2 \, \xit(\br) + f_{\ell}^2 \, \xil(\br)+ f_{\rm n}^2 \, \xin(\br)\\ \nonumber
    &+ \, 2 f_{\ell}(1-f_{\rm tot})\,  \xitl(\br) + 2 f_{\rm n}(1-f_{\rm tot})\, \xitn(\br) + 2 f_{\ell} f_{\rm n} \, \xiln(\br) \, .
\end{align}
\aabf{The validity of this hypothesis in our reference mock catalogues is tested in Sect.~\ref{sec:MCMC-models}, where we directly assess the performance of a model that ignores the radial dependence of the prefactors. As shown in Appendix \ref{sec:AppendixPrefactors}, this dependence can be significant for certain types of interlopers (e.g. \ion{O}{iii}) in specific redshift ranges. Nevertheless, the overall impact remains negligible due to the small amplitude of the corresponding prefactor.}
If more than one population of line interlopers contaminate the catalogue, then Eqs.~\eqref{eq:contam-2pcf-RR} and \eqref{eq:contam-2pcf-const} will include all corresponding auto-correlation functions and the cross-correlations with all other types of objects that were included in the sample. 

Evaluating the prefactors in Eq.~\eqref{eq:contam-2pcf-RR} requires building three random catalogues, where points radially sample the redshift distribution of correct $N_{\rm c}(z)$, line interloper $N_{\ell}(z)$, and noise interloper $N_{\rm n}(z)$ populations. While the random catalogue of the contaminated sample can be generated using the observed redshift distribution of the objects in the EWS, generating the random catalogues of  each object type is less trivial. These could be either modelled or be measured from the samples of interlopers identified in the EDS.

In light of the expectations for the measured 2PCF in the presence of redshift interlopers, the goals of this study are two-fold: (1) to assess the relative amplitude of each term on the right-hand side of Eq.~\eqref{eq:contam-2pcf-RR} with respect to the total signal and relevance of the scale-dependent prefactors; (2) to test our capability of constraining cosmological parameters building a theoretical model of only a subset of those terms.

\section{\label{sec:catalogues}Simulated datasets}

\subsection{EuclidLargeMocks and contamination strategy}
\label{sec:elm-description}

\begin{table}
\caption{Cosmological parameters that define the flat $\Lambda$CDM cosmology used to perform the EuclidLargeMocks parent simulations.}
\centering
\begin{adjustbox}{max width=0.49\textwidth}
\begin{tabular}{|c|c|c|c|c|c|c|c|c|}
\hline
$\Omega_{\rm m}$ & $\Omega_{\rm b}$ & $\Omega_{\rm CDM}$ & $\Omega_{\rm de}$ & $h$  & $\sigma_8$ & $n_s$ & $N_{\rm eff}$ & $m_{\nu} \, \left[\si{eV}\right]$ \\ 
\hline
0.32             & 0.049            & 0.270              & 0.68               & 0.67 & 0.83       & 0.96  & 3.046         & 0         \\ 
\hline
\end{tabular}
\end{adjustbox}
\label{tab:elm-cosmology}
\end{table}

We based our analysis on a set of \num{1000} \Euclid-like simulated mock catalogues, dubbed EuclidLargeMocks
\citep{EP-Monaco1},
which was extracted from a suite of numerical simulations relying on approximated perturbation techniques \citep{Monaco2002,Munari2017}. We list in Table \ref{tab:elm-cosmology} the cosmological parameters used to set up those simulations. The galaxy catalogues extracted from these simulations are lightcones with an angular footprint on the sky of a circle with radius $30\degree$ and spanning the redshift range $0 < z_{\rm true} < 3$. The area of the cone, \num{2763} deg$^2$, is slightly larger than the \num{2500} deg$^2$ expected for the first Data Release (DR1) of the EWS. Moreover, the angular footprint almost encompasses the north and south extents of the DR1 footprint, as planned before launch. These catalogues provide a minimal amount of information for each galaxy: sky coordinates, true redshift including peculiar velocities, and \ha\, line flux. The catalogues are limited to $f_{\ha}>10^{-16}$ erg s$^{-1}$ cm$^{-2}$, that is, half of the fiducial flux limit of the {\Euclid} spectroscopic sample. This is due to the fact that the transition from high to vanishing completeness is not expected to be sharp, so the sample will contain a sizeable fraction of galaxies below the fiducial limit 
(Euclid Collaboration: Granett et al., in prep.).

The measured redshifts have been added to the catalogues using a probabilistic model calibrated on an end-to-end simulation of observations 
(Euclid Collaboration: Granett et al., in prep.).
This pixel-level simulation of 1D spectra has been produced with the FastSpec simulator, processed with the OU-SPE processing function, and eventually used to model the conditional probability distribution function (PDF) $P(z_{\rm meas} | z_{\rm true})$ of the measured redshift $z_{\rm meas}$ given the true one $z_{\rm true}$. This probability is modelled with a mixture of Gaussian PDF with standard deviation of $\sigma_{0,z}=0.001$ for correct galaxies and line interlopers (suitably rescaled for line interlopers), and a broad distribution for noise interlopers. The implementation in the EuclidLargeMocks relies on computing $P(z_{\rm meas} | z_{\rm true})$ at the true redshift of each galaxy and randomly sample the distribution to obtain $z_{\rm meas}$. Galaxies for which $|z_{\rm meas} - z_{\rm true}|<3\,\sigma_{0,z}$ are tagged as correct galaxies, while galaxies whose redshift is within $3\,\sigma_{0,z}$ of the redshift corresponding to a line interloper are tagged as such. The remaining galaxies are tagged as noise interlopers. A close inspection of the redshift PDF reveals that the PDF of noise interlopers overlaps with that of correct galaxies and line interlopers. In particular at a given $z_{\rm true}$, the probability of having noise interloper redshifts within $5\,\sigma_{0,z}$ around $z_{\rm true}$ is not completely negligible.
This contribution could be removed by a more permissive separation of correct galaxies and noise interlopers. Conversely, this approach makes it impossible to separate truly correct galaxies from noise interlopers that happen to have a roughly correct redshift by chance. 

\aabf{It is worth noticing that, in our implementation, all types of interlopers are drawn from a parent sample of ELGs at $z<3$. In reality, noise interlopers should be drawn from the photometric sample of \Euclid galaxies, which are expected to be fainter and therefore less clustered than the brighter ELGs. As a result, drawing noise interlopers from an ELG parent sample overestimates their clustering amplitude and exaggerates their impact on the clustering analysis. This choice, however, provides a deliberately pessimistic scenario to stress-test our interloper models.}
Finally, this choice does not represent the small fraction of stars that are not effectively separated from galaxies and acquire a redshift by chance.
 
Table \ref{tab:frac-zbins_realistic} lists the mean fractions of contaminants in the EuclidLargeMocks for all considered spectroscopic redshift bins. The variation with redshift of the fractions for the different types of interlopers is determined by the corresponding visibility range of the emission line within the NISP wavelength range (see Sect.~\ref{sc:Interlopers}). We elaborate later on the consequence of such differences on the impact on the clustering analysis.

\begin{table}
\caption{Mean fractions of the different interloper types in the four spectroscopic redshift bins in the contaminated EuclidLargeMocks.}
\centering
\begin{adjustbox}{max width=0.49\textwidth}
\begin{tabular}{|l|S|S|S|S|}
\hline
{} &  {$z \in \left[0.9, 1.1\right]$} &   {$z \in \left[1.1, 1.3\right]$} &   {$z \in \left[1.3, 1.5\right]$} &   {$z \in \left[1.5, 1.8\right]$} \\
\hline
$\mathrm{\ion{O}{iii}}$ & 0.03 & 0.12 & 0.09 & 0.01 \\
$\mathrm{\ion{S}{iii}}$ & 0.01 & 0.03 & 0.08 & 0.07 \\
noise & 0.12 & 0.08 & 0.08 & 0.06 \\
\hline
\end{tabular}
\end{adjustbox}
\label{tab:frac-zbins_realistic}
\end{table}

\subsection{Random catalogues}

We used a single set of random catalogues (i.e. one random for each type of galaxy) to characterize the selection function of the sample and compute the 2PCF for all mocks. The radial distribution of random points have been generated by sampling the redshift distribution averaged on the first 100 mocks, in order to smooth out radial fluctuations across individual mock catalogues 
(for details, see Euclid Collaboration: Lee et al., in prep.).
The requirements for \Euclid 2PCF estimation impose that random catalogues should be at least \num{50} times larger than the corresponding galaxy catalogue to minimize the estimator variance \citep{EP-delaTorre}. Therefore, the random catalogue of each population must be at least 50 times larger than the corresponding galaxy catalogues and 
we created random catalogues with $51 \times \bar{N}_{i}$ objects, where $\bar{N}_i$ is the mean number of sources for each galaxy type $i$ averaged across the first 100 mocks.
The random catalogue of the contaminated sample is then obtained by combining the random catalogues of the single populations: correct galaxies, noise interlopers, and the various types of line interlopers.

\subsection{2PCF estimation}

In order to estimate the 2PCF, we made use of the methodology and software developed for estimating the three-dimensional 2PCF within the \Euclid Science Ground Segment \citep{EP-delaTorre}. The latter utilizes the minimum-variance LS estimator and enables the use of the random split method \citep{ElinaKeihanen2019} to speed up the computation. We evaluated the monopole, quadrupole, and hexadecapole moments of the anisotropic 2PCF  using \num{40} equally spaced bins in separation $r \in [0,200]\, \hMpc$ ($\Delta r = 5\, \hMpc$) and \num{200} bins in $\mu$ within $\mu \in [-1,1]$. We computed all terms in Eq.~\eqref{eq:contam-2pcf-RR}, including  both the auto- and cross-correlation functions of the different populations but also the random-random pair counts that appear in the prefactors of Eq.~\eqref{eq:contam-2pcf-RR}. To transform the redshift of the mock galaxies into distance we used the same cosmological model as used to generate the parent simulations. Our analysis focuses on the baseline redshift intervals for the \Euclid galaxy clustering analysis: $z \in \left[0.9, 1.1\right]$, $z \in \left[1.1, 1.3\right]$, $z \in \left[1.3, 1.5\right]$, and $z \in \left[1.5, 1.8\right]$.

\section{\label{sc:2PCFmeas}The contribution of interlopers to the EuclidLargeMocks 2PCF}

\begin{figure}[htbp]
    \centering
    \includegraphics[width=\linewidth]{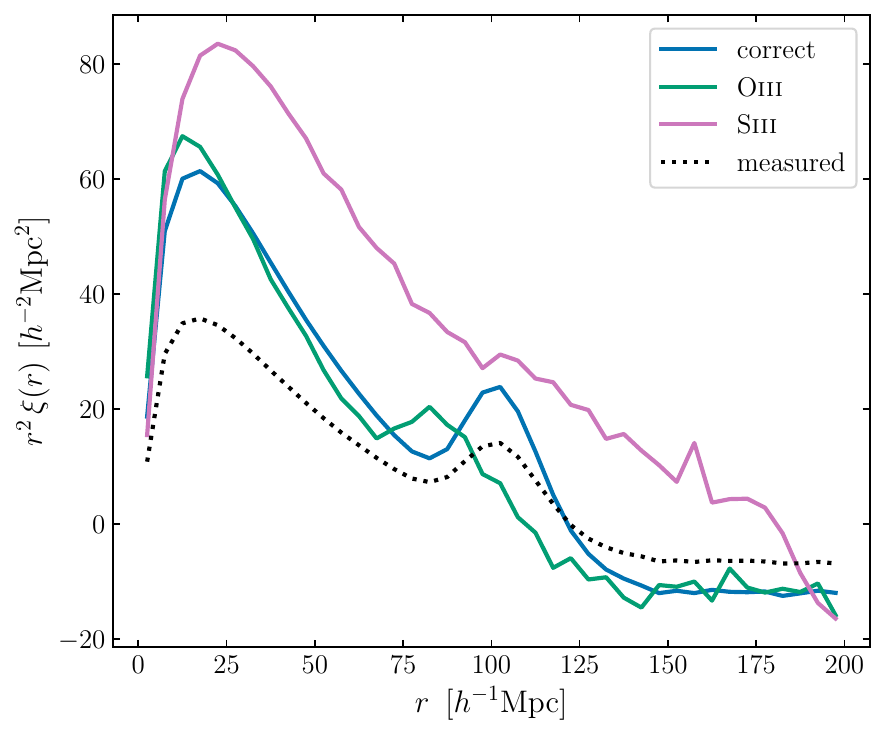}
    \caption{Monopole of the contaminated sample auto-correlation (dashed black line) compared to the intrinsic correct galaxy (blue line) and line interloper (green and pink line) auto-correlations, not weighted by the prefactors in Eq.~\eqref{eq:contam-2pcf-RR}. The 2PCF are averaged over the EuclidLargeMocks in $z \in [1.3, 1.5]$. We can appreciate both the dilution of the clustering amplitude in the presence of contamination and the distortion of the line interlopers' signal, in particular the shift of the BAO peak.}
    \label{fig:2pcf-contam-target-interlopers}
\end{figure}

We use Fig.~\ref{fig:2pcf-contam-target-interlopers} as an example to illustrate how the density contrast introduced in Eq.~\eqref{eq:density-contrast-with-interlopers} in the presence of interlopers translates into the galaxy 2PCF measurements. The 2PCF of correct galaxies (solid blue line) is compared with those of the \ion{O}{iii} and \ion{S}{iii} interlopers (solid green and pink line) and the total measured 2PCF (dotted black line). All 2PCFs are estimated using the mocks presented in Sect.~\ref{sec:catalogues}. The correct galaxies' and line interlopers' 2PCF are the intrinsic auto-correlation functions of the corresponding population, i.e they are not weighted by their prefactors as in Eq.~\eqref{eq:contam-2pcf-RR}. We can see that the different population 2PCFs are characterized by different shapes that cause a broadening of the BAO peak in the resulting measured 2PCF. The 2PCF of line interlopers is shifted and distorted compared to that of correct galaxies. For \ion{S}{iii} interlopers, the 2PCF is broadened towards larger separation scales, while for \ion{O}{iii} interlopers, it is compressed towards smaller scales. This effect is particularly evident when examining the corresponding shifts of the BAO peak position. These results demonstrate the importance of modelling the clustering properties and abundance of all types of interlopers to account for contamination effect on 2PCF measurements. 

In the mock catalogues, we can unambiguously identify and separate all types of objects. This allows us to compute exactly all terms in Eq.~\eqref{eq:contam-2pcf-RR}, both the correlation functions and their prefactors.
This possibility offers the opportunity to evaluate
the contribution of each term to the total correlation function of the contaminated catalogue.
Moreover, we can evaluate the residuals that we obtain if we neglect some terms on the right-hand side of Eq.~\eqref{eq:contam-2pcf-RR}. This evaluation allows us to quantify the most relevant terms that we should include in the theoretical model when the measured signal is fitted to extract cosmological information. 
While the modelling of the autocorrelation of correct galaxies and line interlopers is relatively straightforward, that of the cross-correlations of the various interlopers, characterised by very different redshift distributions, is considerably more challenging. It can be obtained either phenomenologically from  direct measurements in the EDS or theoretically under some simplifying assumptions \citep[see][]{Foroozan_2022}.

For the sake of both simplicity and generality, we only show the results for two specific redshift bins representative of the different types and fractions of interloper galaxies. 
In the first one, $z \in [0.9, 1.1]$, most contaminants are noise interlopers and constitute $10\%$ of the observed catalogue. Line interlopers account for only few per cents. In the second redshift bin, $z \in [1.3, 1.5]$, the fractions of noise and line interlopers are comparable, around 10\% each.

\subsection{Amplitude and shape of the different terms}

\begin{figure*}[htbp]
    \centering
    
    \includegraphics[width=\textwidth]{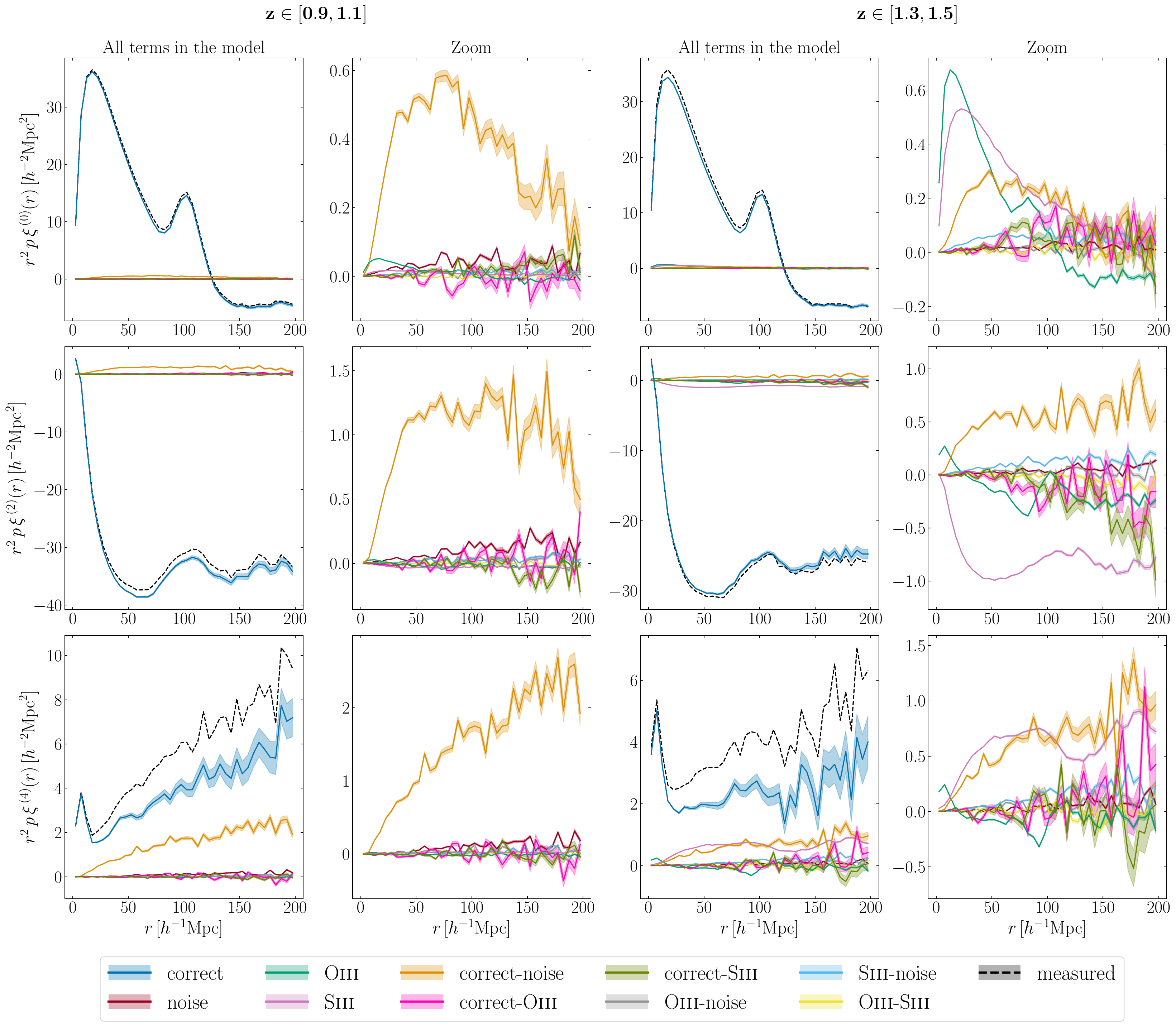}
    \caption{Monopole, quadrupole,  and hexadecapole moments of all terms in Eq.~\eqref{eq:contam-2pcf-RR} averaged over all mock catalogues for $z \in \left[0.9, 1.1\right]$ (\textit{left}) and $z \in \left[1.3, 1.5\right]$ (\textit{right}). All terms comprise the correlation function and the corresponding prefactor. To simplify the visualization of all terms, the rightmost column of each panel shows a zoom-in on the smallest contributions in the corresponding redshift bin.}
    \label{fig:multipoles-z1-z3}
\end{figure*}

Figure \ref{fig:multipoles-z1-z3} shows the monopole (top panels), quadrupole (centre), and hexadecapole (bottom) of all the auto- and cross-correlation functions in Eq.~\eqref{eq:contam-2pcf-RR}, weighted by their corresponding prefactors that we generically denote by $p$ on the $y$-axis label.
The multipole correlation functions have been averaged over all \num{1000} EuclidLargeMocks. 
The panels in the first and third columns show all contributions for the two redshift bins under consideration, as indicated by the labels. For each, a zoomed-in view of the smallest contributions is displayed in the second and fourth columns, highlighting the scale dependence of all interloper contributions.
As expected, the major contribution to the total signal comes from correct galaxies in both redshift bins being the most numerous type of galaxy. 
The other terms are all subdominant, although not negligible. Their relevance depends on the redshift bin considered.

At $z \in [0.9, 1.1]$, as shown on the left panels of 
Fig.~\ref{fig:multipoles-z1-z3}, the most prominent contribution apart from the correct-correct one is the correct-noise correlation signal, which significantly differs from zero. This is not unexpected, since we know that 
the redshift PDF of noise interlopers overlaps with that of correct galaxies (see Sect.~\ref{sec:elm-description}).
In this case, this term is equivalent to the autocorrelation of correct galaxies computed on a sample in which some sources have a larger error on redshift. However, the intensity of the signal ultimately depends on how catastrophic redshift errors are defined with respect to random ones. More details on the origin of this contribution in the EuclidLargeMocks can be found in Appendix \ref{sec:AppendixTargetNoise}. The line interlopers contributions, while characterized by a large intrinsic auto-correlation signal, are damped by the small amplitude of their prefactors, given their small fractions in this redshift interval.

The situation is slightly different at $z \in [1.3, 1.5]$ shown in the right panels of Fig.~\ref{fig:multipoles-z1-z3}-\textit{right}. In this redshift bin, the fractions of noise, \ion{O}{iii}, and \ion{S}{iii} interlopers are comparable. As a consequence, the amplitude of the line interloper auto correlations \aabf{(dark green and violet lines for \ion{O}{iii} and \ion{S}{iii} respectively)} is higher compared to the low-redshift bin and is of the same order as that of the correct-noise correlation. Given the enhancement of the line interlopers' \aabf{auto correlation},
we can appreciate the distortion induced in the shape of their 2PCF, as previously illustrated by the broadening and shifting of the BAO peak in the auto-correlation function of the line interlopers in Fig.~\ref{fig:2pcf-contam-target-interlopers}.
The contribution of line interlopers to the contaminated 2PCF is particularly evident on small scales in the monopole, where they constitute the second most important contribution after correct galaxies.

In both redshift bins, the other terms in Eq.~\eqref{eq:contam-2pcf-RR} either have a negligible amplitude or are very noisy. This is expected for the line-correct and line-line cross-correlation terms, since these populations are very far apart ($\Delta z > 0.6$, or $\Delta r > 846 \,\hMpc$ in terms of comoving distances). The correlation function amplitude of objects characterized by very broad redshift distributions, particularly noise interlopers, is expected to be very small as well. Overall, we cannot appreciate any significant shift or broadening of the BAO peak in the contaminated signal with respect to the correct-correct contribution.

\subsection{
Simplified models for the measured correlation function}
\label{sec:residuals}

In this section, we focus on the residual error obtained when we neglect some terms on the right side of Eq.~\eqref{eq:contam-2pcf-RR}, that is, when
considering an incomplete modelling of the measured correlation function. The first model
considered is one that ignores the specific contamination and only accounts for the correct galaxy contribution attenuated by the prefactor
\begin{equation}
\label{eq:sys-err-onlytarget}
    \xic = (1-f_{\rm tot})^2 \frac{\rm \rt\rt}{\rm \rc\rc} \xit \, .
\end{equation}
In the second model we include the autocorrelation terms for both noise and line interlopers
\begin{equation}
\label{eq:sys-err-onlyauto}
    \xic = (1-f_{\rm tot})^2 \frac{\rm \rt\rt}{\rm \rc \rc} \xit + f_{\rm \ell}^2 \frac{\rm \rl \rl}{\rm \rc \rc} \xil + f_{\rm n}^2 \frac{\rm \rn \rn}{\rm \rc \rc} \xin \, .
\end{equation}
Finally, if we further include the correct-noise cross-correlation term that features prominently in Fig.~\ref{fig:multipoles-z1-z3} we have
\begin{align}
\label{eq:sys-err-auto-and-tn}
    \xic \,=\, & (1-f_{\rm tot})^2  \frac{\rm \rt\rt}{\rm \rc\rc} \xit + f_{\rm \ell}^2 \frac{\rm \rl\rl}{\rm \rc\rc} \xil + f_{\rm n}^2 \frac{\rm \rn\rn}{\rm \rc\rc} \xin \\ \nonumber
    &+ 2 (1-f_{\rm tot})f_{\rm n}\frac{\rm \rt\rn}{\rm \rc\rc} \xitn \, .
\end{align}
For each of the three models, we compared the residuals $(\xic \,-\, \rm{model})$ to the expected statistical uncertainty, $\sigma_{\rm m}$, on the measured 2PCF $\xic$ and looked for the minimal set of terms that brought the systematic error below $\sigma_{\rm m}$ and $10\% \, \sigma_{\rm m}$.

\begin{figure*}[htbp]
\label{fig:2pcf-measured-multipoles-z3}
    \centering
    \includegraphics[width=0.45\textwidth]{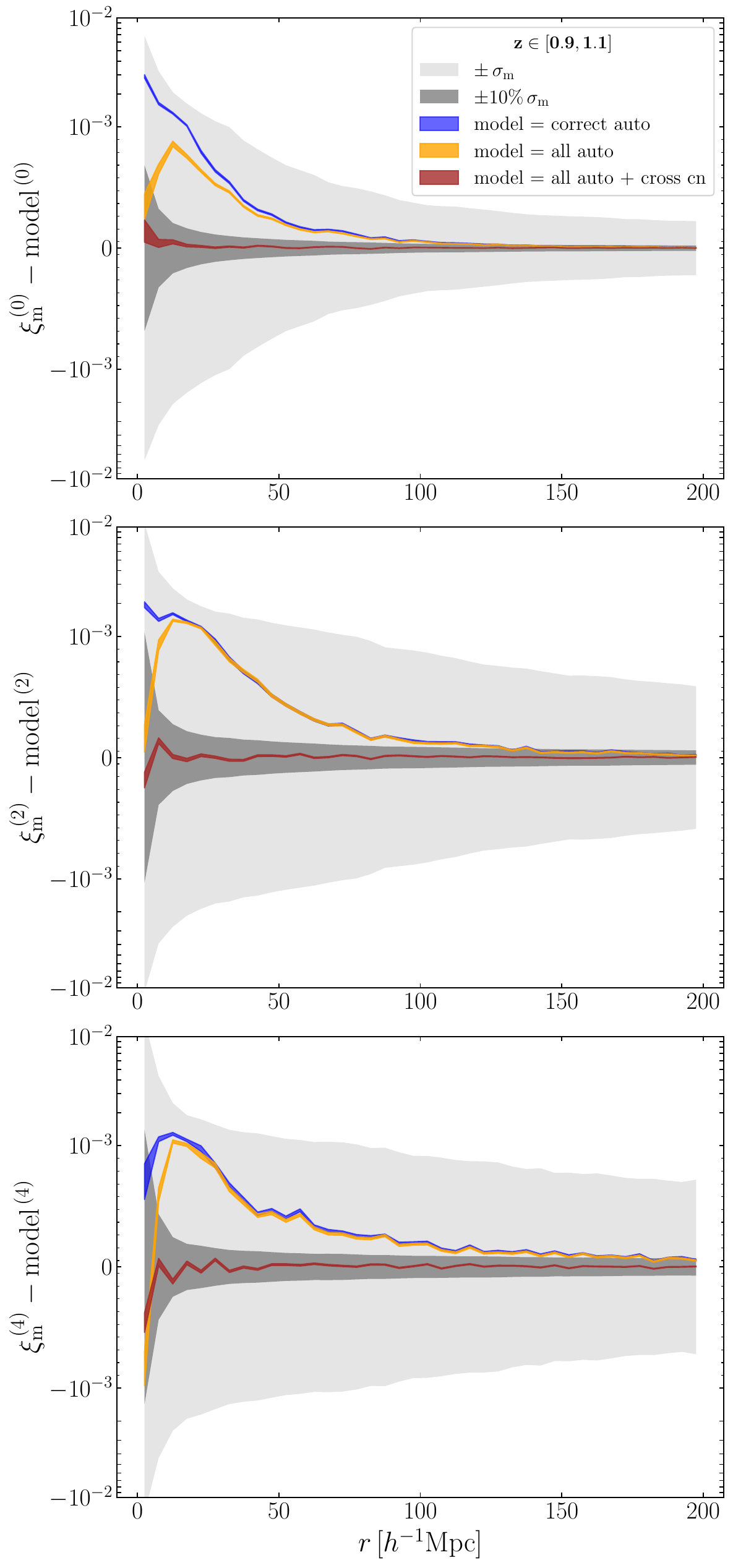}
    \includegraphics[width=0.45\textwidth]{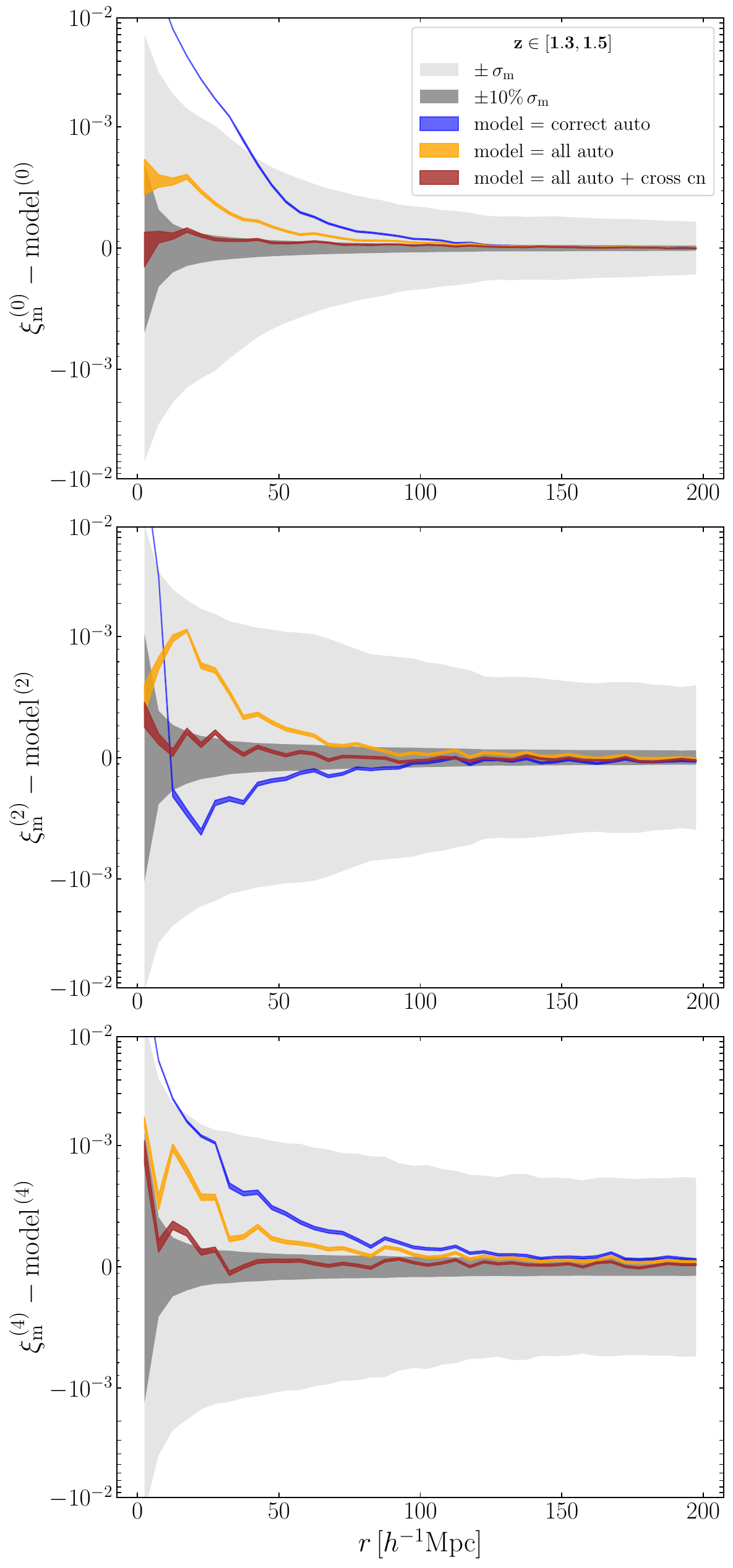}
    \caption{
    Systematic errors in the monopole, quadrupole, and hexadecapole moments in the case of an incomplete parameterization, for $z \in \left[0.9, 1.1\right]$ (\textit{left}) and $z \in \left[1.3, 1.5\right]$ (\textit{right}). The grey bands correspond to the statistical uncertainty $\sigma_{\rm m}$ on the measured monopole and to $10\% \,\sigma_{\rm m}$. The $y$ axis scale is linear between $-10^{-3}$ and $10^{-3}$, and symmetric logarithmic elsewhere.}
    \label{fig:sys-errs-z1-z3}
\end{figure*}

Figure \ref{fig:sys-errs-z1-z3} shows the amplitude of systematic error induced by using the approximate models described by Eq.~\eqref{eq:sys-err-onlytarget} (blue line), Eq.~\eqref{eq:sys-err-onlyauto} (golden line), and Eq.~\eqref{eq:sys-err-auto-and-tn} (brown line), for $z \in [0.9, 1.1]$ (left panel) and $z \in [1.3, 1.5]$ (right panel), in monopole, quadrupole, and hexadecapole correlation functions. Systematic errors, defined as the difference between the measured and modelled quantities, are averaged over the 1000 mocks and the coloured bands around them represent the standard deviation around the mean. The grey bands represent the statistical uncertainty, $\sigma_{\rm m}$, on the measured 2PCF, i.e. the statistical error on a single realization and $10\%$ of its value. The value of $\sigma_{\rm m}$ is obtained from the scatter of $\xi_{\rm m}$ multipoles among mocks realizations, whose area is on the order of the total DR1 area.

At $z \in [0.9, 1.1]$, the simplest modelling including only correct galaxies autocorrelation leads to a systematic error smaller than the expected statistical uncertainty on 2PCF measurements in DR1 at all separations. Adding interlopers auto-correlations has only an impact on the smallest scales, where the line interlopers auto-correlation is highest, but has no effect on scales above 30$\, \hMpc$. The systematic error falls below 10\% of $\sigma_{\rm m}$ at all scales only when including the cross-correlation term between correct galaxies and noise interlopers. The systematic error has a slightly different behaviour at $z \in [1.3, 1.5]$, which is directly linked to the different interloper fractions in this redshift range with respect to the previous one.
The residuals in the monopole when using the simplest model are larger than the statistical uncertainty by up to about $40 \, \hMpc$. In this case, the introduction of the line interlopers auto-correlation is crucial as it brings the systematic error below the statistical one. This is consistent with the monopole amplitudes reported in Fig.~\ref{fig:multipoles-z1-z3}, where we see that the line interlopers signal is prominent at those scales. Despite these differences, the addition of the correct-noise cross-correlation term in this redshift range is required to bring the residuals below $10\%$ of $\sigma_{\rm m}$. Overall, in all models and considered redshift bins, the amplitude of the systematic error decreases with the separation, eventually approaching or dropping below 10\% of the statistical uncertainty beyond $100\, \hMpc\,$.

The adequacy of a given 2PCF model in describing the measured 2PCF in the presence of interlopers must be ultimately evaluated upon its ability to extract unbiased scientific information. The results presented in this section
help us to build an effective model for the measured 2PCF that is both accurate and as simple as possible. In other words, measuring a significant systematic effect at the level of the 2PCF measurements does not imply an equally significant decrease in the precision and accuracy of the estimated cosmological parameters. The ultimate goal of this analysis, which is detailed in the following sections, is to comprehensively assess the impact of interlopers on the inference of some cosmological parameters derived from galaxy clustering measurements.

\section{\label{sc:CosmoFit}The impact of interlopers on RSD parameters}

In the second part of this paper, we aim at evaluating how systematic errors due to adopting an incomplete interlopers model affect the inference of the cosmological parameters.
In this section, we perform a MCMC analysis to sample the posterior distribution of  three key cosmological parameters, namely the growth rate, $f\sigma_8$, the clustering amplitude, $b\sigma_8$, and the pairwise velocity dispersion, $\sigma_{\rm p}$: when included in the clustering model, also the total contamination fraction is let free to vary. We fix all the other parameters to the values adopted in the parent simulations.
Our goal is to identify the simplest theoretical model that accurately provides unbiased estimates of the physical parameter of interest, $f\sigma_8$.
 
We begin by considering different configurations of 
\aabf{a model}
which accounts for the presence of contaminants only through a damping factor \aabf{in front of the correct galaxies 2PCF, like in Eq.~\eqref{eq:sys-err-onlytarget}}. 
Then we test a second model that accounts \aabf{also} for the auto-correlation of the two types of line interlopers but ignores the auto-correlation of the noise interlopers, which has been shown to be negligible.   
The goal is to verify whether the systematic errors induced by ignoring the cross-correlation terms, which are considerably more difficult to model, are small enough to be neglected.
The specific models used to fit the measured 2PCF are detailed in Sect.~\ref{sec:MCMC-models}. 

In our analysis, we are not focused on evaluating the absolute precision with which cosmological parameters can be estimated from DR1 data. Instead, our goal is to assess the impact of systematic errors arising from the adoption of incomplete models for interlopers. To achieve this, we compute the results obtained by fitting a 2PCF measured on the contaminated catalogue to various 2PCF models detailed below: then, we compare these results to those obtained by fitting the 2PCF measured on the pure part of the catalogue (i.e. made of only correct galaxies) with its corresponding model for correct galaxies clustering. We refer to this latter scenario as the `reference case'.
In the following, we detail the models and methodology adopted in this analysis and the corresponding results.

\subsection{Modelling the 2PCF}
\label{sec:pk-model-graeme}

All the 2PCF models adopted in the analysis, presented in Sect.~\ref{sec:MCMC-models}, are derived from the template model for the galaxy power spectrum in redshift space described in \citet{Blanchard-EP7} and generalized in \citet{Addison_2019} to include the modelling of line interlopers
\begin{align}
\label{eq:Pk-graeme}
P\left(k_{\mathrm{obs}}, \mu_{\mathrm{obs}}, z\right)\,=\,&\gamma_{\perp}^2 \gamma_{\parallel}\frac{\left[b\left(z\right) \sigma_8\left(z\right)+f\left(z\right) \sigma_8\left(z\right) \mu_{\mathrm{obs}}^2\right]^2}{1+\left[f\left(z\right) k_{\mathrm{obs}} \, \mu_{\mathrm{obs}} \, \sigma_{\rm p}\left(z\right)\right]^2} \nonumber \\
&\times \frac{P_{\mathrm{dw}}\left(k_{\mathrm{obs}}, \mu_{\mathrm{obs}}, z\right)}{\sigma_8^2\left(z\right)} F_z\left(k_{\mathrm{obs}}, \mu_{\mathrm{obs}}, z\right) \,.
\end{align}
The term $P_{\rm dw}$ is the damped-wiggles matter power spectrum \citep{Ivanov_2018, Blanchard-EP7}
, $f$ is the growth rate, $\sigma_{\rm p}$ is the pairwise non-linear velocity dispersion 
which relates to the relative displacement induced by the peculiar velocity of galaxies 
\citep{Ballinger_1996, Blanchard-EP7},
$\sigma_8$ is the rms density fluctuation at $8 \,\hMpc\,$, and $F_z$ is a Gaussian function to account for the accuracy on the measured spectroscopic redshift (whose rms value $\sigma_{0,z}$ is almost independent of redshift  and equal to $0.001$ in the EuclidLargeMocks).

Since all cosmological parameters (apart from the aforementioned free parameters $f\sigma_8$, $b\sigma_8$, and $\sigma_{\rm p}$) are set equal to those of the simulation, there is no need to model the AP effect: therefore, the values of the gamma parameters are identically equal to one for the power spectrum of the correct galaxies, whereas for line interlopers their values are estimated from Eqs.~\eqref{eq:gamma-perp} and \eqref{eq:gamma-parallel}. Moreover, in the case of line interlopers, the power spectrum measured at redshift $z$ depends on the cosmological parameters evaluated at the true redshift $z_{\rm true}$ of the interloper population \citep{Addison_2019}.
To transform the values of the wave-number modulus and its cosine angle from the true to the observed ones in Eq.~\eqref{eq:Pk-graeme}, we used 
\begin{align}
    k_{\mathrm{obs}} =& \sqrt{\gamma^2_{\perp} k^2_{\perp} + \gamma^2_{\parallel} k^2_{\parallel}} \label{eq:k_obs} \, ,\\
    \mu_{{\mathrm{obs}}} =& \dfrac{\gamma_{\parallel} k_{\parallel}}{\sqrt{\gamma^2_{\perp} k^2_{\perp} + \gamma^2_{\parallel} k^2_{\parallel}}} \label{eq:mu_obs}\, .
\end{align}
Since we worked in configuration space, we started from the anisotropic power spectrum model to obtain the two-dimensional 2PCF model of the correct galaxies and line interlopers auto-correlation terms in Eq.~\eqref{eq:contam-2pcf-RR}.  We then extracted the multipoles by integrating the two-dimensional models weighted by the proper prefactor in front of each correlation function through
\begin{equation}
\label{eq:Pk-xi-multipoles-relation}
    \xi^{\,(\nu)}(s) = \frac{\mathrm{i}^{\nu}}{2\pi^2} \int_0^{\infty} j_{\nu}(ks) \, P^{(\nu)} (k) \, k^2 \, \diff k \, ,
\end{equation}
where $j_{\nu} (ks)$ are the spherical Bessel functions.

\subsection{2PCF phenomenological models with interlopers}
\label{sec:MCMC-models}

We present a set of analyses which involve comparing different 2PCF models, characterized by different sets of free parameters and types of interlopers. Table \ref{tab:MCMC-models} provides a summary of these tests, which are detailed below. We performed these tests in all four \Euclid spectroscopic redshift bins.
For clarity, instances like `A vs. B' should be interpreted as `measurement A fitted against model B'. 

\begin{table*}[]
\centering
\caption{Summary of all tests run in the MCMC, including the reference case (first line).}
\label{tab:MCMC-models}
\resizebox{\textwidth}{!}{%
\begin{tabular}{|l|l|l|l|}
\hline
\textbf{Model}                        & \textbf{Free parameters}                   & \textbf{Priors} & \textbf{n° free parameters} \\ \hline
{\tt correct vs. correct}                & $f\sigma_8, b, \sigma_{\rm p}$               & $\mathcal{U}(0;10)$, $\mathcal{U}(0;20)$, $\mathcal{U}(0;20) \, \hMpc$             & 3                  \\ \hline
{\tt contam vs. correct}                & $f\sigma_8, b, \sigma_{\rm p}$               & $\mathcal{U}(0;10)$, $\mathcal{U}(0;20)$, $\mathcal{U}(0;20)\, \hMpc$             & 3                  \\ \hline
{\tt contam vs. p*correct}             & $f\sigma_8, b, \sigma_{\rm p}$               & $\mathcal{U}(0;10)$, $\mathcal{U}(0;20)$, $\mathcal{U}(0;20)\, \hMpc$             & 3                  \\ \hline
{\tt contam vs. $(1-f_{\rm c})^2$*correct} & $f\sigma_8, b, \sigma_{\rm p}, f_c$          & $\mathcal{U}(0;10)$, $\mathcal{U}(0;20)$, $\mathcal{U}(0;20)\, \hMpc$, $\mathcal{U}(f_{\rm c, min};f_{\rm c, 
 max})$             & 4                  \\ \hline
{\tt contam vs. correct+line} & 
\begin{tabular}[c]{@{}l@{}}$f\sigma_8$, $b$, $m_b$, \\ $\sigma_{\rm p}$, $\sigma_{\rm p}^{\rm \ion{O}{iii}}$, $ \sigma_{\rm p}^{\rm \ion{S}{iii}}$\end{tabular} & 
\begin{tabular}[c]{@{}l@{}}$\mathcal{U}(0;1)$, $\mathcal{U}(0;5)$, $\mathcal{U}(0;3)$\\ $\mathcal{U}(0;20)\, \hMpc$, $\mathcal{U}(0;20)\, \hMpc$, $\mathcal{U}(0;20)\, \hMpc$\end{tabular} & 6 \\ \hline
\end{tabular}
}
\tablefoot{\aabf{Instances like `A vs. B' should be interpreted as `measurement A fitted against model B'.}}
\end{table*}

\subsubsection{The reference case: Correct versus correct}

As mentioned at the beginning of this section, to avoid being sensitive to our choice of a particular power spectrum model, we aim to compare cosmological parameter results across different interloper parameterizations against a reference case that uses the same power spectrum model. In this reference case, we fit the 2PCF measurement of the correct part of sample using the theoretical model for the correct galaxies auto-correlation.
We refer to this case as {\tt correct vs. correct}.\footnote{Sometimes we use the shorter version `ref', especially inside equations.} We fit~\footnote{The superscript `meas' highlights that the 2PCF on the left are the measured ones (in this case performed on the mock catalogues), whereas the ones on the right are theoretical models.}
\begin{align}
    \xit ^{\rm meas} \;\; \mathrm{vs.} \;\; \xit (f\sigma_8, b, \sigma_{\rm p}) \, 
\end{align}
where the parameters $f\sigma_8, b$, and $\sigma_{\rm p}$ are let free in the fit and refer to the correct galaxy population within the measured redshift bin.
Considering this model as reference, in particular the corresponding $f\sigma_8$ value, we can evaluate the improvement induced only by considering more complex and detailed models based on the prefactors parameterization and on the addition of the line interlopers auto-correlation signals to the total theoretical model.  

\subsubsection{A proof-of-concept case: Contaminated versus correct}

The results presented in Sect.~\ref{sc:2PCFmeas} demonstrate that, at first approximation, the contaminated signal can be reproduced by accounting for the contribution of the correct galaxies only, appropriately weighted by the corresponding prefactor. 
We perform a proof-of-concept test in which the measured 2PCF of the contaminated sample is compared to the same correct-only 2PCF model used for the reference case, that is a model which assumes a 100\% pure sample. In this case, we expect that the mismatch in the clustering amplitude will result in an underestimate of the linear bias parameter, $b$. The ultimate scope is to check whether the adoption of this simplified model affects the estimate of the growth rate parameter $f\sigma_8$. 
We refer to this test as {\tt contam vs. correct}. We fit
\begin{align}
    \xi^{\, \rm meas}_{\rm m} \;\; \mathrm{vs.} \;\; \xit (f\sigma_8, b, \sigma_{p}) \, .
\end{align}

\subsubsection{Correct-only modelling with exact prefactor}

This is the simplest realistic model that we used to fit the contaminated signal. As in the previous cases, we account for the auto-correlation of correct galaxies only, but this time weighted by its exact prefactor as in Eq.~\eqref{eq:contam-2pcf-RR} when fitting the contaminated signal. This means that we assume to know exactly the fraction of target galaxies and its scale dependence.

We refer to this test as \texttt{contam vs. p*correct}. We fit
\begin{align}
    \xi^{\, \rm meas}_{\rm m} \;\; \mathrm{vs.} \;\; p_{\rm c} \, \xit (f\sigma_8, b, \sigma_{p}) \, ,
\end{align}
with $p_{\rm c}=(1-f_{\rm tot})^2  \dfrac{\rm \rt\rt}{\rm \rc\rc}$.
The model is very similar to the reference case one, apart from the 2D prefactor in front of the correct galaxies 2PCF. This prefactor is integrated together with $\xit$ when computing the multipoles of the model, which is what we consider in the MCMC analysis. Since the prefactor is exact (because it was measured from the pairs in the random catalogues), the free parameters are the same of the reference case.

\subsubsection{Correct-only modelling with free contamination fraction}

In the real survey, one expects to estimate the fraction of interlopers from the analysis of the EDS. However, it is unlikely that such an analysis will be able to estimate the scale dependence of the contamination in the first stages of the mission. In addition, the total contamination fraction will be measured with some uncertainty.
Therefore, we explore an additional model in which we approximate the contamination fraction $\fc$ by a constant rather than a scale-dependent factor, and we let it free to vary within the interval specified by a uniform prior.

We refer to this test as {\tt contam vs. $(1-f_{\rm tot})^2$*correct}, where the prefactor in this case is scale-independent and only depends on the total contamination fraction $f_{\rm c}$ (see Eq.~\ref{eq:contam-2pcf-const}). We fit
\begin{align}
    \xi^{\, \rm meas}_{\rm m} \;\; \mathrm{vs.} \;\; (1 - f_{\rm tot})^2 \xit (f\sigma_8, b, \sigma_{p}) \, .
\end{align}
In this case, we have one more free parameter with respect to the previous tests, which is $\fc$. The prior on this and on the other parameters are discussed in the dedicated Sect.~\ref{sec:methodology}. We expect that a large prior of $\fc$ may cause, on one hand, a strong degradation of shape parameters like $f\sigma_8$ and $b\sigma_8$ due to natural degeneracies of the model. On the other hand, the $\beta = f/b$ parameter should be insensitive to the choice of this prior.

\subsubsection{Correct galaxies and line interlopers modelling}

This is the most complete model we present in this paper. In addition to the correct galaxies contribution, we include the \ion{O}{iii} and the \ion{S}{iii} line interlopers auto-correlation terms in the theoretical model. The complete 2PCF model \aabf{(described in Eq.~\eqref{eq:model-target+line-exactpref}, after introducing the set of approximations we adopted)} is therefore the sum of three contributions, all derived from the corresponding power spectrum models as in Eq.~\eqref{eq:Pk-graeme}. We do not include the noise interlopers auto-correlation since in Sect.~\ref{sc:2PCFmeas} we have shown that it is expected to be negligible.
Despite this simplification, the model still depends on a large number of free parameters, some of which are highly degenerate.  To reduce the number of degenerate parameters while maintaining a focus on estimating
$f\sigma_8$ and $b\sigma_8$, we have adopted several simplifying assumptions, which are detailed below.
 
First, in analogy with the \texttt{p*correct} model, we assume that the contamination fractions of correct galaxies and line interlopers can be estimated from the data. We also assume that not only their average values but also their scale dependence is known. The impact of this second assumption is expected to be negligible, since, as shown in Appendix \ref{sec:AppendixPrefactors}, the scale dependence of the prefactors is either mild or, when it is not, the magnitude of the prefactor itself is small.
In summary, we fix the prefactors for the three auto-correlation terms included in the model.
Second, we leverage both physical and empirical considerations to build a redshift-dependent model for the growth factor and bias of the interlopers, as detailed in the following.

We assume the cosmological model of the parent simulations to constrain the redshift dependence of the $f\sigma_8$ value accordingly. With this assumption, only a single free parameter, that is the correct galaxies growth rate $f\sigma_8$ measured at the observed redshift $z$, is needed to characterize the growth rate, since the value of $f\sigma_8$ at the redshift of the line interlopers \aabf{$z_{\rm int}$} is uniquely determined.
Furthermore, since in $\Lambda$CDM (which is the EuclidLargeMocks cosmology) the function $f\sigma_8 \left(z\right)$ is nearly linear within the redshift interval of interest, we adopt a simplified linear model that best fits the exact relation
\begin{align}
\label{eq:fs8-redshift-linear-relation}
    f\sigma^{\rm int}_8 (z_{\rm int}) = f\sigma_8 + m \, (z_{\rm int} - z_{\rm t}) \, ,
\end{align}
where $m=-0.09$ is the slope of the relation derived from theory, assuming an error of $0.02$
on the $f\sigma_8$ values.

Conversely, the redshift dependence of the bias cannot be inferred from theory, as it is related to the selection function of the spectroscopic sample and to the physical properties of the different types of observed galaxies. In principle, the three population of objects that contribute to the measured 2PCF have different bias values that should be treated as independent free parameters in the model. However, to reduce the number of free parameters and find a proper relation to link the interlopers bias to the correct galaxies bias, we calibrate the bias dependence on redshift directly on the mock measurements. For each redshift bin, we separately fit the correct galaxies, \ion{O}{iii} interlopers, and \ion{S}{iii} interlopers signal with a model accounting for the exact measured prefactors in front of the 2PCF.

\begin{figure}
    \centering
    \includegraphics[width=0.48\textwidth]{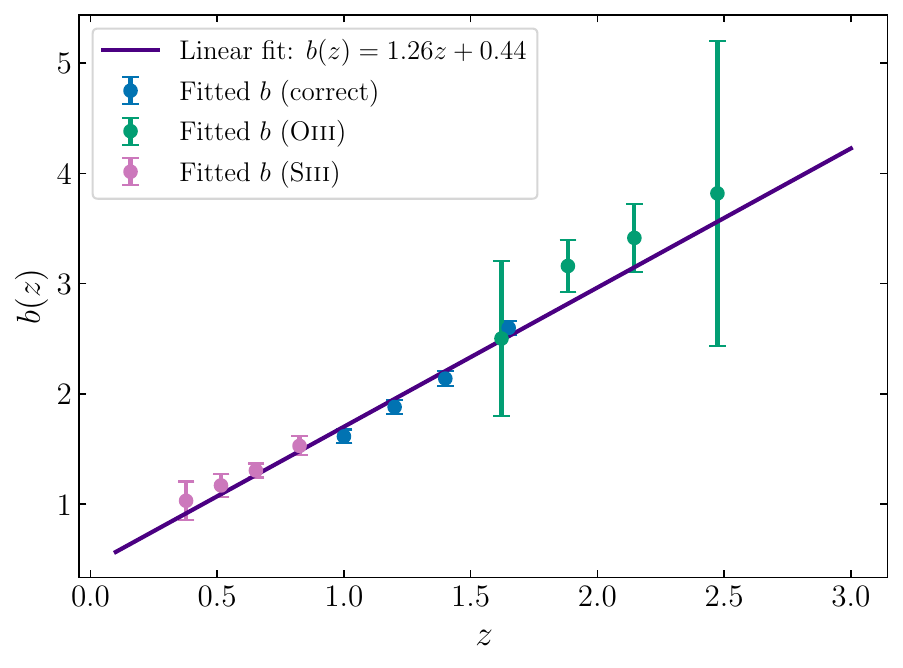}
    \caption{Redshift dependence of the bias in the EuclidLargeMocks, obtained by fitting the 2PCF of each population in the \Euclid spectroscopic bins. 
    Each colour refers to a different population and reveals the true redshifts of the line interlopers compared to correct galaxies.
    On top of the bias values, we plot the linear relation we derived to interpolate the bias' redshift dependence.}
    \label{fig:linear-bias-model}
\end{figure}

Figure \ref{fig:linear-bias-model} shows the bias values obtained fitting all populations separately in every measured redshift bin. 
The corresponding redshifts in the $b(z)$ relation are the centres of the measured redshift bins for correct galaxies, while for line interlopers they are the true original redshifts related to the measured redshift bin through Eq.~\eqref{eq:zmeas-vs-ztrue}. We interpolate the overall dependence on redshift with a linear relation to find the angular coefficient $m_{b}$\footnote{As a caveat, galaxy bias depends on luminosity and interlopers do not have the same luminosity as the corresponding \ha\, galaxies, so it is not guaranteed that the bias can be fit by a single relation, as happens to be true in this case.}. When running chains, we vary the correct galaxies bias $b$ and $m_{b}$ and then we compute the interlopers bias with the simple deterministic relation
\begin{align}
\label{eq:linear-bias-model}
    b(z_{\rm int}) = b + m_b \, (z_{\rm int} - z_{\rm t})\, .
\end{align} 
We refer to this test as {\tt contam vs. correct+line}. We fit
\begin{align}
\label{eq:model-target+line-exactpref}
    \xi^{\, \rm meas}_{\rm m} \;\; &\mathrm{vs.} \;\; \,p_{\rm c} (r) \, \xit (f\sigma_8, b, \sigma_{p}, m_b) \nonumber \\
    &\,+\, p_{\rm \ion{O}{iii}} (r) \, \xi_{\rm \ion{O}{iii}} \left(\sigma_{1,2} ^{\rm \ion{O}{iii}} \,|\, \gamma^{\rm \ion{O}{iii}}_{\perp}, \gamma^{\rm \ion{O}{iii}}_{\parallel}\right) \\
    &\,+\, p_{\rm \ion{S}{iii}} (r) \, \xi_{\rm \ion{S}{iii}} \left(\sigma_{1,2} ^{\rm \ion{S}{iii}} \,|\, \gamma^{\rm \ion{S}{iii}}_{\perp}, \gamma^{\rm \ion{S}{iii}}_{\parallel}\right) \, . \nonumber
\end{align} 
We leave the velocity dispersions $\sigma_{\rm p}$ free for all populations since we do not have a physical model for them, and we treat them as nuisance parameters. The line interloper factors $\gamma_{\parallel}$ and $\gamma_{\perp}$ are computed at a redshift $z$ corresponding to the centre of the observed redshift interval under study via Eqs.~\eqref{eq:gamma-perp} and \eqref{eq:gamma-parallel}, and they are reported in Table~\ref{tab:gammatable}: we use Eq.~\eqref{eq:zmeas-vs-ztrue} to derive the original effective redshift of line interlopers.

\subsection{Methodology}
\label{sec:methodology}

To estimate the free parameters of the models, we sampled their posterior probability distribution using the MCMC sampler {\tt emcee} \citep{emcee2013}.
For this, we assumed a Gaussian 
likelihood for the data, which is explicitly expressed as
\begin{equation}
\label{eq:likelihood}
\ln{\mathcal{L}} \propto -\frac{1}{2}\chi^2 \, ,
\end{equation}
where $\chi^2$ is defined as
\begin{align}\label{eq:Chi2} 
\chi^2 =  \left[\vec{\xi}(r)-\vec{\xi}^\mathrm{meas}(r)\right]^\mathrm{T}\tens{C}^{-1}\left[\vec{\xi}(r)-\vec{\xi}^\mathrm{meas}(r)\right] \, 
\end{align}
and $\tens{C}$ is the data covariance matrix.
Our data vector $\vec{\xi}^\mathrm{meas}(r)$ is  made of the monopole and quadrupole of the measured 2PCF $\xi^{\, \rm meas}_{\rm m}$ averaged over the full set of mock catalogues for all tests, apart from the reference case in which we fit the average of the correct galaxies $\xit ^{\rm meas}(r)$ data vector. 
The covariance matrix $\tens{C}$, instead, is that of a single realization, since we are interested in assessing the precision with which these parameters will be estimated in the DR1 \Euclid survey.
We limited our fit to the range of separations $r = \left[40, 200\right] \hMpc$ in order to exclude the smallest scales, which cannot be properly described by a tree-level theoretical model of the power spectrum (more details can be found in Appendix \ref{sec:AppendixSmallScales}). We report the mean of the posterior probability as the best estimate for the cosmological parameters, and the $1 \sigma$ of the marginalized posterior as uncertainty. In the triangle plots, the coloured bands in the marginalized 1D posteriors correspond to the just mentioned $1\sigma$ uncertainty; in the 2D posteriors, we report contours corresponding to $68\%$ and $95\%$ confidence levels.

In Table ~\ref{tab:MCMC-models}  we list the uniform priors we used in the different tests. 
In all cases explored except the last one, we adopted non-informative priors over very broad intervals. For the last model, which accounts for line interloper contamination, we set our priors based on physically motivated constraints. For example, the upper limit $f\sigma_8 < 1$  is consistent with assuming a $\Lambda$CDM  model, whereas the upper limit on the bias parameter $b\leq5$ is consistent with the linear $b\left(z\right)$ model that we have adopted
(see Fig.~\ref{fig:linear-bias-model}). Moreover, we decided to adopt a wide [0,3] prior on $m_b$.

When testing the correct-only model with the total contamination fraction free to vary, we tested different uniform priors on $\fc$. In particular, we tested a symmetric $\pm 1\%$ and $\pm 10\%$ uniform prior  around the true values of $f_{\rm tot}$ derived from the fractions of interlopers averaged over the 100 mock catalogues. This corresponds to the condition $f_{\rm tot, min} < \fc < f_{\rm tot, max}$, with 
$f_{\rm tot, min} = (1-0.01)\,\fc$ 
and
$f_{\rm tot, max} = (1+0.01)\,\fc$ in the case of a $1\%$ prior. We rely on the ability to estimate the fractions of the various interlopers by analysing the EDS, and we assume that we are able to measure these fraction with a precision in the range 1--10\%, as reflected by the chosen priors. We only show an example  where $\fc$ is allowed to vary freely within its physical limits $\left[0,1\right]$. This extreme case illustrates the `worst' pessimistic scenario where no external constraints are put on $\fc$, dramatically impacting the results of the analysis. Given the peculiarity of the test, we do not explicitly compare it with the other cases.

We used the MCMC acceptance rate and the integrated auto-correlation time as diagnostic to decide whether or not the chain was long enough to have converged. For the tests we show in this paper, we found that a configuration with 20\,000 steps and 40 walkers was adequate, providing a number of effectively independent samples greater then 100 for all redshifts, tests, and model parameters.
To compare our results to those obtained in Sect.~\ref{sc:2PCFmeas}, aimed at testing the individual contribution of the interlopers to the measured 2PCF, we show the results in the same redshift bins, i.e. $\rm{z1} = \left[ 0.9, 1.1\right]$ and $\rm{z3} = \left[ 1.3, 1.5 \right]$.

\subsection{Results of the amplitude fits}
\label{sc:CosmoFitResults}

\begin{figure}[!htbp]
    \centering
    \includegraphics[width=0.5\textwidth]{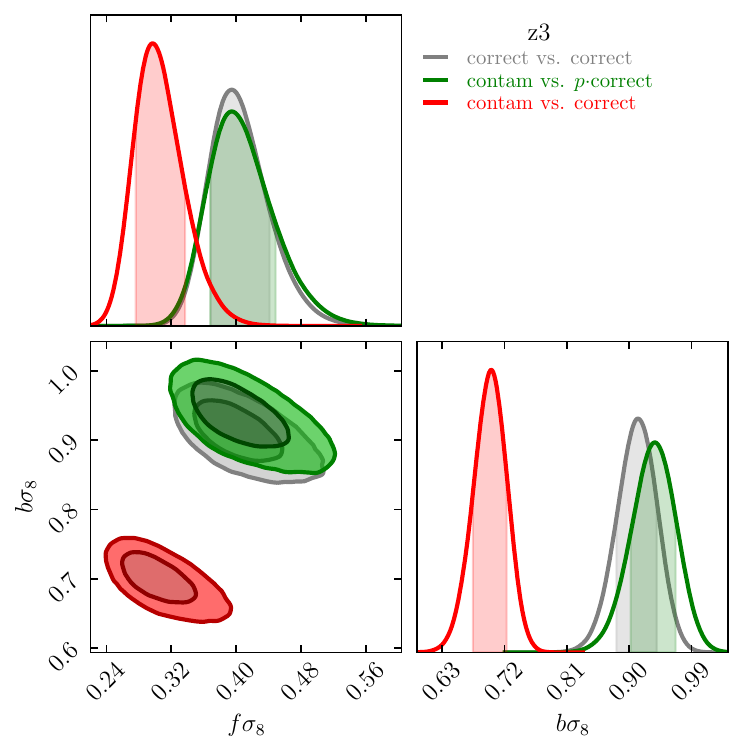}
    \caption{Comparison between the reference case (grey line), a model including interlopers (the minimal one, green line), and a model with contributions from correct galaxies only without the proper weighting (red line) in $z \in [1.3, 1.5]$.}
    \label{fig:contam-vs-target}
\end{figure}

The contour plot in Fig.~\ref{fig:contam-vs-target} 
shows the 2D and 1D marginalized posterior probability contours for the parameters $f\sigma_8$ and $b\sigma_8$ obtained from three different analyses (in $\rm{z3}$ as an example).
The grey contours show the \texttt{correct vs. correct} reference case. We compare them with the results of the \texttt{contam vs. correct} case with no allowance for interlopers contamination (red curves) and that of the \texttt{contam vs. p*correct} case (green curves), in which the correct prefactor is used to account for interlopers contamination.
In the \texttt{contam vs. correct} scenario, the systematic error on the $f\sigma_8$ and $b\sigma_8$ values is significantly larger than the statistical uncertainty, and a simpler rescaling of the fitted galaxy bias value is not enough to recover the correct $f\sigma_8$ value. However, when adding the exact prefactor of the correct galaxies contribution to the model, the contour plots (in green) overlap with those of the reference case. 
This comparison highlights the systematic error deriving from ignoring the presence of interlopers altogether and assuming that the sample is 100\% pure. An additional degree of freedom is required to account  for the overall decrease in the clustering amplitude, which is the main effect induced by interloper galaxies.

\begin{figure*}[htbp]
    \centering
    \includegraphics[width=\textwidth]{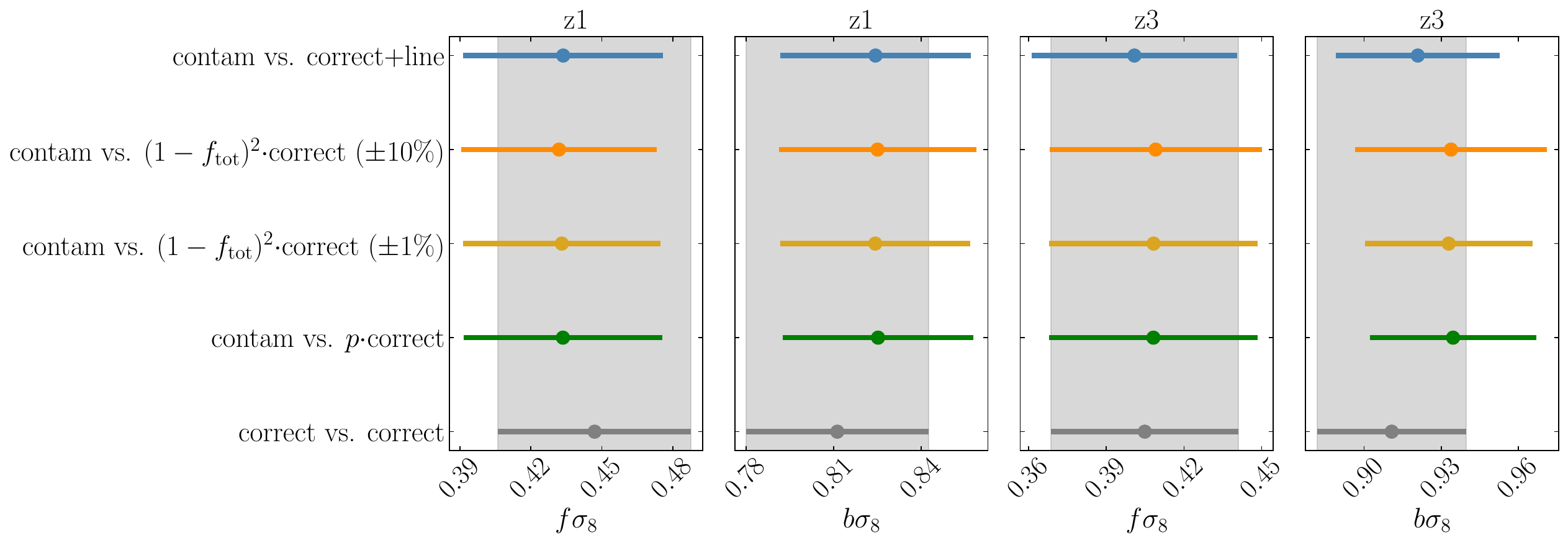}
    \caption{Inferred $f\sigma_8$ and $b\sigma_8$ values and $1\,\sigma$ uncertainties for all tests in the two reference redshift bins. They grey band corresponds to $1 \sigma$ around the values inferred in the reference case.}
    \label{fig:errorbars-parameters}
\end{figure*}

Figure \ref{fig:errorbars-parameters} shows the inferred values of $f\sigma_8$ and $b\sigma_8$ determined in each chain in the two reference redshift intervals, along with their uncertainties. A grey band indicating the $1\,\sigma$ uncertainty in the reference case is displayed to facilitate the comparison between different models.

\begin{figure}[!htbp]
    \centering
    \includegraphics[width=0.5\textwidth]{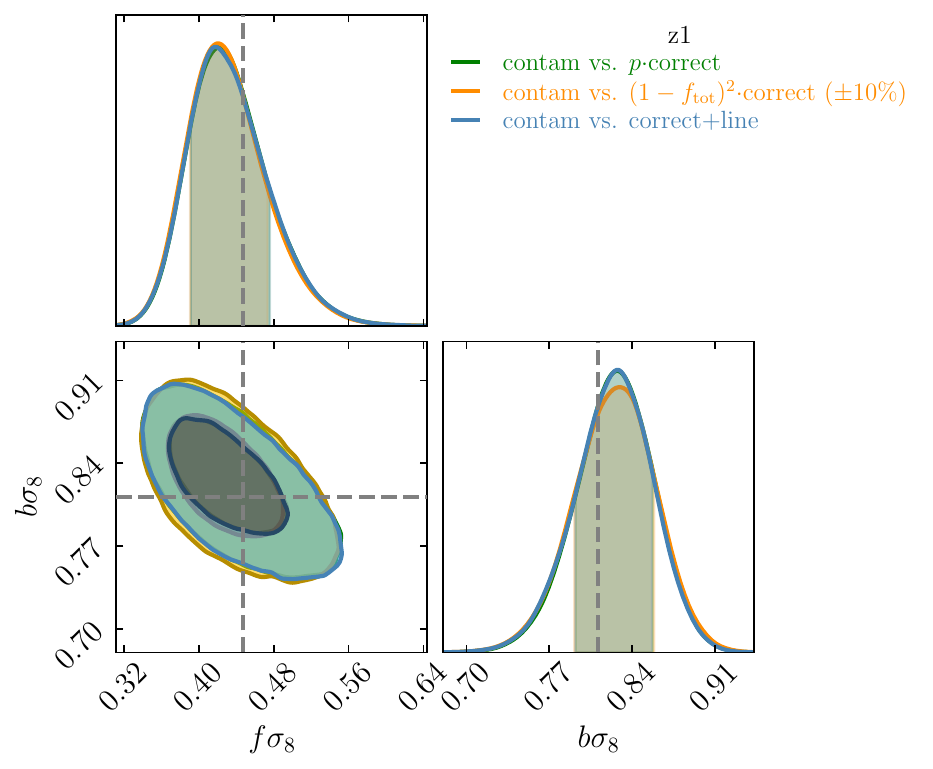}
    \hfill
    \includegraphics[width=0.5\textwidth]{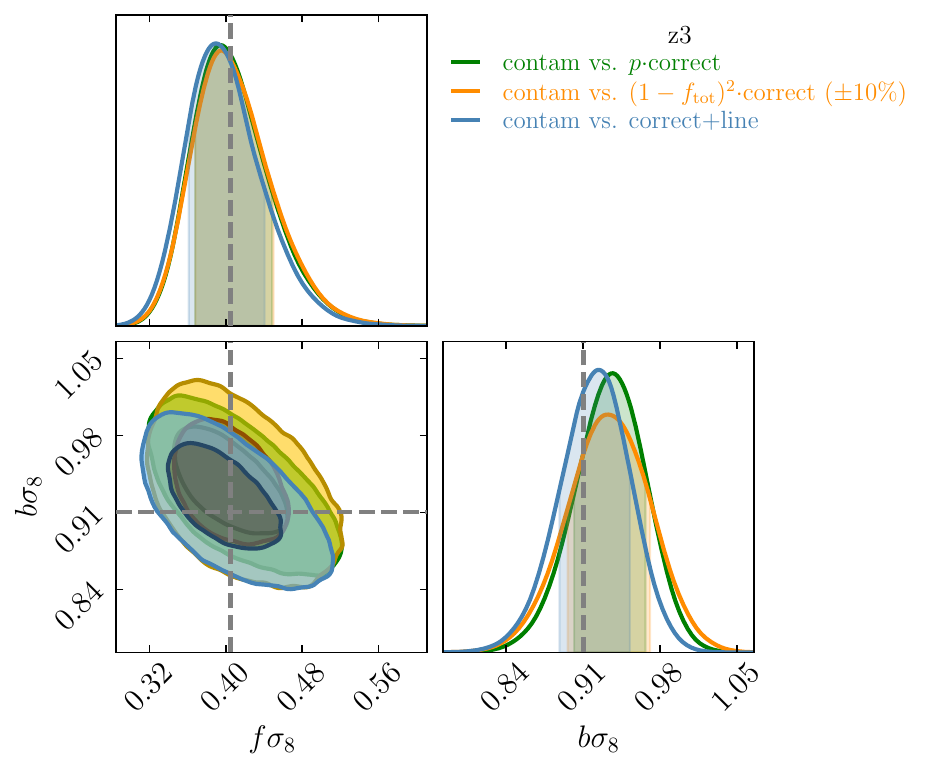}
    \caption{Contour plots for $f\sigma_8$ and $b\sigma_8$ derived from three representative tests in $z \in [0.9, 1.1]$ (\textit{top}) and $z \in [1.3, 1.5]$ (\textit{bottom}). The dashed grey lines indicate the mean values of the reference case posterior distribution.}
    \label{fig:fs8b_contour_all}
\end{figure}

Let us now focus on the models that include only the correct galaxies contribution. We considered three cases: one in which we assumed to know exactly the prefactor of the correct galaxies auto-correlation and its scale dependence (green curves in Fig.~\ref{fig:errorbars-parameters} and Fig.~\ref{fig:fs8b_contour_all}); one in which we treated the prefactor $(1-f_{\rm tot})^2$ as a free parameter (with no scale dependence), with a strong symmetric $\pm 1\%$ uniform prior centred on a reference value for $f_{\rm tot}$ estimated by averaging over the actual fraction of contaminants inserted in the mocks (golden line in Fig.~\ref{fig:errorbars-parameters}); and one in which a milder prior $\pm 10\%$ on $f_{\rm tot}$ was assumed (orange lines in Fig.~\ref{fig:errorbars-parameters} and Fig.~\ref{fig:fs8b_contour_all}).

\begin{figure}[htbp]
    \centering
    \includegraphics[width=0.5\textwidth]{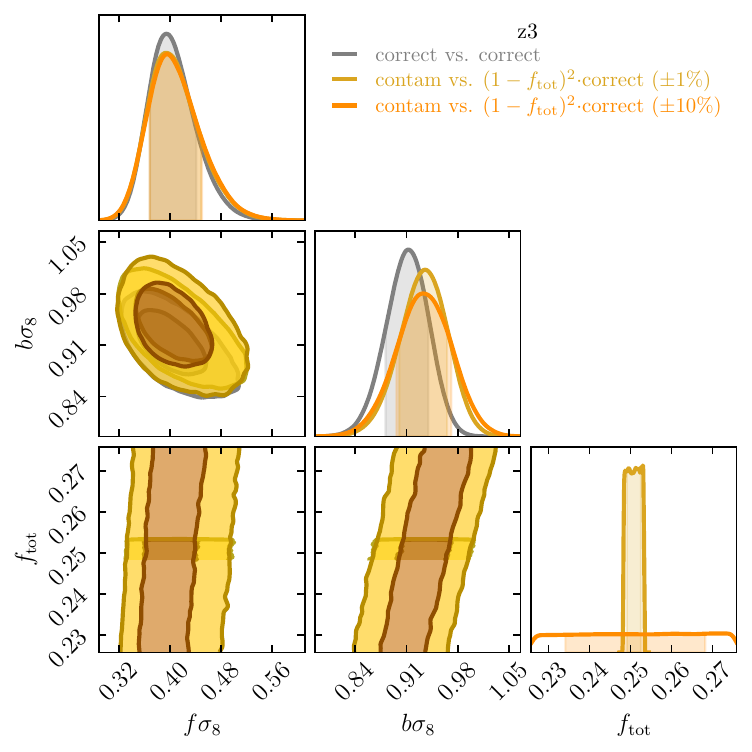}
    \caption{Comparison between the cosmological parameters' constraints obtained with a correct-only constant-prefactor model (\texttt{contam vs. $(1-f_{\rm tot})^2$*correct}) with an uncertainty of 1\% and 10\% on the total contamination fraction $f_{\rm tot}$. As an example, we show the results for z3.}
    \label{fig:contour-different-frac-1-10}
\end{figure}

\begin{figure}[htbp]
    \centering
    \includegraphics[width=0.5\textwidth]{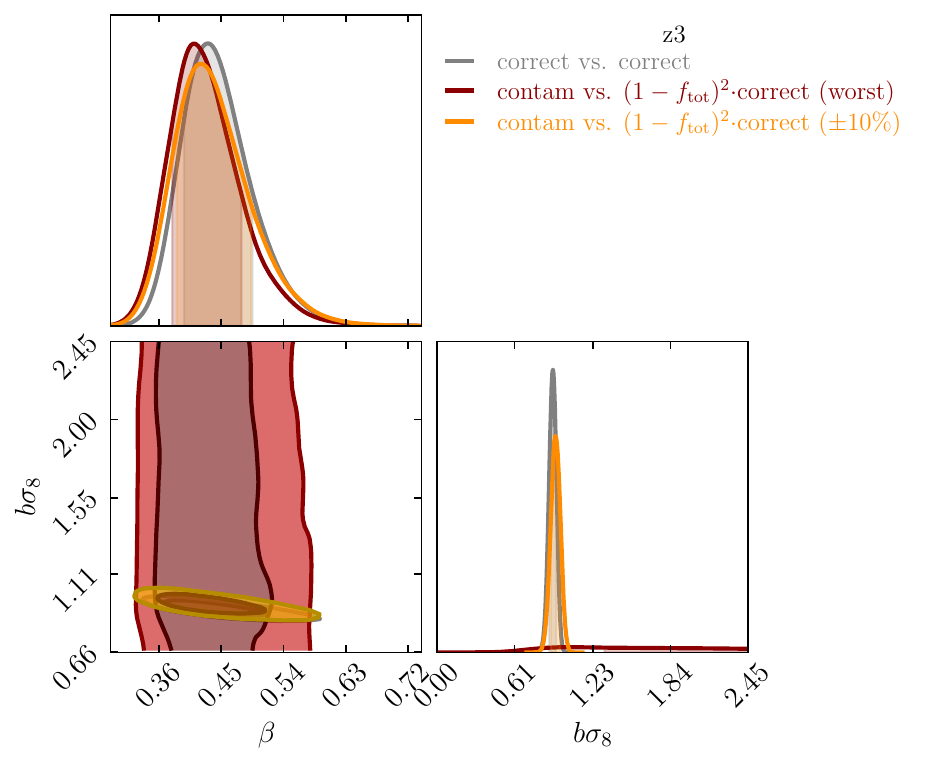}
    \caption{Independently from the width of the prior on the total contamination fraction $f_{\rm tot}$, the $\beta$ parameter is always constrained and its uncertainty does not change in a correct-only constant-prefactor model.}
    \label{fig:contour-beta}
\end{figure}

Looking at the error bars shown in Fig.~\ref{fig:errorbars-parameters}, we immediately notice that there is no substantial difference between the exact prefactor case and the constant one.
Focusing on the centre-left panel of Fig.~\ref{fig:contour-different-frac-1-10}, we observe that the probability contours for the $f\sigma_8$ and $b\sigma_8$ parameters remain largely unaffected by the assumed interloper fraction, provided this is known a priori with 10\% precision. However, this stability is lost when the contaminating fraction is allowed to vary freely between 0 and 1, i.e. when no prior information about the contamination is available. In this scenario, the probability contours for these parameters broaden significantly, as expected.
It is noteworthy, though, that even in this pessimistic yet unlikely case (since we expect to measure the sample purity through analyses of the EDS), our ability to estimate the distortion parameter $\beta$ remains relatively unaffected (as shown in Fig.~\ref{fig:contour-beta}). This resilience is due to the fact that 
$\beta$ can be measured from the ratio of the monopole to the quadrupole of the galaxy 2PCF, a calculation in which the contaminating fraction cancels out.

\begin{figure}[h]
    \centering
    \includegraphics[width=0.49\textwidth]{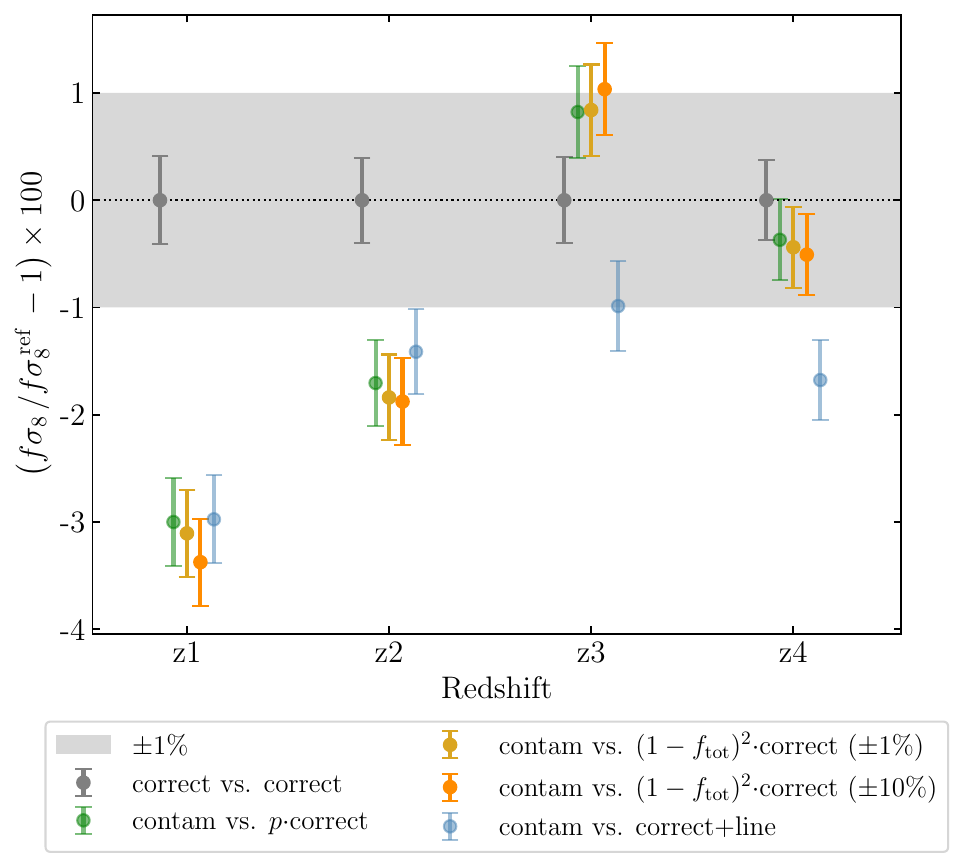}
    \caption{Percent difference between the $f\sigma_8$ value estimated in each chain and the one inferred in the reference case. Here, the errors on the percent difference are derived by considering the errors on the mean value of $f\sigma_8$, i.e. the uncertainties given by the chains divided by the square root of the number of mocks. }
    \label{fig:perc-diff-fs8-all}
\end{figure}

The most realistic model we show in this study is the one including also the contribution of line interlopers (see Sect.~\ref{sec:MCMC-models}), which corresponds to the 
blue curves in Fig.~\ref{fig:errorbars-parameters} and in the triangle plots of Fig.~\ref{fig:fs8b_contour_all}, where we show the 1D a 2D marginalized posterior distributions for $f\sigma_8$ and $b\sigma_8$ in three representative chains. The dashed grey lines indicate the mean values of the reference case marginalized posterior distributions.
The results show that including the contribution of line interlopers in the models does not significantly modify the posterior distributions. Consequently, this inclusion does not affect the precision with which these parameters are estimated, as shown by the error bars in Fig.~\ref{fig:errorbars-parameters}.

\begin{figure*}[h]
    \centering
    \includegraphics[width=0.49\textwidth]{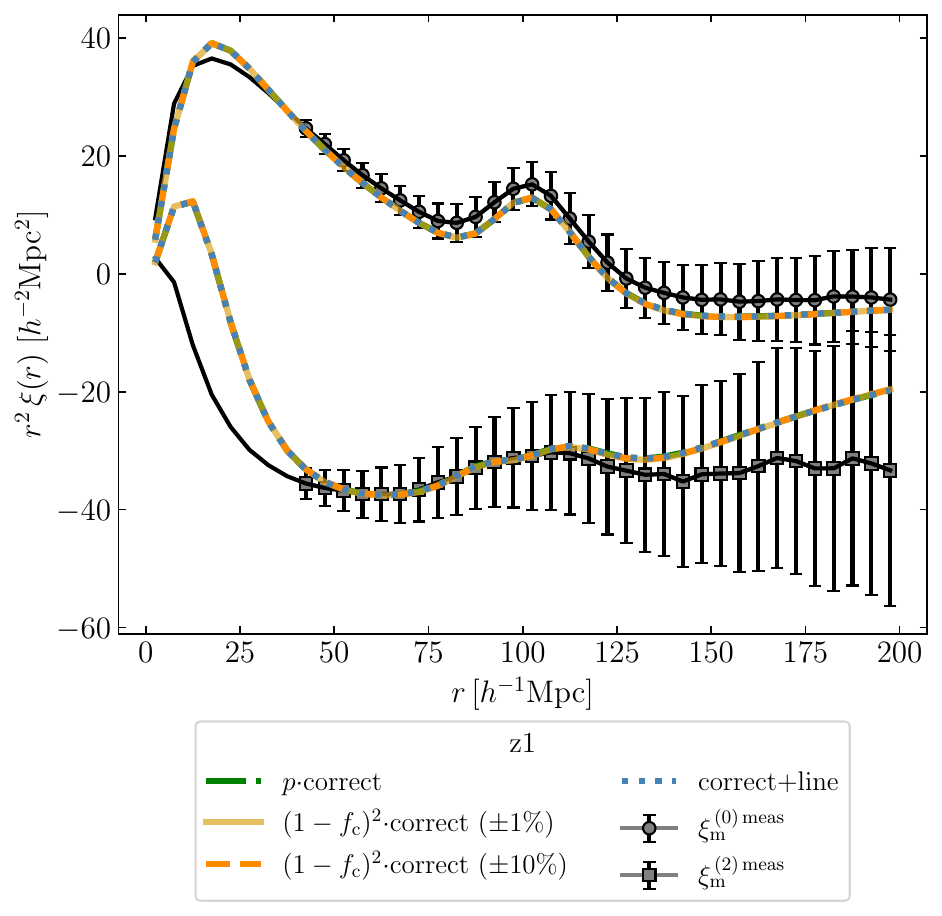}
    \hfill
    \includegraphics[width=0.49\textwidth]{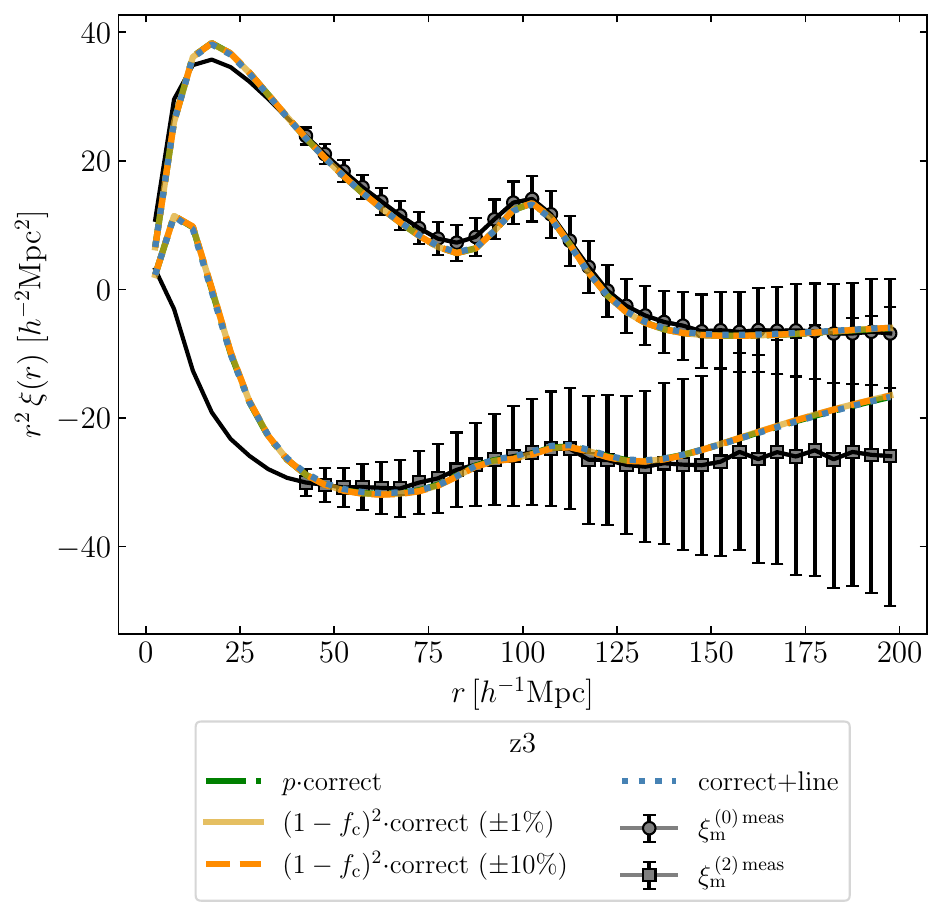}
   
    \caption{Agreement between the measured 2PCF $\xi^{\,\rm meas}_{\rm m}$ multipoles and all tested models. The error bars are shown in the fit range and correspond to the diagonal elements of the covariance matrix used to sample the posterior probability in the chains, i.e. the single-measurement DR1-like covariance. \textit{Left}: $z \in [0.9, 1.1]$; \textit{right}: $z \in [1.3, 1.5]$.
    }
    \label{fig:xi_models_from_chain}
\end{figure*}

On the other hand, the accuracy with which these parameters are estimated varies with redshift.
Focusing on $f\sigma_8$, in Fig. \ref{fig:perc-diff-fs8-all} we show the percent difference between its estimated value in all tests (coloured dots) and the one fitted in the reference case (grey dots). Different colours are used consistently with the previous plots and indicate the different model used.
The error bars were computed propagating the error on the mean inferred $f\sigma_8$ values, i.e. dividing the $1\,\sigma$ uncertainty of the marginalized $f\sigma_8$ posterior by the square root of the number of mocks, in order to quantify the systematic error in the estimate of the parameter.
The grey band indicates a reference 1\% difference.
All models underestimate the $f\sigma_8$ value with respect to the reference case in the nearest redshift bins. However, the discrepancy is small, decreasing from 3\% at z1  to 1\% at z2.  In the data, this mismatch corresponds to a systematic underestimate of the absolute amplitude of the  2PCF monopole and quadrupole moments, as shown in Fig.~\ref{fig:xi_models_from_chain}-\textit{left}. At the  higher redshifts z3 and z4, the more sophisticated model that includes line interlopers continue to underestimate the parameter, though the mismatch remains minor. In contrast, the predictions of all other models agree to within 1\% with expectations. 

The redshift dependence of the mismatch of the models may have different causes. One is the inadequacy of our power spectrum model, used to predict the galaxy 2PCF, in accounting for non-linear effects. We expect its impact to be small since we limit our analysis on separations larger than $40\,\hMpc$. Moreover, we compared the performance of the models to a reference case that also relies on the same matter power spectrum model. However, the  model of the measured 2PCF at a given redshift accounts for the contribution of all interlopers at various redshifts, and its value at a given separation  may include contributions from various type of pairs, including those at smaller separations and thus probing non-linear scales. One hint that this is a plausible explanation is the fact that the mismatch decreases with the redshift for most of the models explored.

A distinguishing feature of the first redshift bin compared to the others is the presence of a high fraction of noise interlopers and an almost complete absence of line interlopers. As a result, the cross-correlation between correct galaxies and noise interlopers becomes the most significant contribution after the correct galaxy signal, as shown in the left panel of Fig.~\ref{fig:multipoles-z1-z3}. However, although this contribution is larger in z1 than in z3, its magnitude contributes but is not sufficient to account for the observed difference in $f\sigma_8$. To verify this, we compared the results of fitting the correct-only model to both the measured 2PCF, i.e., the total measured signal, and the 2PCF after subtracting the correct-noise cross-correlation contribution. The fit results do not change significantly in either z1 or z3.
This suggests that, at least with the contaminant fractions present in the EuclidLargeMocks, the cross-correlation between correct galaxies and noise interlopers does not play a role in the results of the first redshift bin (at least when focusing on $r > 40\,\hMpc$). If the fraction of noise interlopers were to increase, the relevance of the cross-correlation signal with correct galaxies would correspondingly rise. In such a case, we are confident in our ability to model this contribution if necessary (see Appendix \ref{sec:AppendixTargetNoise}).

\begin{figure}[h]
    \centering
    \includegraphics[width=0.5\textwidth]{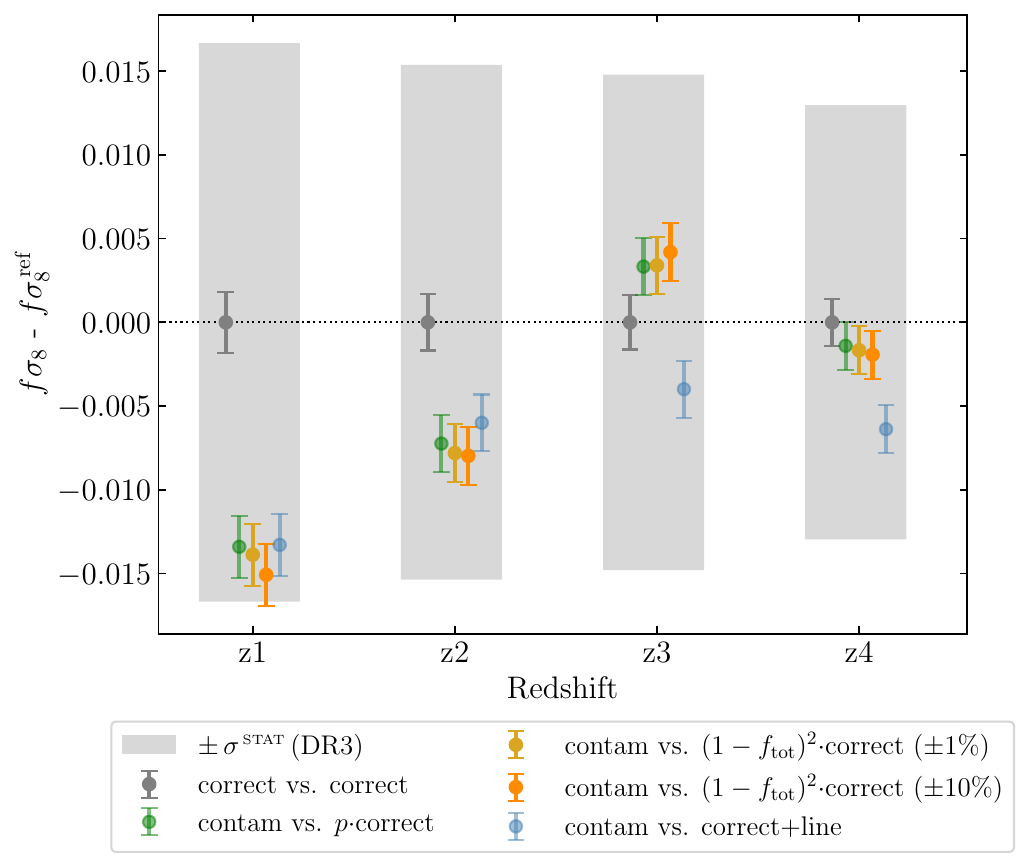}
    \caption{Systematic bias on $f\sigma_8$ of all tests with respect to the reference case.  
    The grey bands represent an estimate of the precision on $f\sigma_8$ at the end of the mission (DR3), when the observed  volume will be six times bigger than DR1. As in Fig.~\ref{fig:perc-diff-fs8-all}, the error bars were derived using the errors on the mean values of the inferred $f\sigma_8$.}
    \label{fig:diff-fs8-all-DR3}
\end{figure}

Figure \ref{fig:diff-fs8-all-DR3}
shows the systematic bias between the $f\sigma_8$ value inferred from all models relative to the reference case, as a function of redshift. To better quantify this bias, we computed its uncertainty by dividing the uncertainties on $f\sigma_8$ in the chains by the square root of the number of mocks, as we did in Fig.~\ref{fig:perc-diff-fs8-all}. The systematic errors for the different models are compared to the statistical uncertainty $\sigma^{\, \sfont{STAT}}$ on $f\sigma_8$ in the reference case (grey bands) when considering a statistics comparable to the \Euclid final data release DR3 (i.e. the uncertainty on $f\sigma_8$ from the reference case chain has been divided by $\sqrt{6}$, since the DR3 volume will be about six times that of DR1). 
All redshift bins show systematic errors below or comparable to the statistical error regardless of the 2PCF model tested. In particular, moving towards high redshifts, the systematic errors tend to be smaller.
This suggests that, with DR3-level sensitivity, we must be more careful in assessing the quality of our models and check whether this discrepancy arises from the assumption of an overly simplistic model, or whether the inadequacy of the interlopers’ modelling becomes significant with such high statistics. On the other hand, a simple interloper model appears to be more than sufficient when working with DR1-like sensitivity.

\section{\label{sc:BAOFit}The impact of interlopers on BAO parameters}

In this section, we focus on modelling the BAO signal in the measured 2PCF. In this, we parameterise the deviation between the measured and fiducial BAO scale, along and across the line-of-sight, using the AP parameters $\alpha_\perp,\alpha_\parallel$, defined as 
\begin{eqnarray}\label{eq:alphas}
    \alpha_\perp(z) &=& \frac{D_\sfont{A}^\mathrm{true}(z)r^\mathrm{fid}_\mathrm{s}}{D_\sfont{A}^\mathrm{fid}(z)r^\mathrm{true}_\mathrm{s}} \label{eq:alpha_perp} \, ,\\
    \alpha_\parallel(z) &=& \frac{H^\mathrm{fid}(z)r^\mathrm{fid}_\mathrm{s}}{H^\mathrm{true}(z)r^\mathrm{true}_\mathrm{s}} \, .
    \label{eq:alpha_par}
\end{eqnarray}
Here the superscripts $^\mathrm{true}$ and $^\mathrm{fid}$ mark the true and fiducial values of the Hubble parameter $H(z)$, the comoving angular diameter distance $D_\sfont{A} (z)$, and the sound horizon scale $r_\mathrm{s}$, as estimated using, respectively, the true cosmological model and the fiducial cosmological model used to convert redshifts into distances.

The AP parameters enter the 2PCF templates as a dilation of the measured radial and angular galaxy pair separations, in the same way as the \(\gamma_{\perp,\parallel}\) parameters, defined in Eqs.~\eqref{eq:gamma-perp} and \eqref{eq:gamma-parallel}, are incorporated into the clustering template to account for line interlopers (Eq.~\ref{eq:Pk-graeme}; \citealt{Ross2017}). Therefore, when analysing contaminated data, we expect these two sets of parameters to be highly degenerate.
In particular, when employing the state-of-the-art model for the BAO peak  
(Euclid Collaboration: Sarpa et al., in prep.),
which does not explicitly model systematic effects -- neither for the redshift errors nor the interlopers signal included in Eq.~\eqref{eq:Pk-graeme} -- we anticipate that the AP parameters will fully absorb the effects of the \(\gamma\) parameters, ultimately degrading the BAO constraints.

In the following, we quantify the amplitude of this degradation by applying the standard BAO model to the contaminated signal at different redshifts. \aabf{The analysis presented in this article does not take into account the effects of interlopers on the BAO reconstruction, as this exploration is addressed in a companion paper (Euclid Collaboration: Sarpa et al., in prep.).}

\subsection{Methodology}
We analysed the full set of EuclidLargeMocks introduced in Sect.~\ref{sec:catalogues}, distinguishing between the correct and contaminated cases (correct + line + noise). By matching the fiducial cosmology to the true cosmology of the mocks when mapping redshift coordinates into distances, we expect the estimated values of the AP parameters to be consistent with unity in both scenarios.

Given that the model does not incorporate any source  of  systematic errors, we fit both the correct and measured 2PCF multipoles using the same template \citep{Sarpa2021}
\begin{equation}\label{eq:xi_model_bao}  
\xi^{\,(\nu)}(r;\bar{z})=b^2\xi^{\,\mathrm{Ph}, \,(\nu)}(r,\alpha_\perp,\alpha_\parallel,b,f,\Sigma_\perp,\Sigma_\parallel;\bar{z}) + \mathrm{BB}_{\nu}(r) \,,
\end{equation}
where \(\xi^{\,\mathrm{Ph}, \,(\nu)}\) encapsulates the physical properties of the signal, modulated by the linear bias, \(b\), the growth rate of structures, \(f\), and the phenomenological parameters \(\Sigma_\perp\) and \(\Sigma_\parallel\) describing the anisotropic damping of BAO in the parallel and transverse directions to the line of sight. The polynomial broadband term
\begin{equation}
\mathrm{BB}_{\nu}(r;A_{\nu,0},A_{\nu,1},A_{\nu,2}) = A_{\nu,0} + \frac{A_{\nu,1}}{r} + \frac{A_{\nu,2}}{r^2}
\end{equation}
was included to model high-order non-linear effects (beyond first-order perturbation theory) and systematic features not explicitly captured in the physical model. 

There are five main characteristics of the contaminated signal that are not accounted for in Eq.~\eqref{eq:xi_model_bao}: the relative fraction of different mass tracers (correct galaxies and interlopers), their respective biases, the dilation of the BAO scale induced by line interlopers, the noise contribution from noise interlopers, and redshift errors. Noting that the biases and fractions of contaminants modulate the amplitude of the 2PCF signal, we expect their effects to be reabsorbed by the estimated values of the linear bias parameter, $b$. As previously discussed,
since \(\gamma_\perp\) and \(\gamma_\parallel\) are degenerate with \(\alpha_\perp\) and \(\alpha_\parallel\), we used only the latter two as free parameters in the analysis.
Finally, the broadband term \(\rm{BB}\) accounts for both the noise background level and the redshift errors.

\begin{table}
\caption{Prior distributions for 2PCF model parameters of the BAO analysis (Eq.~\ref{eq:xi_model_bao}).}
\centering
\begin{tabular}{|c|l|}
\hline
\textbf{Parameter} & \textbf{Prior Distribution} \\
\hline
\(\alpha_\perp\) & $\mathcal{U}(0.8, 1.2)$ \\
\(\alpha_\parallel\) & $\mathcal{U}(0.8, 1.2)$ \\
\(b\) & $\mathcal{U}(0.0, 5.0)$ \\
\(f\) & $\mathcal{U}(0.0, 2.0)$ \\
\(\Sigma_\perp\ [h^{-1}\,\mathrm{Mpc}]\) & $\mathcal{U}(0.0, 20.0)$ \\
\(\Sigma_\parallel\ [h^{-1}\,\mathrm{Mpc}]\) & $\mathcal{U}(0.0, 20.0)$ \\
\(\{A_{\nu,0}\}_i\) & $\mathcal{U}(-20.0, 20.0)$ \\
\(\{A_{\nu,1}\}_i\ [h^{-1}\,\mathrm{Mpc}]\) & $\mathcal{U}(-20.0, 20.0)$ \\
\(\{A_{\nu,2}\}_i\ [h^{-2}\,\mathrm{Mpc}^2]\) & $\mathcal{U}(-20.0, 20.0)$ \\

\hline
\end{tabular}
\label{tab:priors_BAO}
\end{table}
We validated our hypothesis on the effect of interlopers on the AP parameters by defining a Gaussian likelihood for 
the data (as shown in Eq.~\ref{eq:likelihood}) and sampling the model's parameter posterior, assuming flat priors, as detailed in Table \ref{tab:priors_BAO}. 
To better understand the impact of interlopers on the BAO scale, we focused on modelling the mean monopole, quadrupole, and hexadecapole of the 2PCF averaged over the full set of mocks, utilizing the same mock covariance matrix as in the previous section. In this context, the mean 2PCF multipoles served as the theory data vector, enabling us to assess the performance of the state-of-the-art BAO template in the presence of contaminated signals.
Finally, we performed the likelihood sampling using the \textsc{BAOFitter} Python package\footnote{\url{https://gitlab.com/esarpa1/BAOFit}}, restricting the fitting range to \([50,150]\,h^{-1}\mathrm{Mpc}\).

\subsection{Results of the BAO fit}

\begin{figure}[!htbp]
    \centering
\includegraphics[width=\linewidth]{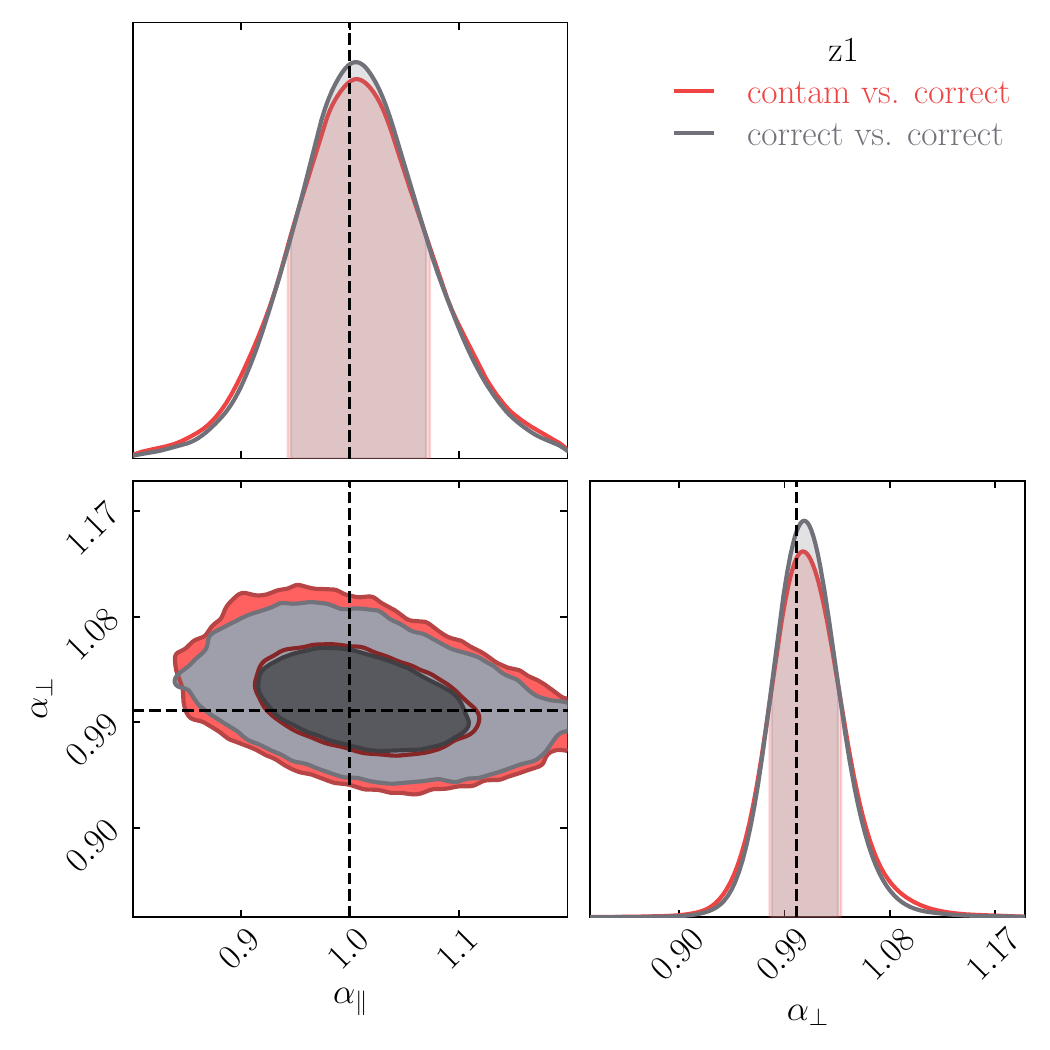}
    \caption{Alcock--Paczynski parameters in the redshift range $z\in[0.9,1.1]$.}
    \label{fig:alphas_z1}
\end{figure}

\begin{figure}[!htbp]
    \centering
    \includegraphics[width=\linewidth]{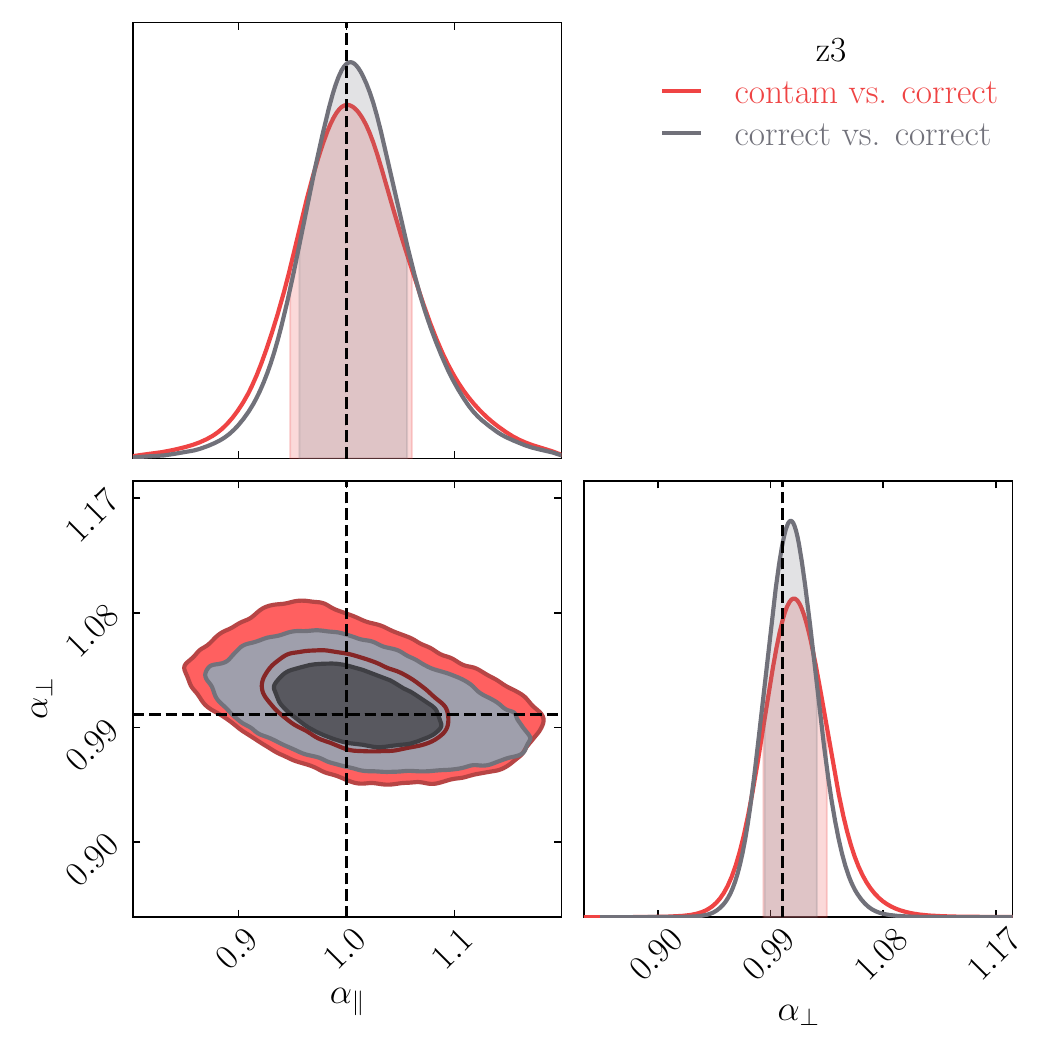}
    \caption{Alcock--Paczynski parameters in the redshift range $z\in [1.3,1.5]$.}
    \label{fig:alphas_z3}
\end{figure}

\begin{figure}
    \centering
    \includegraphics[width=\linewidth]{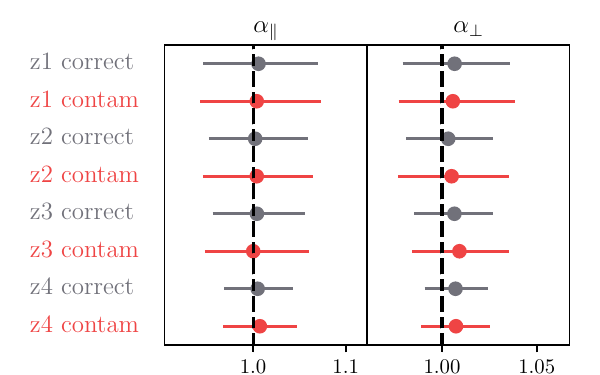}
    \caption{Summary plot of the AP parameters in all the redshift bins, both in the correct and contaminated case.}
    \label{fig:AP_pars_summary}
\end{figure}

Figures~\ref{fig:alphas_z1} and \ref{fig:alphas_z3} show the \(1\sigma\) and \(2\sigma\) posterior probability for the AP parameters as estimated from the correct (black) and contaminated (red) catalogues for the two reference bins used in this analysis, z1 and z3, respectively. At z1, where noise interlopers dominate over line interlopers, the contaminated constraints are in almost perfect agreement with correct galaxy results, showing no bias and a slight increase of \(9\%\) in the uncertainty of \(\alpha_\perp\) and \(4\%\) in \(\alpha_\parallel\).
At z3, where the fraction of noise interlopers, \ion{O}{iii}, and \ion{S}{iii} line interlopers are comparable, we detect a significant increase in the uncertainties, corresponding to \(21\%\) and \(14\%\) of the target prediction on \(\alpha_\perp\) and \(\alpha_\parallel\), respectively. Similarly to z1, their estimated value is not  biased by the presence of contaminants.
The best fit values for the AP parameters for z1 and z3, as well as for the other bins are shown in Table ~\ref{tab:AP_params_shifted}, together with relative errors and the reduced $\tilde{\chi}^2$ values, and visualized in Fig.~\ref{fig:AP_pars_summary}.

Our results indicate that the presence of noise interlopers neither biases nor significantly degrades the BAO estimates, as their effects are effectively accounted for by the polynomial broadband term. In contrast, line interlopers result in a substantial enlargement of the probability contours. This enlargement can be attributed to the intrinsic clustering signal of these interlopers contributing to the measured 2PCF. Specifically, as discussed in Sect.~\ref{sec:damping-shift-interlopers}, the \ion{S}{iii} and \ion{O}{iii} line interlopers exhibit BAO features 
located, respectively, at smaller and larger separations than that of the BAO peak of the correct galaxies 2PCF.
Consequently, the BAO peak of the contaminated sample is broader than that of a pure sample, since it could be roughly modelled as the superposition of three different Gaussian curves, leading to larger uncertainties on the position of the centroid.
\renewcommand{\arraystretch}{1.3}
\begin{table*}[th]
\caption{Best fit, percent relative error, and $\chi^2$ values for the AP parameters in all the redshift bins for both correct and contaminated catalogues, with $\alpha_\parallel$ and $\alpha_\perp$ values shifted by subtracting 1.}
\label{tab:BAO_pars_summary}
    \centering
    \begin{tabular}{|lccccc|}
        \hline
                \ & $\alpha_\parallel - 1$ & $\alpha_\perp - 1$ & $\sigma_{\alpha_\parallel}/\alpha_\parallel [\%]$ & $\sigma_{\alpha_\perp}/\alpha_\perp [\%]$ & $\tilde{\chi}^2$\\ 
                \hline
                z1 correct & $0.006^{+0.065}_{-0.060}$ & $0.008^{+0.028}_{-0.029}$ & 6.17 & 2.80 & 0.99\\ 
                z1 contam & $0.004^{+0.069}_{-0.062}$ & $0.006^{+0.033}_{-0.029}$ & 6.51 & 3.06 & 1.06\\ 
        z2 correct & $0.002^{+0.058}_{-0.051}$ & $0.003^{+0.023}_{-0.022}$ & 5.40 & 2.25 & 0.98 \\ 
                z2 contam & $0.002^{+0.061}_{-0.057}$ & $0.005^{+0.030}_{-0.028}$ & 5.89 & 2.87 & 0.98\\ 
                z3 correct & $0.004^{+0.052}_{-0.048}$ & $0.007^{+0.021}_{-0.021}$ & 4.97 & 2.07 & 1.00\\ 
                z3 contam & $0.000^{+0.061}_{-0.053}$ & $0.009^{+0.026}_{-0.025}$ & 5.67 & 2.51 & 1.06\\
        z4 correct & $0.007^{+0.040}_{-0.040}$ & $0.007^{+0.018}_{-0.018}$ & 3.93 & 1.78 & 1.08\\ 
                z4 contam & $0.006^{+0.036}_{-0.037}$ & $0.007^{+0.018}_{-0.016}$ & 3.66 & 1.65 & 1.01\\ 
                \hline
    \end{tabular}
    \label{tab:AP_params_shifted}
\end{table*}
Similar results hold for the other redshift bins (see Fig.~\ref{fig:AP_pars_summary}).

\begin{figure}
    \centering
    \includegraphics[width=\linewidth]{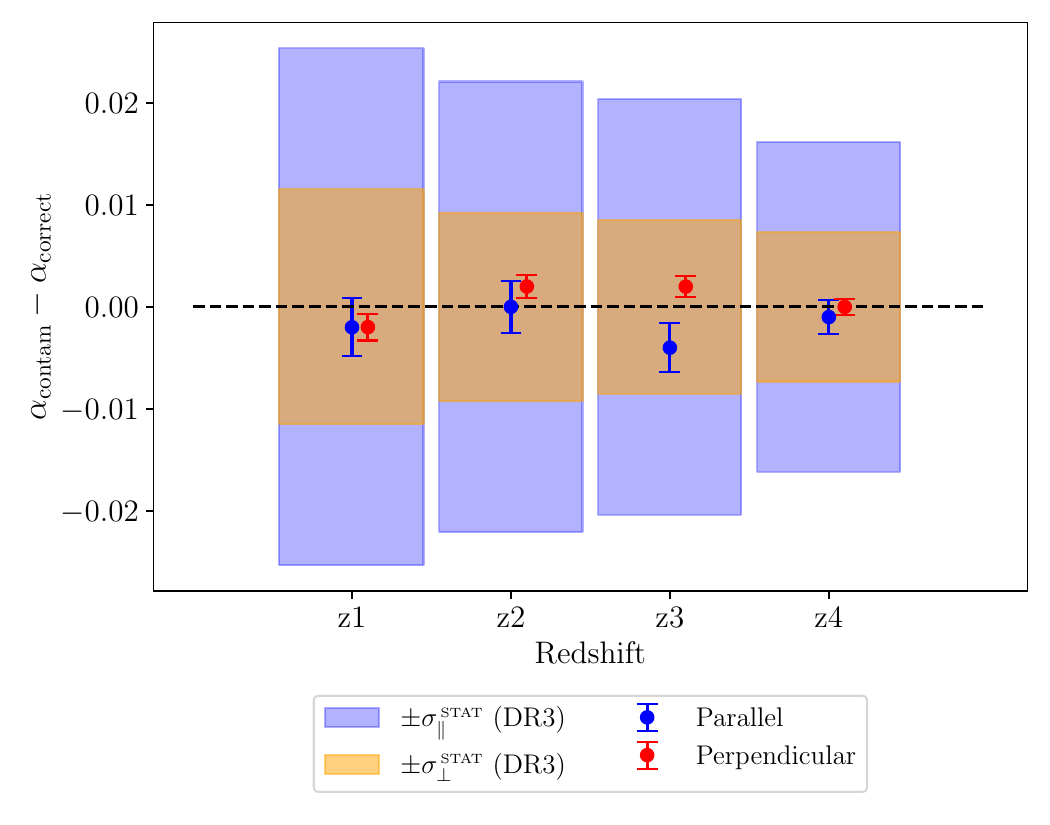}
    \caption{Comparison between the systematic bias on the two AP parameters with respect to the reference case (correct). }
    \label{fig:systematic_bias_AP_pars}
\end{figure}

Figure \ref{fig:systematic_bias_AP_pars} concludes our analysis by illustrating the systematic bias between the contaminated and correct cases across all redshift bins. As discussed in Sect.~\ref{sec:methodology}, we quantify this uncertainty as $\alpha_{\parallel,\perp}^\mathrm{contam}-\alpha_{\parallel,\perp}^\mathrm{correct}$ (blue and red dots). The error bars represent the error on the mean, estimated by dividing the standard deviations of the \(\alpha_\parallel\) and \(\alpha_\perp\) posterior distributions by the square root of the number of mocks.
For comparison, the coloured bands represent the standard deviations of \(\alpha_\parallel\) (blue) and \(\alpha_\perp\) (yellow) in the correct galaxy case
rescaled by a factor \(\sqrt{6}\) to account for the six-fold larger area covered by the complete \Euclid survey in DR3. In all instances, the systematic bias is significantly smaller than the statistical uncertainty.

All these results are in agreement with the fact that residuals between the correct and the measured 2PCF are very small at the BAO scale, as shown in Sect.~\ref{sec:damping-shift-interlopers}.
In conclusion, we find that a minimal 2PCF model which does not include the interloper galaxies contribution does not introduce significant systematic errors in the estimate of the AP parameters when applied to contaminated catalogues, and therefore it could be safely adopted in the analysis of the \Euclid DR1 survey.

\section{\label{sc:Conclusions}Conclusions}

In this work we investigated the impact of interloper galaxies on the \Euclid 2PCF and the effect of this impact on cosmological parameter inference during the early phase of the mission. Our forecast is based on the set of 1000 EuclidLargeMocks catalogues that simulate the survey area of \Euclid DR1 and incorporate the realistic types and fractions of galaxy contaminants with incorrect measured redshifts.
We introduced the formal expression for the measured 2PCF in Eq.~\eqref{eq:contam-2pcf-RR}, which specifies the individual contributions of each type of interloper. Using this expression, 
we evaluated the relative contributions of each population to the measured 2PCF signal. 

In Sect.~\ref{sc:2PCFmeas}, 
we showed the amplitude of all the different terms compared to the measured signal. These results allowed us to quantify the relevance of each term.
The dominant contribution comes from the correct galaxy population, followed by the auto-correlation of line interlopers and the cross-correlation of the correct galaxies with noise interlopers. The relative importance of these latter contributions changes with redshift. All other contributions, including cross-correlation terms and the noise-noise auto-correlation, are negligible in the expected \Euclid spectroscopic sample. All relevant contributions exhibit a distinct scale dependence. The measured 2PCF is not merely a rescaled version of the true underlying one but instead has its own shape. As a result, the amplitude of systematic errors induced by neglecting some terms in the modelling of the measured signal shows a scale dependence as well. In general, these systematic errors tend to decrease when going towards large separations. For example, for $z \in [1.3, 1.5]$, when all interloper contributions are ignored and only the correct galaxies contribution is considered, the mismatch with the measured 2PCF is comparable to the statistical error expected for the DR1 \Euclid sample below $30\,\hMpc$. This mismatch decreases down to 20\% of the statistical error at the BAO scale and is further reduced on larger scales. We conclude that the sensitivity to interlopers is larger when the 2PCF analysis is extended to smaller scales. In all considered redshift bins, modelling the correct-noise term always lowers the systematic error below 10\% of the statistical uncertainty on the single measurement of the measured 2PCF at all scales.

In principle, these results do not rule out the possibility that it may be necessary to develop a model for the correct-noise cross-correlation and include it in the complete theoretical model for the cosmological analysis, especially if we are interested in studying the smallest non-linear scales in view of DR3.
In fact, the assessment of the impact of interlopers in the \Euclid mission will evolve as the survey progresses. As the survey area expands, sample variance and Poisson noise will decrease and the attained depth of the EDS will increase. This should lead to improved calibration and reduced systematic errors, including better control over interloper types and fractions. However, as long as we limit to DR1-like uncertainties, a modelling that only accounts for correct galaxies and line interlopers auto-correlations leads to systematic errors smaller than the expected statistical uncertainty. It should also be kept in mind that the amplitude and significance of each contribution and systematic effect depend on the considered redshift interval via the types and fractions of interlopers that we expect to have. This means that we are only able to draw precise final conclusions when the true fraction of interlopers in the \Euclid survey is determined, which will occur after comparing EWS and EDS observations.

To evaluate the impact of interlopers on the inference of cosmological parameters, we conducted two analyses comparing the measured 2PCF with a series of 2PCF models that incorporate interloper contributions with increasing complexity. In the first analysis, we used the full 2PCF shape to constrain the  clustering amplitude in redshift space, quantified by the normalized growth rate of structure $f\sigma_8$ and the $b\sigma_8$ parameter (Sect.~\ref{sc:CosmoFit}). In the second one, we focused on the BAO feature and extracted the AP parameters $\alpha_{\parallel}$ and $\alpha_{\perp}$ (Sect.~\ref{sc:BAOFit}). Both studies indicated that for \Euclid DR1, a minimal 2PCF model that accounts for the attenuation of the clustering signal, independent of the type of contaminants, is sufficient to accurately extract cosmological parameters in the presence of the expected interloper types and fractions. 

In the full-shape analysis, we adopted a simple model for the 2PCF that accounts for a fixed redshift error, a damping of the BAO, linear RSD, and a Fingers-of-God damping on small scales \citep{Addison_2019}. We tested different phenomenological models of increasing complexity in light of the results obtained in Sect.~\ref{sc:2PCFmeas}, while working at a fixed cosmology. To disentangle the systematic effects due to a specific interloper modelling from the suitability of the theoretical model chosen for the 2PCF, we always compared our results to a reference case in which only the non-contaminated part of the catalogue is used, so that the theoretical model is applied to a sample without interlopers. 
Our results showed that a simple 2PCF model, which only accounts for the attenuation of the correct galaxies clustering, is sufficient to estimate $f\sigma_8$ with 1\%–3\% accuracy with respect to the reference case, depending on the redshift. The systematic error induced by an incomplete modelling of the interloper galaxies is well below the statistical uncertainty that we expect for DR1, in all tested models. For the DR3 analysis, in which the statistical error is expected to decrease by a factor of about $\sqrt{6}$, the effect of interlopers will require a more elaborate model than a simple attenuation in the clustering amplitude.
 
In the BAO analysis, we used the 2PCF template BAO model proposed by 
Euclid Collaboration: Sarpa et al. (in prep.)
to fit the measured 2PCF. We quantified the systematic error induced by ignoring the presence of all interlopers altogether in the theoretical modelling. We used the same model to fit both the 100\% pure and contaminated samples. 
Our results showed that the BAO analysis is extremely robust to the presence of interlopers. Specifically, the estimated values for $\alpha_{\parallel}$ and $\alpha_{\perp}$ obtained when ignoring the contribution of any type and fraction of interloper are almost identical to those from the analysis of the 100\% pure sample, and the systematic bias between the two cases is significantly smaller than the statistical error expected not just for DR1 but also for DR3 spectroscopic catalogues of \Euclid. 

The results presented here 
\aabf{form the basis for a companion paper}, where the impact of interlopers on the two-point statistics is examined in Fourier space rather than in configuration space 
(Euclid Collaboration: Lee et al., in prep.).
The primary difference between the two studies lies in the power spectrum model adopted. In the companion paper, the model is based on Effective Field Theory of Large-Scale Structure (EFT-of-LSS) predictions, which are expected to better capture non-linear effects.
\aabf{ 
Although EFT models can in principle yield more precise cosmological inferences by extending the analysis to smaller scales and capturing non-linear effects,
in this work our primary goal is different. As stated at the beginning of Sect.~\ref{sc:CosmoFit} and reiterated here in the Conclusion, we aim to disentangle the impact of interloper mis-modelling from the suitability of the theoretical description adopted for the 2PCF. 
For this reason, we consistently focus on the shifts in cosmological parameters between the reference case without interlopers and the test cases with specific interloper modelling, always at a fixed theoretical model for the 2PCF. 
This strategy makes our analysis effectively model independent and allows us to rely on a simpler theoretical framework with fewer nuisance parameters to marginalize over. 
As a validation of this approach, the results obtained in both configuration and Fourier space analyses are fully consistent.
}

A final aspect that we have not considered in this work is the so-called BAO reconstruction. State-of-the-art BAO analyses are performed on reconstructed catalogues, which are obtained from a backward non-linear transformation of the redshift space positions of the observed galaxies \citep{Eisenstein_2007, Padmanabhan_2009} to remove the effect of non-linear evolution leading to the damping of the BAO feature. However, the quality of the reconstruction depends on the characteristics of the sample and, potentially, on the presence of interlopers. For this reason, we plan to further investigate this in the future: a key question is whether the BAO template model needs to be updated or it is flexible enough to capture all possible spurious effects that interlopers can introduce into the reconstruction. We plan to address this issue, along with other systematics, in a dedicated paper that analyses the first set of unblinded \Euclid data (Euclid Collaboration: Sarpa et al., in prep.).

The interloper fraction assumed in this work, which never exceeds 20\%, aligns with the expected contamination level by the end of the survey. This assumption has guided the construction of the simulated galaxy catalogues used to assess the impact of interlopers on clustering statistics.
However, it is not guaranteed that this target contamination level will be achieved by the time of the first data release. The analysis of the recently obtained Euclid Quick Data Release (Q1, \citealt{Q1-TP001}) data suggests that the contamination level is significantly higher than the one adopted in this work \citep{Q1-TP007}. While it is important to emphasize that a higher contamination level does not diminish the scientific value of the Q1 data -- whose primary focus is astrophysical rather than cosmological and does not necessarily require a high level of purity -- this finding serves as a warning that the 20\% target contamination level may not be reached in DR1. If that is the case, we will need to update our simulated catalogues and repeat the analysis presented in this paper to reassess the impact of a higher interloper fraction on the BAO and full-shape analysis of the galaxy 2PCF.

Another crucial aspect to consider is our ability to accurately estimate the contamination level and characterize the nature of the contaminants. The baseline strategy adopted in the Euclid data analysis pipeline involves estimating the sample’s purity and completeness by comparing galaxy catalogues extracted at full depth with those obtained at the depth of the wide survey, using data from the EDS (Granett et al., in prep.). Since this method depends on the availability of full-depth EDS, which will only be achieved by DR3, it may not be optimal for the first data release.
For this reason, alternative approaches are being explored. These include comparisons of redshift measurements for known sources against external, reliable datasets \citep{cosmos2020}, self-calibration techniques based on galaxy-galaxy correlations across redshift bins \citep{Peng_2023}, and clustering redshifts (d’Assignies et al., in prep.). For DR1 in particular -- where interloper quantification remains tentative -- it will also be crucial to perform statistical tests to validate our contamination modelling, following approaches such as that described in \citet{loopit}.

\begin{acknowledgements}

The authors acknowledges support from MIUR, PRIN 2022 (grant 2022NY2ZRS 001).
Simulations and computations in this work have been run at the computing facilities of INFN, Sezione di Genova: the authors wish to thank the INFN IT personnel in Genova for their precious and constant support. 
P.M. acknowledges support from Italian Research Center on High
Performance Computing Big Data and Quantum Computing (ICSC), by the
Fondazione ICSC National Recovery and Resilience Plan (PNRR) Project ID
CN-00000013 and by the PRIN 2022 PNRR project (code no. P202259YAF)
funded by ``European Union – Next Generation EU'', Mission 4, Component
1, CUP J53D23019100001. We acknowledge usage of Pleiadi system of INAF
\citep{Taffoni2020,Bertocco2020}.

\AckEC  
\end{acknowledgements}

\bibliographystyle{aa}
\bibliography{mybib}

\begin{appendix}
  \onecolumn 
  
\section{Scale dependence of the 2PCF prefactors}
\label{sec:AppendixPrefactors}

Equation \eqref{eq:contam-2pcf-RR} is exact when measuring all 2PCFs through the LS estimator. 
All terms in the equation are two-dimensional quantities depending on both the separation scale $r$ and the cosine of the angle with respect to the line of sight $\mu$. 
This holds for the prefactors in front of each term as well, since the pair numbers in the random catalogues depend on the window function and on the redshift distribution of each galaxy population. However, we expect that at first order those prefactors are linked to the fractions of interlopers in the catalogue. This comes naturally from Eq.~\eqref{eq:density-contrast-with-interlopers}, which says that the total density contrast is the weighted sum of the density contrasts of the single populations: the more abundant a population is, the greater its contribution to the total signal. The crucial point is that the fractions in Eq.~\eqref{eq:density-contrast-with-interlopers} depend on the window function of the corresponding galaxy population.

Contrary to the correlation function, the scale dependence of the prefactors cannot be modelled from first principles but requires a specific knowledge of the survey geometry and selection function. It is therefore crucial to understand if this scale dependence is actually significant or if it can be ignored when modelling the terms on the right hand side of Eq.~\eqref{eq:contam-2pcf-RR}.

\begin{figure*}[h]
    \centering
    \includegraphics[width=\textwidth]{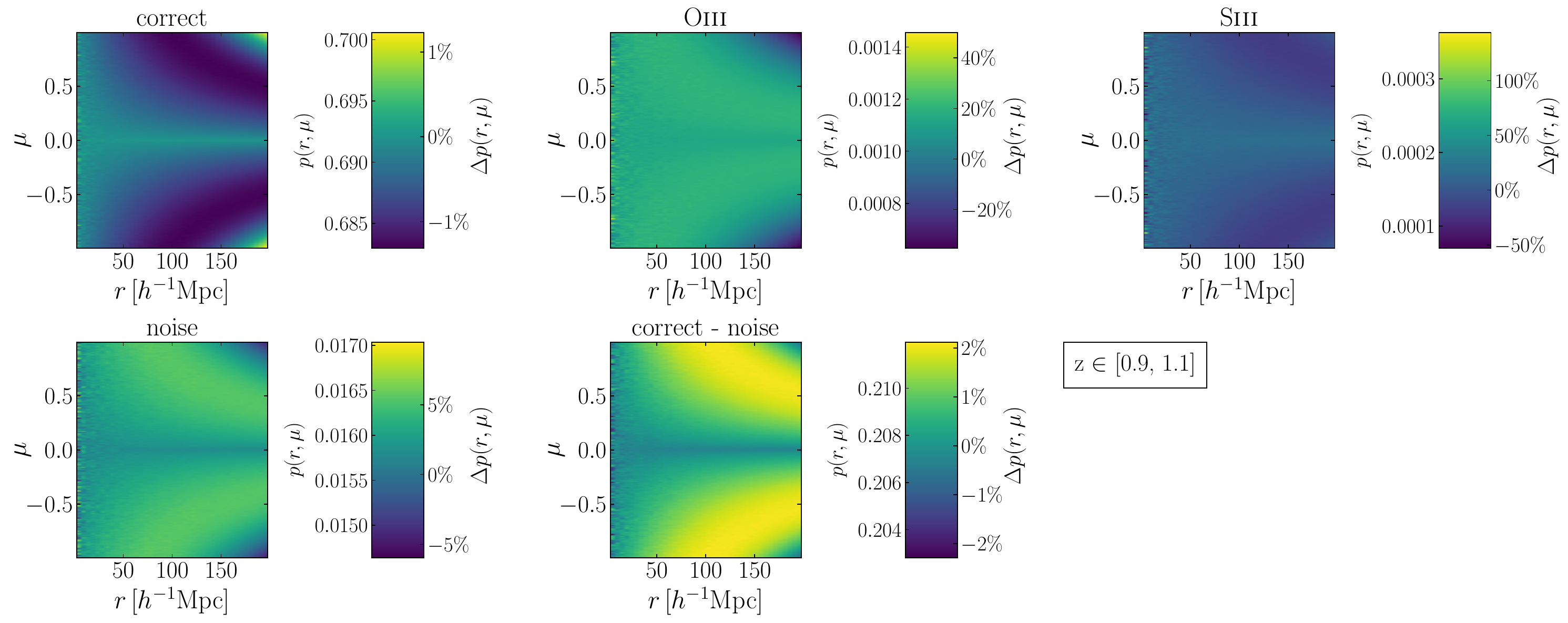}
    \caption{Two-dimensional prefactors $p(r, \mu)$ (\textit{left} side of the colour bar scale) and percent difference $\Delta p (r, \mu)$ (\textit{right} side) for a subset of the total terms in Eq.~\eqref{eq:contam-2pcf-RR} for $z \in [0.9, 1.1]$.}
    \label{fig:prefactors-2D-z1}
\end{figure*}

Figure \ref{fig:prefactors-2D-z1} shows the two-dimensional prefactors $p(r, \mu)$ for the correct galaxies, all populations of interlopers, and the correct-noise cross-correlation term in the low-redshift bin (see the \textit{left} scale in the colour bar). Both an angular and radial dependence arise that are different for each term.  To quantify the relevance of this scale dependence, we compare each constant prefactor (dubbed $c_{i}$, where the subscript identifies the various prefactors) appearing in Eq.~\eqref{eq:contam-2pcf-const}, with the corresponding exact value $p_i$ derived from the pair count ratios. We show the percent difference between the two, defined as
\begin{equation}
\label{eq:percent-diff}
    \Delta p_{i} = \frac{p_i - c_i}{c_i} \times 100 \,.
\end{equation}

The \textit{right} scale of the colour bar in Fig.~\ref{fig:prefactors-2D-z1} shows the two-dimensional percent difference $\Delta p(r, \mu)$ of the showed two-dimensional prefactors. We see that, whereas the correct galaxy prefactor shows a mild sub-percent difference with respect to the constant prefactor, there are some prefactors (like the \ion{O}{iii} one) which show a tens of percent variation across the measured $r$ and $\mu$ interval.

\begin{figure*}[h]
    \centering
    \includegraphics[width=\textwidth]{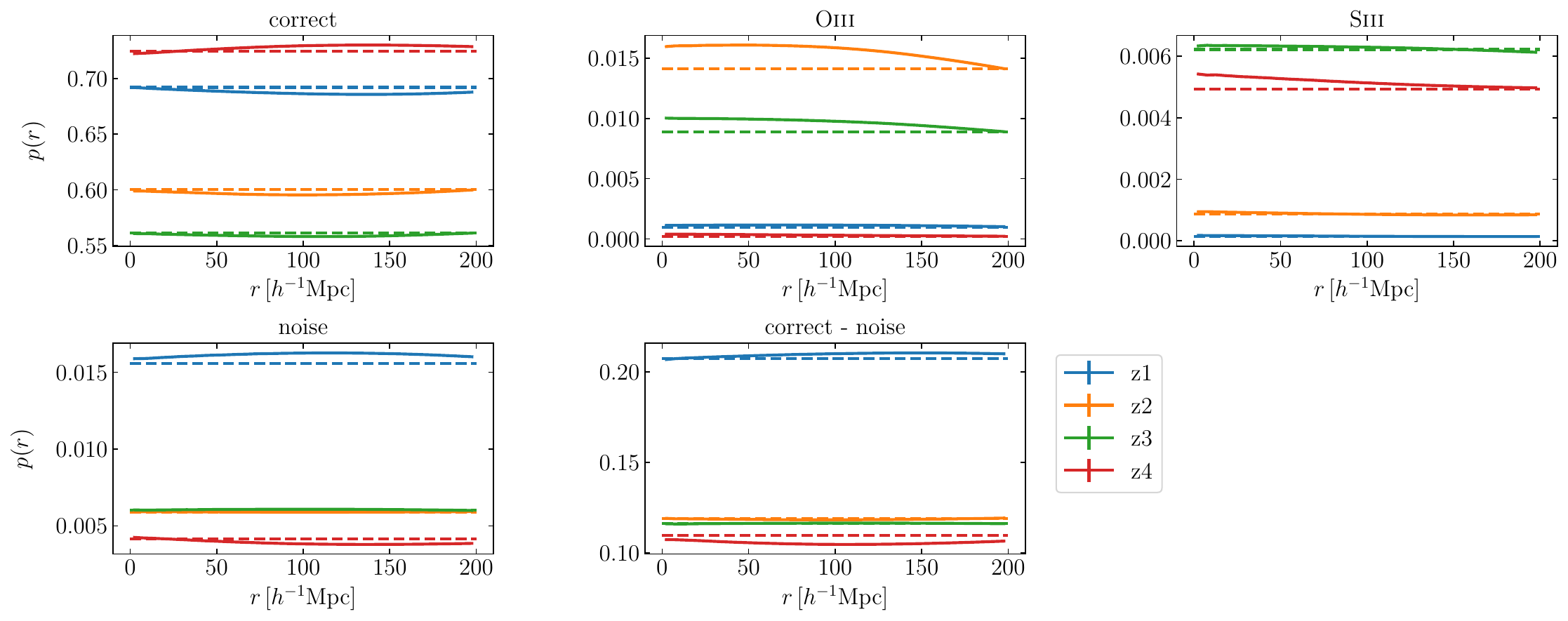}
    \caption{One-dimensional $r$-dependent prefactors $p(r)$ for a subset of terms in Eq.~\eqref{eq:contam-2pcf-RR}, compared to their constant counterpart (dashed lines) for all redshift bins (different colours).}
    \label{fig:prefactors-allz}
\end{figure*}

\begin{figure*}[h]
    \centering
    \includegraphics[width=\textwidth]{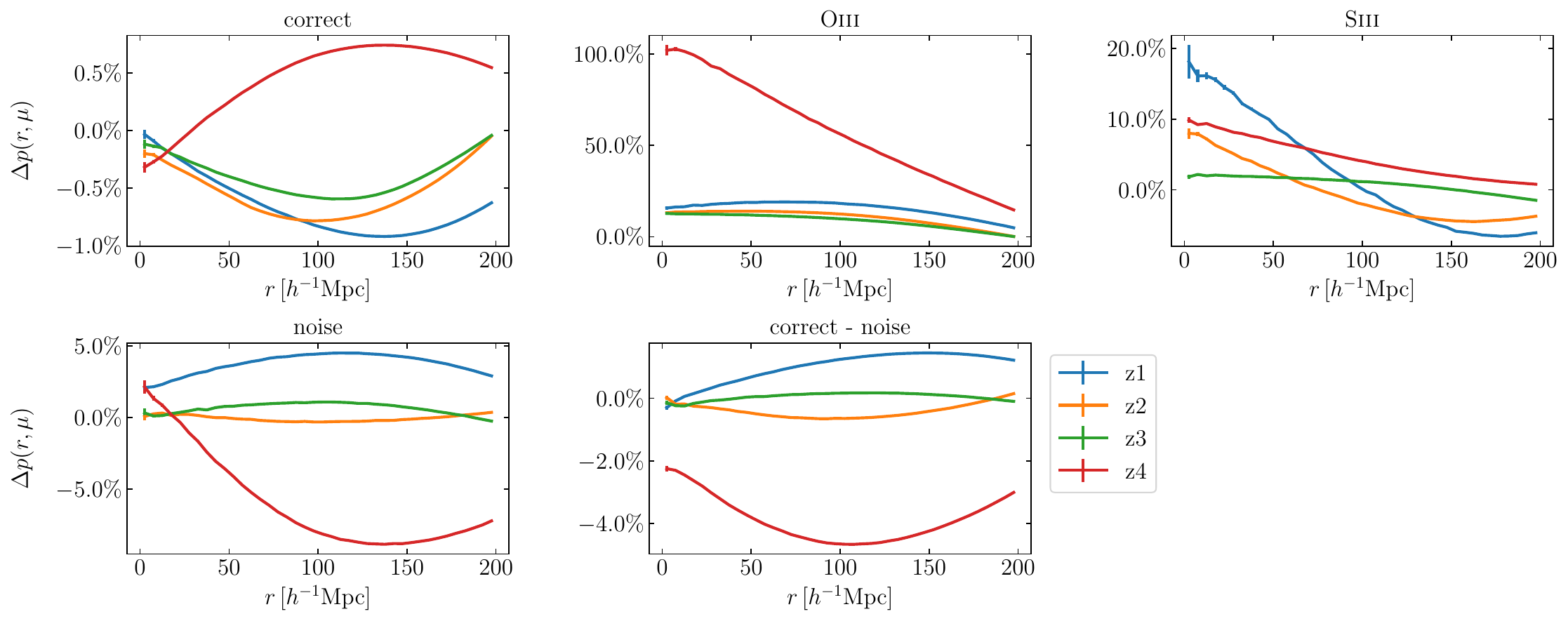}
    \caption{One-dimensional $r$-dependent percent difference $\Delta p(r)$ between scale-dependent and constant prefactors for a subset of terms in Eq.~\eqref{eq:contam-2pcf-RR}, for all redshift bins (different colours).}
    \label{fig:prefactors-percdiff-allz}
\end{figure*}

To allow an easier visualization and interpretation of these results, it is useful to look at the same quantities just shown in one single dimension, i.e. considering the scale dependence on the separation $r$ only. Figure \ref{fig:prefactors-allz} shows the $r$-dependent version $p(r)$ of the prefactors in Fig.~\ref{fig:prefactors-2D-z1}, i.e. the two-dimensional prefactors averaged over $\mu$, compared to their constant counterpart (dashed lines) for all redshift bins. 

Two features are worth noticing. First of all, the curves are systematically offset from the dashed lines. Offsets can be positive or negative, but in all cases they largely exceed the Poisson error derived from the pair counting (not even visible in the plot). 
The reason for the mismatch is that number counts used to estimate these prefactors only consider pairs within the maximum separation used to compute the 2PCF, in this case 200 $\hMpc$, ignoring objects separated by larger distances.
This systematic offset is therefore an artefact that could be eliminated  by considering all the pair counts in the sample, which, however, would be too expensive from a computational point of view. 

The second feature is that, while in some cases the coloured curves are reasonably constant, for some samples they are not and, instead, exhibit a very significant $r$-dependence. To make this statement more quantitative, we plot 
in Fig.~\ref{fig:prefactors-percdiff-allz}
the percent difference  
$\Delta p(r)$ of the prefactors with respect to the corresponding constant value, i.e. Eq.~\eqref{eq:percent-diff} considering $p(r)$ instead of $p(r, \mu)$. The results in all redshift bins are shown together. 
The percent difference shown on the $y$ axis quantifies the mismatch between the fraction of objects estimated in whole sample and those actually used in estimating the counts, as previously discussed. 
The maximum percent variation across the $r$ interval, on the other hand, quantifies the magnitude of the $r$-dependence and the prefactors' genuine departure from a constant behaviour. Here we can see much more clearly that, depending on the redshift bin and on the term under analysis, the scale dependence can be significant, largely exceeding the $1\%$ level. 

The specific spatial dependencies for each prefactor arise from the different redshift distributions $N(z)$ of the various objects in the contaminated sample. Objects in all random catalogues  have a uniform angular distribution, but different redshift distributions matching those of the corresponding object type. 
As a result, the random pair counts will feature a dependence on both $r$ and $\mu$ that is different for all the random samples.
This $r$ and $\mu$ dependence in the counts of the various random catalogue propagates into the 2PCF prefactors and originate the departure from the constant prefactor model, depending on the survey window function and on the redshift distribution of the sources.

Ultimately, the scale dependence of the prefactors was not significant for our purposes. In modelling the measured 2PCF in Sect.~\ref{sec:MCMC-models}, we use both the exact prefactor and a constant approximation in front of the correct correlation. The latter is employed when fitting the unknown contamination fraction of the sample, which is reasonable given the correct prefactor's very mild scale dependence across all redshift bins. For line interlopers, we choose a conservative approach by using the exact prefactors in front of the line correlations. Although the line interloper prefactors exhibit a stronger scale dependence (up to tens of percent), their absolute value remains small due to the low fraction of these interlopers compared to correct galaxies. Therefore, while scale dependence exists, it does not significantly affect the measured 2PCF at DR1-like precision.

\section{Minimum separation scale for the amplitude parameters fit}
\label{sec:AppendixSmallScales}

As evidence that our theoretical model for the power spectrum is unable to yield realistic results when fitting smallest separation scales, we compare the outcomes obtained by considering a minimum scale of $r_{\rm min} = 20 \,\hMpc$ and $r_{\rm min} = 40 \,\hMpc$ in the fit. 

\begin{figure}[!htbp]
    \centering
    \includegraphics[width=0.49\textwidth]{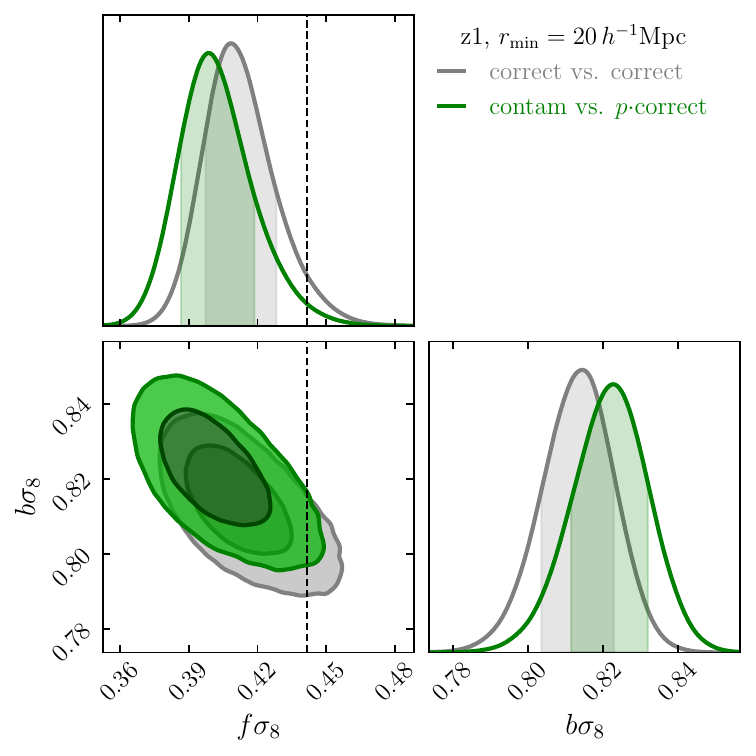}
    \hfill
    \includegraphics[width=0.49\textwidth]{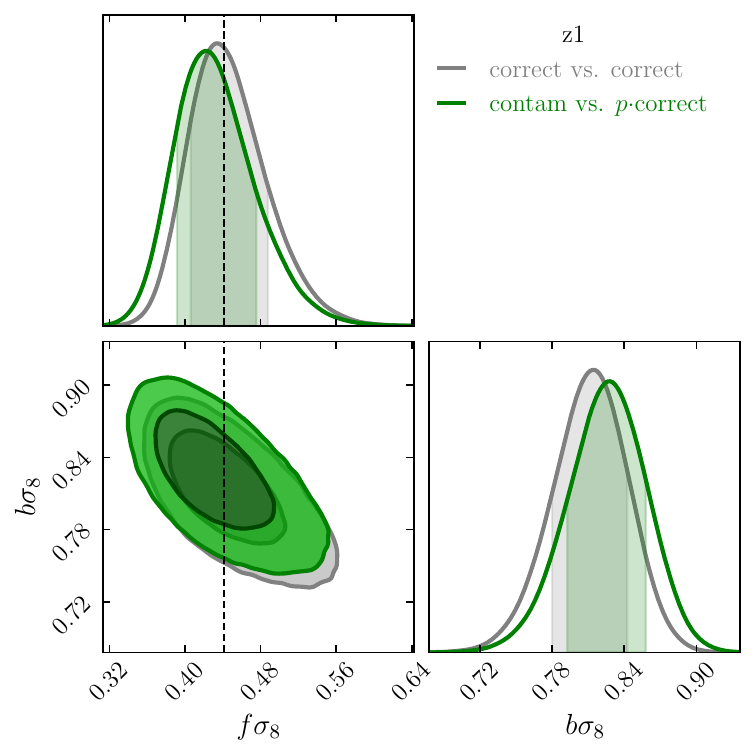}
    \medskip
    \includegraphics[width=0.49\textwidth]{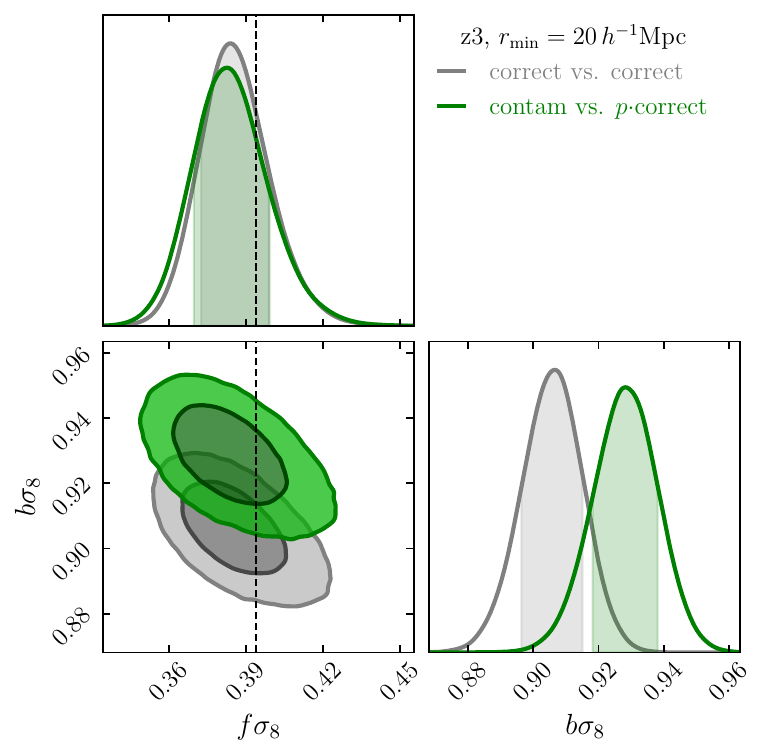}
    \hfill
    \includegraphics[width=0.49\textwidth]{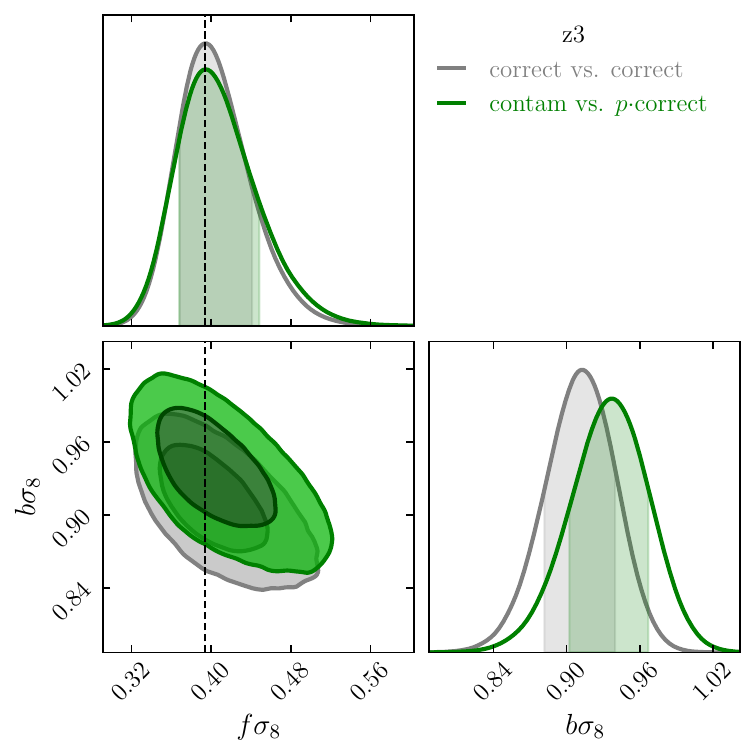}
    \caption{Contour plot of the correct galaxy parameters for the reference case (grey contour) and the minimal correct-only model with the exact prefactor (green contour) in $z \in [1.3, 1.5]$, running chains with $r_{\rm min} = 20 \, h^{-1}\,$Mpc (\textit{left}) and $r_{\rm min} = 40 \, h^{-1}\,$Mpc (\textit{right}).}
    \label{fig:contour-z1-z3-diff-ref-r20-r40}
\end{figure}
Figure \ref{fig:contour-z1-z3-diff-ref-r20-r40} shows the contour plots in z1 and z3 for the reference case and for the minimal correct-only model with the exact prefactor for the correct galaxies contribution. 
When we include scales up to a minimum separation scale of $20\, \hMpc$ in the fit, the bias between the inferred value of $f\sigma_8$ and its true value derived from cosmology in z1 exceeds the statistical uncertainty with which we expect to determine $f\sigma_8$ in DR1. 
Conversely, this bias is smaller than the uncertainty we foresee for DR1 when considering a minimum separation scale of $40 \,\hMpc$ in the fit (same figure, \textit{right} panels), independently on the redshift bin.
Considering smaller scales lowers the statistical uncertainty, since we are adding signal both in the monopole and quadrupole, and therefore we become more sensitive to systematic effects related to the choice of the power spectrum model.
This demonstrates that the simple power spectrum model we have decided to adopt in this analysis (see Sect.~\ref{sec:pk-model-graeme}) cannot be considered reliable when dealing with scales smaller than $40 \,\hMpc$, even in a context comparable to DR1 and in absence of contamination (grey contours).

The bias between the returned values in the reference case and the theoretical ones quantifies the inadequacy of the chosen model to represent the power spectrum. On the other hand, the bias between the results of various tests we performed to fit the contaminated signal, with or without considering certain types of interlopers, and those obtained in the reference case quantifies what we aim to determine: the impact of more or less accurate modelling of interlopers on the estimation of cosmological parameters at different redshifts. This bias graphically corresponds to the shift between the grey contours (or dots, depending on the plot) and the coloured ones, and it is analysed in Sect.~\ref{sc:CosmoFitResults}.  

\section{The correct-noise cross-correlation in the EuclidLargeMocks}
\label{sec:AppendixTargetNoise}

\begin{figure}
    \centering
    \includegraphics[width=\linewidth]{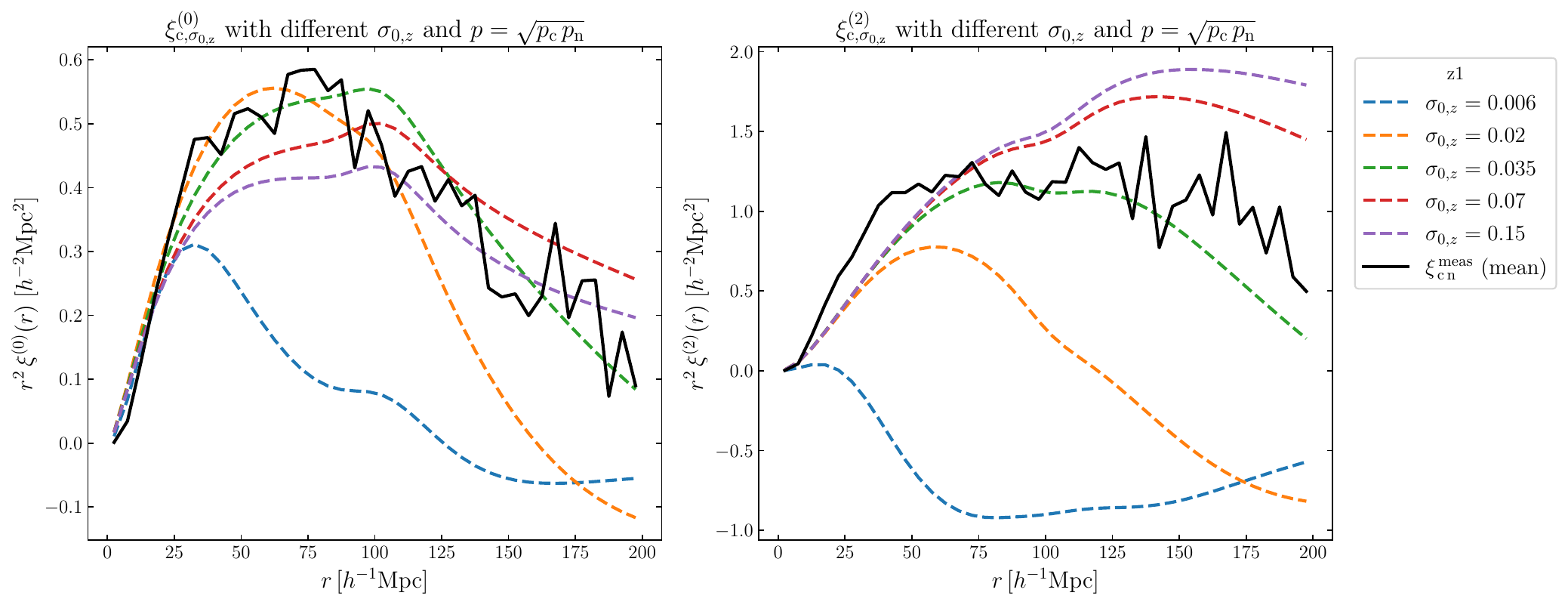}
    \caption{Monopole (\textit{left}) and quadrupole (\textit{right}) of the correct-noise cross-correlation measurements and models in $z \in [0.9, 1.1]$. The solid black line represents the average of the $\xi_{\rm cn}$ measurements over the mock catalogues. Dashed lines correspond to different models for the signal (see Eq.~\ref{eq:large-z-err-tn}), each colour representing a different value for the redshift error in the 2PCF model.}
    \label{fig:z1-large-z-errors}
    \label{LastPage}
\end{figure}

The separation of correct galaxies and noise interlopers based on $|z_{\rm meas} - z_{\rm true}|$ is affected by the intrinsic ambiguity due to the fact that galaxies without emission lines can acquire a roughly correct redshift by chance. This induces some level of correct-noise cross-correlation. In our mocks, this is emphasized by the assumed cut at $3\,\sigma_{0,z} = 0.003$ of the redshift residuals when selecting correct galaxies.
However, looking at Fig.~\ref{fig:ztrue-vs-zmeas-2D}, we can appreciate that noise interlopers have a tendency to cluster around the correct galaxies and line interlopers lines (light blue shaded area around the straight lines) and to correlate with them up to distances of many $\sigma_{0,z}$. This has an impact on the correct-noise cross-correlation, which is contributed by all those noise interlopers which where artificially shifted along the line of sight by $\Delta r < 200\,\hMpc$ (the maximum scale we consider when computing the 2PCF). In the EuclidLargeMocks, these sources make up the 10\%--20\% of noise contaminants, depending on the redshift bin under analysis. To prove that our interpretation is correct,
we attempted to model the cross-correlation signal between these two populations assuming that a fraction 
of the noise interlopers is made by correct galaxies with a larger (Gaussian) redshift error:
\begin{align}
\label{eq:large-z-err-tn}
    \xi_{\rm cn} \simeq (1-f_{\rm tot}) f_{\rm n} \, \langle \delta_{\rm c} \delta_{\rm n} \rangle \equiv (1-f_{\rm tot}) f_{\rm n} \, \xi_{\rm c, \sigma_{0, z}} \, .
\end{align}
We tested various values for the Gaussian redshift error $\sigma_{0,z}$.
Figure \ref{fig:z1-large-z-errors} shows the results of this test. Each coloured curve represents the results of the fit using the correct-only model (see Sect.~\ref{sec:MCMC-models}) with a different redshift error standard deviation value. In this case, the parameterisation in Eq.~\eqref{eq:large-z-err-tn} appears to work particularly well for an effective $\sigma_{0,z} = 0.035$, a value $35$ times higher than the instrumental error expected for \Euclid. If we consider as an example a noise interloper originally located at $z_{\rm true}=1$, whose redshift was mistaken by one effective $\sigma_{0,z}$ so that $z_{\rm meas}=1.035$, the corresponding shift along the line of sight is  $\Delta r \approx 57 \,\hMpc < 200 \,\hMpc$. One may wonder if, in view of final \Euclid results, an explicit modelling of these cross terms would not be necessary at some point. However, because a separation of the two classes of galaxies is not possible in the EWS, and because the effect of noise interlopers is significant only when the galaxies are close to their true redshift (noise interlopers with large redshift errors do not correlate any more with correct galaxies), this modelling can be absorbed in the modelling of the tails of the $P(z_{\rm meas}|z_{\rm true})$ PDF. We know from simulations and preliminary measurements that the shape of this PDF shows heavy tails, which would further increase the number of correct galaxies misinterpreted as noise interlopers with a $3\,\sigma_{0,z}$ criterion. Therefore, the treatment of the tails of the redshift random error PDF is clearly a point to be further investigated.

\end{appendix}

\end{document}